\documentclass[12pt]{article}
\usepackage{amsfonts,amsthm,amsmath,amssymb,upgreek,bm}
\usepackage{hyperref}
\usepackage[paper=letterpaper,margin=.85in]{geometry}
\usepackage{graphicx}
\usepackage{units}
\usepackage{float}
\usepackage[export]{adjustbox}
\usepackage[dvipsnames]{xcolor}
\usepackage[shortlabels]{enumitem}
\usepackage[mathscr]{euscript}
\usepackage[normalem]{ulem}
\usepackage{bm}
\usepackage{tikz}
\usepackage{tikzsymbols}
\usepackage{pgfplots}
\usepackage{ytableau}
\usepackage{slashed}
\usepackage{mathtools}
\usepackage{bbold}
\usetikzlibrary{decorations.pathmorphing}
\usetikzlibrary{arrows.meta}
\usetikzlibrary{calc}
\usepgfplotslibrary{fillbetween}
\usepackage[nameinlink]{cleveref}
\pgfplotsset{compat=1.18}

\usetikzlibrary{decorations.pathmorphing,intersections}
\usetikzlibrary{arrows.meta,decorations.markings}

\theoremstyle{definition}

\newcommand{\lr}[1]{\langle #1 \rangle}

\renewcommand{\Im}{\operatorname{Im}}

\newcommand{\be}{\begin{equation}}
\newcommand{\ee}{\end{equation}}
\newcommand{\bea}{\begin{eqnarray}}
\newcommand{\eea}{\end{eqnarray}}

\renewcommand{\i}{\mathrm{i}}
\renewcommand{\d}{\mathrm{d}}

\numberwithin{equation}{section}

\def\tr{\text{Tr}}
\def\Tr{\text{Tr}}

\begin{document}
\thispagestyle{empty}

\vspace*{2.5cm}
\begin{center}

{\bf {\LARGE Negative shocks versus static patch holography}}

\begin{center}

\vspace{1cm}

{Yiming Chen${}^{1,2}$, Douglas Stanford${}^1$, Haifeng Tang${}^1$, and Zhenbin Yang${}^{3}$}\\
 \bigskip \rm

\bigskip 

{${}^1$  Leinweber Institute for Theoretical Physics, Stanford University, Stanford, CA 94305, USA}
\,\vspace{0.5cm}\\
{${}^2$  Department of Physics, Cornell University, Ithaca, NY 14853, USA}
\,\vspace{0.5cm}\\
{${}^3$ Institute for Advanced Study, Tsinghua University, Beijing 100084, China}

\rm
  \end{center}

\vspace{2.5cm}
{\bf Abstract}
\end{center}
\begin{quotation}
\noindent

We study a version of de Sitter static patch holography in which the Euclidean gravitational path integral, with an observer worldline included, is conjectured to compute a trace. Motivated by recent evidence for this conjecture from the sphere path integral, we test it further by inserting operators along the observer worldline and computing two-point functions and out-of-time-ordered four-point correlators (OTOCs). We extend an earlier OTOC calculation by Kolchmeyer and Liu using the shockwave formalism, incorporating both observer recoil and gravitational backreaction. We find that the OTOC conflicts with two basic properties of a trace in a Hilbert space: cyclicity and positivity. The signaling feature of the shockwave geometry gives rise to two distinct resummations of the perturbative eikonal expansion, which are interchanged by cyclicity. Positivity is violated by the fact that the leading perturbative contribution causes the (regularized) OTOC to increase.

\end{quotation}

\setcounter{page}{0}
\setcounter{tocdepth}{2}
\setcounter{footnote}{0}

\newpage

\setcounter{page}{2}
\tableofcontents

\newpage

\section{Introduction}
De Sitter space is arguably the simplest model of cosmology. It is homogeneous and isotropic, and appears static to local observers. In Euclidean signature, it becomes a round sphere $S^d$. Throughout this paper we set its radius to one.

If an observer worldline is included, the isometry group $SO(d,1)$ is spontaneously broken to the subgroup $ SO(d-1) \times \mathbb{R}$ that preserves the so-called ``static patch.'' The static patch contains the observer at its center, surrounded by a cosmic horizon. In four dimensions, the horizon area is given by $A=4\pi -8\pi G m + O(G^2m^2)$, where $m$ is the mass of the observer measured with respect to their own proper time. In many respects, the de Sitter horizon resembles the black hole horizon: the observer can throw particles towards the horizon and measure Hawking radiation. At least for certain experiments, the results will be compatible with modeling the rest of the system as a chaotic thermodynamic system \cite{gibbons1977cosmological,Susskind:2011ap}, and toy models along these lines have been studied in \cite{Banks:2003cg,Parikh:2004wh,Banks:2006rx,Anninos:2011af,Susskind:2021esx,Narovlansky:2023lfz}.\footnote{See~\cite{Alishahiha:2004md,Anninos:2017hhn,Coleman:2021nor,Banihashemi:2026mje,Collier:2025lux} for some other approaches to de Sitter holography.}

In \cite{Chandrasekaran:2022cip}, it was shown how to define a von Neumann algebra generated by bulk operators dressed to a de Sitter observer's clock, together with the clock energy. The background independence of this  operator product algebra was emphasized in \cite{Witten:2023xze}, which also proposed that the Hartle-Hawking state is a maximum entropy tracial state for that algebra. In \cite{Maldacena:2024spf}, it was shown how to incorporate the observer in the Euclidean gravity path integral formalism, as an unstable saddle point with a massive particle moving on a great circle. 

An appealing extrapolation from these results is the conjecture to be stated shortly. Unfortunately, the conclusion of our paper is that a rather weak version of this conjecture is not correct. We nevertheless find it useful to review some checks that the conjecture passes, and to explore its failure in detail. Perhaps a modified version can be rescued in future work.

\par\medskip
\phantomsection\label{conj:static-patch-holography}
\noindent\textbf{Static patch holography conjecture:} Correlation functions along an observer worldline, as computed by the Euclidean gravitational path integral, are equal to a trace,
\par
 \begin{equation}\label{eq:necklace}
  \begin{tikzpicture}[scale=1, baseline=(current bounding box.center)]
  \def\R{1.35}
  \def\Ry{0.4}
  % round sphere (disk cross-section with shading)
  \shade[inner color=gray!0, outer color=gray!20] (0,0) circle (\R);
  % great circle on the sphere (vertical): back arc dashed, front arc solid
  \draw[thick, gray!55, dashed] (90:{\Ry} and {\R}) arc (90:270:{\Ry} and {\R});
  \draw[thick] (90:{\Ry} and {\R}) arc (90:-90:{\Ry} and {\R});
  % observer necklace on the great circle
  \fill (90:{\Ry} and {\R}) circle (0.08);
  \node[above=1pt] at (90:{\Ry} and {\R}) {$\varphi_{2}$};
  \fill (0:{\Ry} and {\R}) circle (0.08);
  \node[right=5pt] at (0:{\Ry} and {\R}) {$\varphi_{1}$};
  \fill (-90:{\Ry} and {\R}) circle (0.08);
  \node[below=1pt] at (-90:{\Ry} and {\R}) {$\varphi_{n}$};
  %\fill[gray] (180:{\Ry} and {\R}) circle (0.08);
  %\node[left=5pt, gray] at (180:{\Ry} and {\R}) {$\mathcal{O}_{n-1}$};
  \node[anchor=south west] at (30:{\Ry} and {\R}) [xshift=-2pt, yshift=-2pt] {$m_2$};
  \node[anchor=south west] at (-55:{\Ry} and {\R}) [xshift=2pt, yshift=2pt] {$m_1$};
  \node[anchor=north east, gray] at (-135:{\Ry} and {\R}) [xshift=-2pt, yshift=12pt] {$m_n$};
  \node[anchor=south east, gray] at (155:{\Ry} and {\R}) [xshift=0pt, yshift=-14pt] {$\cdots$};
  \end{tikzpicture}
  \hspace{8pt}\stackrel{?}{=}\hspace{5pt}
\Tr\left(\mathcal{O}_n\Pi_{m_n}...\mathcal{O}_1\Pi_{m_1}\right).
\end{equation}
On the LHS, we have a Euclidean gravity problem specified by the observer's worldline with fixed energy and a sequence of operator insertions. For positive cosmological constant, the sphere topology shown will be the leading contribution and the only one we will consider in this paper. On the RHS, one could imagine conjectures of different strengths. We will find a problem with a rather weak version where the RHS is only assumed to be cyclic $\tr(ab) = \tr(ba)$ and positive $\tr(a^\dagger a) \ge 0$, together with an analyticity assumption that is automatic if for finite $G$ the trace is over a Hilbert space of finite dimension $e^{S_{\rm dS}}$.\footnote{We also assume that the map $\varphi \to \mathcal{O}$ preserves Hermiticity. Later we will use the same symbol for $\varphi$ and $\mathcal{O}$.} At least perturbatively, cyclicity is baked into the expression on the LHS. Positivity is more obviously subtle, because of the conformal mode problem in Euclidean quantum gravity \cite{Gibbons:1978ac}.

\medskip

This conjecture can be tested against calculations in a recent paper by Kolchmeyer and Liu \cite{Kolchmeyer:2024fly}, who studied out-of-time-order correlators (OTOCs) in a non-gravitational ``recoil'' limit that will be described in detail below. Our interpretation of their result is that a combination of cyclicity and positivity is violated. The purpose of our paper is to study this violation further, and to generalize the computation to include gravitational effects.

OTOCs are useful in this context because they are sensitive to an important difference between the de Sitter horizon and black hole horizons\footnote{It might seem ill-motivated to compute OTOCs in de Sitter, given that the most straightforward way to measure them involves reversing the direction of time evolution -- something that is out of the question for an observer in cosmology. But whether or not such correlation functions can actually be measured by an observer in de Sitter, they are part of the mathematical conjecture (\ref{eq:necklace}).}
\cite{Anninos:2018svg,Verheijden2018,Aalsma:2020aib,Narovlansky:2025tpb}. This difference has to do with the time delay associated to positive-energy shock waves:
\begin{equation}
\begin{tikzpicture}[
    scale=1.35,
    line cap=round,
    line join=round,
    >=stealth,
    baseline=(current bounding box.center)
]
    % Force the visible bounding box so the two side-by-side figures align
    \path[use as bounding box] (-1.28,-1.42) rectangle (1.28,1.22);

    % Auxiliary paths for intersections.
    % The option overlay prevents these invisible paths from changing the bounding box.
    \path[overlay, name path=topcurve]
        (-1.05,1.05) .. controls (-0.35,0.82) and (0.35,0.82) .. (1.05,1.05);

    \path[overlay, name path=bottomcurve]
        (-1.05,-1.05) .. controls (-0.35,-0.82) and (0.35,-0.82) .. (1.05,-1.05);

    % 45-degree ray going up-left from (.5,.5)
    \path[overlay, name path=blueray]
        (.5,.5) -- ++(-2,2);

    % Intersection of the ray with the top curve
    \path[overlay, name intersections={of=topcurve and blueray, by=topblue}];

    % Left and right edges
    \draw[thick] (-1.05,-1.05) -- (-1.05,1.05);
    \draw[thick] (1.05,1.05) -- (1.05,-1.05);

    % Top and bottom wiggly edges
    \draw[
        thick,
        decorate,
        decoration={snake, amplitude=1.2pt, segment length=7pt}
    ]
        (-1.05,1.05) .. controls (-0.35,0.82) and (0.35,0.82) .. (1.05,1.05);

    \draw[
        thick,
        decorate,
        decoration={snake, amplitude=1.2pt, segment length=7pt}
    ]
        (-1.05,-1.05) .. controls (-0.35,-0.82) and (0.35,-0.82) .. (1.05,-1.05);

    % Diagonals
    \draw[gray!55, thick] (-1.05,-1.05) -- (1.05,1.05);
    \draw[gray!55, thick] (-1.05,1.05) -- (1.05,-1.05);

    % Red horizon line
    \draw[red, line width=2.4pt] (-1.05,-1.05) -- (1.05,1.05);

    % Blue lines
    \draw[blue, line width=1.8pt] (topblue) -- (.5,.5);
    \draw[blue, line width=1.8pt] (0.2,0.2) -- (1.05,-0.65);

    \node[below=4pt, fill=white, inner sep=1.4pt] at (0,-1.05)
        {Black hole horizon};
\end{tikzpicture}
\qquad\qquad
\begin{tikzpicture}[
    scale=1.35,
    line cap=round,
    line join=round,
    >=stealth,
    baseline=(current bounding box.center)
]
    % Same forced bounding box as the first figure
    \path[use as bounding box] (-1.28,-1.42) rectangle (1.28,1.22);

    % Square
    \draw[thick] (-1.05,-1.05) rectangle (1.05,1.05);

    % Diagonals
    \draw[gray!55, thick] (-1.05,-1.05) -- (1.05,1.05);
    \draw[gray!55, thick] (-1.05,1.05) -- (1.05,-1.05);

    % Red horizon line
    \draw[red, line width=2.4pt] (-1.05,-1.05) -- (1.05,1.05);

    % Blue lines
    \draw[blue, line width=1.8pt] (0.2,0.2) -- (1.05,-0.65);
    \draw[blue, line width=1.8pt] (-.1,-.1) -- (-1.05,.85);

    \node[below=4pt, fill=white, inner sep=1.4pt] at (0,-1.05)
        {Cosmic horizon};
\end{tikzpicture}
\label{eq:Gao-Wald}
\end{equation}
At left we show the time delay of an infalling particle (blue) after crossing some matter on the horizon (red). For black hole horizons, the positive time delay is protected by the averaged null energy condition \cite{Morris:1988tu,Friedman_1995,Hochberg:1998ii}  and ensures that signals sent from the right boundary cannot reach the left. This is essential for the ER $=$ EPR interpretation of the two-sided black hole \cite{Maldacena:2001kr,Maldacena:2013xja}. At right, we show the time advance associated to a homogeneous shock wave on the de Sitter horizon.\footnote{The same effect is also sometimes described as de Sitter space getting ``taller". See \cite{Leblond:2002ns,Aalsma:2021kle,Batra:2024qju,Banihashemi:2026mje} for some further discussion of this effect.} Unlike the time delay, which tends to decrease OTOCs, this advance leads to an increase. We will show that this increase conflicts with the conjecture (\ref{conj:static-patch-holography}).

\medskip

The plan for the paper is as follows. In the rest of the introduction, we review how including the observer fixes a positivity problem with the sphere partition function. We also discuss the two point function along the observer's worldline, for which there does not appear to be any problem with cyclicity and positivity. 

In \hyperref[sec:OTOC]{\textbf{section two}}, we compute the OTOC in the eikonal approximation. In order to set this up, we discuss shock waves in de Sitter with the observer included. A nice technical result is that the gravitational shock wave includes both the effect of gravitational backreaction and observer recoil. This is possible because the equation for the shock wave in pure de Sitter has pure gauge $\ell = 1$ zero modes that are lifted by the presence of the observer. These become physical modes that describe observer recoil. This means that gravitational backreaction and observer recoil can be studied on the same footing, and it is not necessary to distinguish their effects.

Using the shock wave solution, one can then compute the eikonal OTOCs using the procedure in \cite{Shenker:2014cwa}, and we carry this out in detail for the case of dS$_3$. Compared to OTOCs in black holes or ordinary large $N$ thermal systems, there are two (related) differences. First, the sign of the first perturbative correction is such that the regularized OTOC increases above its disconnected early time value \cite{Aalsma:2020aib,Kolchmeyer:2024fly}. Second, there are two different resummations of the asymptotic perturbation series (denoted $\mathcal{F}_{12}$ and $\mathcal{F}_{14}$ below) and these are exchanged by cyclicity.

We proceed to study the issue of cyclicity in more detail in the pure recoil limit \cite{Kolchmeyer:2024fly} with $G = 0$, where semiclassically the OTOC is captured by saddle points associated to tetrahedron graphs on $S^d$. Here the issue of cyclicity is associated to a Stokes phenomenon where the integration contours for $\mathcal{F}_{12}$ and $\mathcal{F}_{14}$ pick up the tetrahedron saddle point when the masses are in different regimes. This clarifies how the naive cyclicity property of the Euclidean problem can be broken by a choice of integration contour.

In \hyperref[sec:chaosbound]{\textbf{section three}} we emphasize that the perturbative OTOC computation of the previous section is incompatible with positivity. This conclusion is reached using the argument of the chaos bound, which constrains the sign of the first exponentially growing correction in OTOCs. 

In \hyperref[app:static_patch_jt_gravity]{\textbf{appendix A}}, we discuss a toy model of static patch holography, the static patch de Sitter JT gravity. This model reduces to an integral over immersions of a disk on the round $S^2$, with an action given by minus the length of the boundary. We couple this model to a 2d BCFT, and study the perturbative gravitational contribution to the four point function. We also establish that in the eikonal approximation, the OTOC in this model is given by the usual AdS JT gravity OTOC, with a negative coupling constant $C\to -C$.

In \hyperref[app:recoil in higher d]{\textbf{appendix E}} we generalize the eikonal pure recoil computation in \cite{Kolchmeyer:2024fly} to dS$_d$. 

The other appendices contain technical details or extensions.

In the rest of the introduction, we review two examples that are encouraging for the static patch holography conjecture.

\medskip

\textbf{Note added:} While this work was in progress, we became aware of several other groups \cite{Cui_Kolchmeyer_ToAppear,Harlow_Ying_ToAppear,Milekhin_Narovlansky_Xu_ToAppear} working on closely related topics. We've coordinated to submit our papers on the same day. 

 \subsection{Sphere partition function}
\label{sec: sphere partition function}
The sphere partition function with an observer was computed in \cite{Maldacena:2024spf}. This can be regarded as a test of (\ref{eq:necklace}) for the case where no operators are inserted along the observer's worldline:
\be
Z_{\text{sphere}}(m)\equiv
\begin{tikzpicture}[scale=1, baseline=(current bounding box.center)]
  \def\R{1.35}
  \def\Ry{0.4}
  % round sphere (disk cross-section with shading)
  \shade[inner color=gray!0, outer color=gray!20] (0,0) circle (\R);
  % great circle on the sphere: back arc dashed, front arc solid
  \draw[thick, gray!55, dashed] (90:{\Ry} and {\R}) arc (90:270:{\Ry} and {\R});
  \draw[thick] (90:{\Ry} and {\R}) arc (90:-90:{\Ry} and {\R});
  % observer necklace on the great circle
  \node[anchor=south east] at (135:{\R} and {\Ry}) [xshift=35pt, yshift=-10pt] {$m$};
  \end{tikzpicture} \hspace{10pt} \stackrel{?}{=} \hspace{5pt}
  \Tr (\Pi_m).
  \label{fig: Z sphere (m)}
\ee
The saddle point for the Euclidean gravity path integral on the LHS consists of a particle worldline with fixed mass $m$ wrapping around a great circle of $S^d$. A constraint from the RHS is that the answer should be positive. This is nontrivial because the Euclidean gravity path integral without the observer has a dimension-dependent phase \cite{Polchinski:1988ua}, due to the need to Wick-rotate most but not all of the conformal modes. The observer introduces new unstable modes that also require Wick rotation, and the final one-loop determinant turns out to be real \cite{Maldacena:2024spf} and positive (at least for the integration order discussed in \cite{Chen:2025jqm}):
\be\label{eqn:sphere}
Z_{\text{sphere}}(m)\approx (\text{positive})\cdot m^{d-1}\cdot e^{S_{\rm dS}-2\pi m}.
\ee
An important aspect of this computation is that the length of the observer's worldline $\beta$ is integrated over an inverse Laplace transform contour, parallel to the imaginary axis, with measure $\frac{\d\beta}{2\pi\i}$. Note that $Z_{\text{sphere}}(m)$ grows as $m$ becomes smaller, and for some purposes it is more natural to define the energy $E$ to be $-m$ and interpret the sphere partition function as a spectral density:
\be
 \rho(E)\equiv Z_{\rm sphere}(-E).
\ee
Here $E = -m$ can be understood as the energy of everything {\it except} the observer.

In \cite{Yang:2025lme}, a connected contribution to $\langle \rho(E)\rho(E')\rangle$ was computed from a de Sitter double cone wormhole with topology $S^1\times S^{d-1}$. This contribution is consistent with spectral rigidity for the microscopic spectrum of $E$.

\newcommand{\setSpherePoint}[3]{
  \pgfmathsetmacro{\spherePointX}{\sphereR*((cos(#2)*\axisdepth-sin(#2)*cos(#1)*\equatordepth)*cos(\axisangle)
   -sin(#2)*sin(#1)*sin(\axisangle))}
  \pgfmathsetmacro{\spherePointY}{\sphereR*((cos(#2)*\axisdepth-sin(#2)*cos(#1)*\equatordepth)*sin(\axisangle)
   +sin(#2)*sin(#1)*cos(\axisangle))}
  \coordinate (#3) at (\spherePointX,\spherePointY);
}
\newcommand{\drawSphereMeridian}[2]{%
  \pgfmathsetmacro{\horizonAngle}{atan2(\equatordepth,-\axisdepth*cos(#1))}
  \draw[#2, dashed]
    plot[domain=0:\horizonAngle, samples=50, variable=\a]
    ({\sphereR*((cos(\a)*\axisdepth-sin(\a)*cos(#1)*\equatordepth)*cos(\axisangle)
     -sin(\a)*sin(#1)*sin(\axisangle))},
     {\sphereR*((cos(\a)*\axisdepth-sin(\a)*cos(#1)*\equatordepth)*sin(\axisangle)
     +sin(\a)*sin(#1)*cos(\axisangle))});
  \draw[#2]
    plot[domain=\horizonAngle:180, samples=50, variable=\a]
    ({\sphereR*((cos(\a)*\axisdepth-sin(\a)*cos(#1)*\equatordepth)*cos(\axisangle)
     -sin(\a)*sin(#1)*sin(\axisangle))},
     {\sphereR*((cos(\a)*\axisdepth-sin(\a)*cos(#1)*\equatordepth)*sin(\axisangle)
     +sin(\a)*sin(#1)*cos(\axisangle))});
}
\subsection{Two point function}\label{sec:2pt} 
Next, we discuss the observer's two point function, in the simplifying limit where Newton's constant $G$ is zero and the observer mass $m$ is large. Let $W$ be a Hermitian field operator for a free field of mass 
\begin{equation} 
m_W^2 = \nu^2 + \left(\tfrac{d-1}{2}\right)^2,
\label{eq: m in terms of nu}
\end{equation}
and let $\bm{\mathcal{G}}(m_1,m_2)$ be the sphere partition function with an observer and two $W$ insertions:
\begin{equation}\label{twoptdefenergy}
\bm{\mathcal{G}}(m_1,m_2) \equiv \hspace{10pt}\begin{tikzpicture}[scale=1, line cap=round, line join=round, baseline=(current bounding box.center)]
  \def\R{1.35}
  \def\Ry{0.4}
  \shade[inner color=gray!0, outer color=gray!20] (0,0) circle (\R);
  \draw[thick, gray!55, dashed] (90:{\Ry} and {\R}) arc (90:270:{\Ry} and {\R});
  \draw[thick] (90:{\Ry} and {\R}) arc (90:-90:{\Ry} and {\R});
  \fill[blue] (-15:{\Ry} and {\R}) circle (0.07) node[right] {${\color{blue} W}$};
  \fill[blue!50] (160.3:{\Ry} and {\R}) circle (0.07) node[left] {$W$};
  \node[right] at (35:{\Ry} and {\R}) {$m_2$};
  \node[left = 3pt] at (-40:{\Ry} and {\R}) {$m_1$};
\end{tikzpicture}\hspace{10pt} \stackrel{?}{=} \tr\!\left({\color{blue} W}\Pi_{m_2}{\color{blue} W}\Pi_{m_1}\right).\hspace{5pt} 
\end{equation}
This is defined as a gravity path integral, with the lengths of the observer segments $\beta_1,\beta_2$ integrated over inverse Laplace transform contours. 

Before trying to compute anything, it will be useful to think about what types of time-domain correlators one might build from this function. To this end, consider the dependence on the average mass $m = \frac{m_1 + m_2}{2}$. Essentially, this comes from the behavior of the sphere partition function with the observer, $\propto e^{-2\pi m}$. This should be understood as an entropic effect: the density of states decreases exponentially with $m$, or equivalently grows exponentially with $E = -m$. In other words, we have a Hagedorn spectrum with a cutoff $m > m_0$ associated to the ground state energy of the observer. Because of this cutoff, it makes sense to consider infinite temperature (tracial) correlators without any need for Boltzmann factors. These are the correlators discussed in \cite{Chandrasekaran:2022cip,Witten:2023xze}: 
\begin{equation}\label{Gtracialdef}
\bm{\mathcal{G}}_{\rm tracial}(\i t)\equiv \int_{m_0}^\infty \d m_1\d m_2 e^{\i (m_1-m_2)t}\bm{\mathcal{G}}(m_1,m_2).
\end{equation}
Because of the exponential pressure $e^{-2\pi m}$ from the density of states, the integrals will be dominated near the $m_0$ endpoint, rather than a saddle point. The resulting correlator does not approximate the QFT answer in de Sitter space.

Another option is to include Boltzmann factors $e^{-\beta_i E_i} = e^{+ \beta_i m_i}$ with $\beta_1 + \beta_2 = 2\pi$ to cancel the pressure from the density of states. This doesn't quite lead to a good ensemble, because the average mass $m$ would be a zero mode. But we can fix the average mass by hand, defining\footnote{Our notation here and in the rest of the paper is that bold quantities contain the sphere partition function. Un-bold quantities have it divided out.}
\begin{align}\label{GGrel}
\mathcal{G}(\theta|m) &\equiv \frac{e^{-2\pi m}}{Z_{\text{sphere}}(m)}\int \d m_1\d m_2\,\delta\!\left(m-\tfrac{m_1+m_2}{2}\right)e^{\theta m_1 + (2\pi-\theta) m_2}\bm{\mathcal{G}}(m_1,m_2).
\end{align}
One can think about this as a problem where $\beta_1 + \beta_2$ is integrated over an inverse Laplace transform contour in order to set $\frac{m_1 + m_2}{2} = m$, while $\beta_1 - \beta_2$ is fixed to $2(\theta-\pi)$. There is a saddle point where $\beta_1 + \beta_2 = 2\pi$, so the observer follows a great circle, with operator insertions separated by $\theta$ along this circle. In our approximation of small $G$ and large $m$, the saddle point is sharply peaked, and so the function $\mathcal{G}(\theta|m)$ will just reduce to the QFT two point function for this configuration of points, times the sphere partition function. We divided out by the sphere partition function in the definition, so
\begin{equation}
\mathcal{G}(\theta|m) = G_\nu(\theta)
\end{equation}
where $G_\nu(\theta)$ is the free-field propagator on the sphere. This can be inserted into (\ref{GGrel}) in order to determine $\bm{\mathcal{G}}(m_1,m_2)$: 
\begin{align}\label{GGmm}
\bm{\mathcal{G}}(m_1,m_2) &= Z_{\rm sphere}(\tfrac{m_1+m_2}{2})\int_{-\infty}^\infty \frac{\d t}{2\pi} e^{\i t(m_2-m_1)}G_\nu(\pi + \i t).
\end{align}
And we can plug this into (\ref{Gtracialdef}) to compute the tracial correlator,
\begin{align}\label{Git1stline}
\bm{\mathcal{G}}_{\rm tracial}(\i t)
&=Z_{\rm sphere}(m_0)\int_{m_0 + |\omega|/2} \d m \int \d \omega e^{\i\omega t}e^{-2\pi(m-m_0)}\int_{-\infty}^\infty \frac{\d t'}{2\pi}e^{-\i \omega t'}G_\nu(\pi + \i t')\\
&=Z_{\rm sphere}(m_0) \int  \frac{\d\omega\d t'}{(2\pi)^2} e^{-\pi|\omega| + \i\omega(t-t')}G_\nu(\pi + \i t')\\
&= Z_{\rm sphere}(m_0)\int \frac{\d t'}{2\pi} \frac{G_\nu(\pi + \i t')}{\pi^2+ (t-t')^2}.\label{tracialans}
\end{align}
In the second line, we changed variables to 
\begin{equation}
m_1 = m + \tfrac{\omega}{2}, \hspace{20pt} m_2 = m - \tfrac{\omega}{2}.
\end{equation}
Because $Z_{\rm sphere}(m)$ exponentially prefers lower observer energy, $m$ will be close to $m_0$ and we approximated $Z_{\rm sphere}(m) \approx Z_{\rm sphere}(m_0) e^{-2\pi(m-m_0)}$. In the third and fourth lines we simply did the $m$ and $\omega$ integrals.\footnote{Equation \eqref{tracialans} shows that the tracial two point function obeys the KMS condition with zero Euclidean period $\beta=0$, since $\bm{\mathcal{G}}_{\rm tracial}(\i t)=\bm{\mathcal{G}}_{\rm tracial}(-\i t)$.
Moreover, since $G_{\nu}(\pi +\i t')$ is holomorphic in the strip $|\Im(t')|\leq \pi$, $\bm{\mathcal{G}}_{\rm tracial}(\i t)$ has an extended domain of holomorphicity on a strip that is \textit{twice} the width of $G_{\nu}(\pi +\i t')$: it is holomorphic in the region of $|\Im(t)|\leq 2\pi$. 
Strictly speaking, the operator $e^{+\tau m}$ is unbounded since $m$ is not bounded from above for $G = 0$. The extended region of holomorphicity exists thanks to a combination of entropy and matrix element effects -- both $Z_{\rm sphere}$ and the $\Gamma$ function factor in (\ref{GGmmsecondline}) decay exponentially $e^{-\pi m_1}$ as $m_1\to\infty$ with $m_2$ fixed. The decay of $Z_{\rm sphere}$ is due to the decreasing entropy, and the decay of the $\Gamma$ functions is due to decreasing matrix elements. The width of this strip defines a notion of a fake inverse temperature.
In the de Sitter case, we see that the fake inverse temperature is twice the inverse temperature seen by the QFT correlators, $\beta_{\text{fake}}=4\pi.$} This reproduces the answer from \cite{Chandrasekaran:2022cip,Witten:2023xze}. Note that as promised, the answer does not approximate $G_\nu(\i t)$, but rather a version that is smeared out in time.\footnote{As pointed out in \cite{Witten:2023xze, Kolchmeyer:2024fly}, one can get correlators that approximately agree with the QFT answers if $m_1$ is fixed to a large value $m_* \gg m_0$. In other words, the observer was assumed to be in a highly excited initial state. The effect of this is that the energy difference $\omega = m_1 - m_2$ is almost unconstrained, so the answer approximates the QFT one. One can generalize this by fixing a weighted average of the observer energies. In particular, if we impose $m_* = x m_1 + (1-x) m_2$ then one finds $G_\nu(2\pi x + \i t)$, so the effective position along the Euclidean circle depends on the linear combination of energies that is fixed. In particular, fixing $m_1$ gives a one-sided correlator, and fixing $m_1 + m_2$ gives a two-sided correlator.}

So far, we have not used the explicit formula for the free-field propagator on the sphere. The explicit formula and the Fourier transform (\ref{GGmm}) are
\begin{align}
G_\nu(\theta) &= \frac{\Gamma(\Delta_+)\Gamma(\Delta_-)}{(4\pi)^{d/2}\Gamma(d/2)}{}_2F_1\left(\Delta_+, \Delta_-,\frac{d}{2},\frac{1+\cos(\theta)}{2}\right), \ \ \ \ \text{where } \Delta_\pm = \frac{d-1}{2}\pm\i \nu\\
\bm{\mathcal{G}}(m_1,m_2)&=Z_{\rm sphere}(\tfrac{m_1+m_2}{2})\frac{\Gamma\!\left(\frac{d-1}{4} \pm\i \frac{ \nu}{2} \pm\i\frac{m_1-m_2}{2}\right)}{16\pi^{\frac{d+3}{2}}\Gamma(\frac{d-1}{2})}
.\label{GGmmsecondline}
\end{align}
Here and below, a Gamma function with $\pm$ signs denotes the product over all independent sign choices. Note that the answer has the positivity $\bm{\mathcal{G}}(m_1,m_2)\ge 0$ and cyclicity $\bm{\mathcal{G}}(m_1,m_2) = \bm{\mathcal{G}}(m_2,m_1)$ properties of the RHS of (\ref{twoptdefenergy}).\footnote{A candidate formula for $\bm{\mathcal{G}}(m_1,m_2)$ at finite mass incorporating recoil effect will be recorded elsewhere~\cite{chen_hartman_stanford_tang_wip}. It appears that positivity and cyclicity continue to hold.}

\section{Four-point OTOC}\label{sec:OTOC}
\subsection{Setup}\label{sec:otocsetup}
We now discuss the setup for the OTOC four point function
\begin{equation}\label{otocdefenergy}
\bm{\mathcal{F}}(m_1,m_2,m_3,m_4) \equiv \hspace{10pt} \begin{tikzpicture}[scale=1, line cap=round, line join=round, baseline=(current bounding box.center)]
  \def\R{1.35}
  \def\Ry{0.4}
  \shade[inner color=gray!0, outer color=gray!20] (0,0) circle (\R);
  \draw[thick, gray!55, dashed] (90:{\Ry} and {\R}) arc (90:270:{\Ry} and {\R});
  \draw[thick] (90:{\Ry} and {\R}) arc (90:-90:{\Ry} and {\R});
  \fill[blue!50] (140:{\Ry} and {\R}) circle (0.08) node[left] {$ W$};
  \fill[blue] (-40:{\Ry} and {\R}) circle (0.08) node[left] {${\color{blue} W}$};
  \fill[red] (20:{\Ry} and {\R}) circle (0.08) node[left] {${\color{red} V}$};
  \fill[red!50] (-160:{\Ry} and {\R}) circle (0.08) node[left] {$ V$};
  \node[anchor=east] at (173.3:{\Ry} and {\R}) [xshift=-2pt] {${\color{black!50}m_1}$};
  \node[anchor=east] at (-141.5:{\Ry} and {\R}) [xshift=2pt, yshift=-4pt] {${\color{black!50}m_2}$};
  \node[right] at (34.3:{\Ry} and {\R}) [xshift=0pt, yshift=4pt] {$m_4$};
  \node[right] at (-3.3:{\Ry} and {\R}) [xshift=2pt, yshift=-4pt] {$m_3$};
\end{tikzpicture} \hspace{10pt} \stackrel{?}{=} \tr\!\left({\color{blue} W}\Pi_{m_4}{\color{red} V}\Pi_{m_3}
{\color{blue} W}\Pi_{m_2}{\color{red} V}\Pi_{m_1}\right).\hspace{5pt} 
\end{equation}
We will work in the eikonal approximation where the mass of the observer $m$ is large and the gravitional coupling $G$ is small, {\bf but} we will retain corrections that grow exponentially after transforming to Lorentzian time, i.e.~powers of $G e^{\pm t}$ and $(1/m)e^{\pm t}$. In practice, this means that we can ignore most types of fluctuations in the geometry and observer trajectory, but we will have to keep track of the shock wave modes that lead to the exponentially growing effects. In particular, we will be able to ignore fluctuations in the total length of the observer trajectory.

As in the case of the two point function, it will be convenient to do the bulk calculations in a mixed representation, where the overall mass of the observer is held fixed, and operators are inserted at positions $\theta_i$ along the worldline:\footnote{One can also consider an infinite temperature tracial version of the OTOC, see appendix \ref{app: tracial}.}
\begin{align}\label{FFrel}
\mathcal{F}({\color{red}\theta_1},{\color{blue}\theta_2}, {\color{red}\theta_3},{\color{blue}\theta_4}|m)
&\hspace{5pt}=\hspace{5pt}\frac{1}{Z_{\text{sphere}}(m)}\hspace{5pt}\times \hspace{5pt}
\begin{tikzpicture}[scale=1, line cap=round, line join=round, baseline=(current bounding box.center)]
  \def\R{1.35}
  \def\Ry{0.4}
  \shade[inner color=gray!0, outer color=gray!20] (0,0) circle (\R);
  \draw[thick, gray!55, dashed] (90:{\Ry} and {\R}) arc (90:270:{\Ry} and {\R});
  \draw[thick] (90:{\Ry} and {\R}) arc (90:-90:{\Ry} and {\R});
  \fill[blue!50] (140:{\Ry} and {\R}) circle (0.08) node[left] {$ \theta_4$};
  \fill[blue] (-40:{\Ry} and {\R}) circle (0.08) node[left] {${\color{blue} \theta_2}$};
  \fill[red] (20:{\Ry} and {\R}) circle (0.08) node[left] {${\color{red} \theta_3}$};
  \fill[red!50] (-160:{\Ry} and {\R}) circle (0.08) node[left] {$ \theta_1$};
\end{tikzpicture}\\
&= \frac{e^{-2\pi m}}{Z_{\text{sphere}}(m)}\int \d^4 m_i \delta(m - \tfrac{1}{4}\textstyle\sum_i m_i)e^{(\theta_{i}-\theta_{i-1})m_i}\bm{\mathcal{F}}(m_i).
\end{align}
Here $\theta_0 \equiv \theta_4 - \beta(m)$ where $\beta(m)$ is the total length of the observer's worldline. In dS$_3$, $\beta(m) = 2\pi$ even including backreaction from the observer worldline. For $d > 3$, $\beta(m) = 2\pi - O(Gm)$. 

Within the approximation described above, the worldline of the observer will be a piecewise geodesic trajectory on a shock wave geometry sourced by the $W$ and $V$ quanta. In the next section, we will present the relevant shock wave metric. In the following section, we will use it to compute the OTOC.

\subsection{The shock wave solution}\label{sec:shockwave}
The key ingredient in the eikonal computation of the OTOC is a shock wave sourced by a particle with large null momentum moving on the de Sitter horizon. Such shock waves were studied in \cite{Hotta:1992qy,Sfetsos:1994xa,Aalsma:2020aib}. Compared to black holes, de Sitter shock waves have some unusual features. We will review this and then generalize the computation to include the effect of the observer. 

The metric ansatz for a shock wave on a Killing horizon is
\begin{equation}\label{shockans}
\d s^2 = -A(x^+x^-) \d x^+ \d x^- + B(x^+ x^-) \d s_Y^2 + A(0)X^+(y) \delta(x^-) (\d x^-)^2.
\end{equation}
The $x^\pm$ coordinates are Kruskal coordinates for the geometry without the shock wave. Time translations of the static patch act as boosts $x^\pm \to \lambda^{\pm 1} x^\pm$. The meaning of the funny-looking third term in (\ref{shockans}) is that we have glued together two copies of the original geometry across the $x^- =0$ horizon, with a null shift $x^+ \to x^+ + X^+(y)$:
\begin{equation}
\begin{tikzpicture}[scale=1.35, line cap=round, line join=round, >=stealth, baseline=(current bounding box.center)]
    \draw[thick] (-1.05,-1.05) rectangle (1.05,1.05);
    \draw[gray!55, thick] (-1.05,-1.05) -- (1.05,1.05);
    \draw[gray!55, thick] (-1.05,1.05) -- (1.05,-1.05);
    \draw[red, line width=2.4pt] (-1.05,-1.05) -- (1.05,1.05);
    \draw[blue, line width=1.8pt] (0,1) -- (.5,.5);
    \draw[blue, line width=1.8pt] (0.2,0.2) -- (1.05,-0.65);
    \node[below=4pt, fill=white, inner sep=1.4pt] at (0,-1.05) {$X^+(y)>0$};
\end{tikzpicture}
\qquad \qquad
\begin{tikzpicture}[scale=1.35, line cap=round, line join=round, >=stealth, baseline=(current bounding box.center)]
    \draw[thick] (-1.05,-1.05) rectangle (1.05,1.05);
    \draw[gray!55, thick] (-1.05,-1.05) -- (1.05,1.05);
    \draw[gray!55, thick] (-1.05,1.05) -- (1.05,-1.05);
    \draw[red, line width=2.4pt] (-1.05,-1.05) -- (1.05,1.05);
    \draw[blue, line width=1.8pt] (0.2,0.2) -- (1.05,-0.65);
    \draw[blue, line width=1.8pt] (-.1,-.1) -- (-1.05,.85);
    \node[below=4pt, fill=white, inner sep=1.4pt] at (0,-1.05) {$X^+(y)<0$};
\end{tikzpicture}
\label{eq:shock-wave-gluing}
\end{equation}
Here the red line is the shock wave along the $x^- = 0$ horizon, and the blue line is a continuous null trajectory. The ``$--$'' component of the Ricci tensor for the metric (\ref{shockans}) is
\begin{equation}\label{eq:shock-ricci-component}
    R_{--} = \delta(x^-)\left[-\frac{A(0)}{2B(0)}\nabla_Y^2  - \frac{d-2}{2}\frac{B'(0)}{B(0)}\right]X^+(y).
\end{equation}
Assuming we start with a solution to Einstein's equations {\it without} the shock wave, then with the shock wave present the metric will be a solution if 
\begin{equation}
R_{--} = 8\pi G T_{--}.
\end{equation}

For the case of dS$_3$ with $d = 3$,\footnote{See appendix \ref{app:shock_waves_in_higher_dimensions} for general $d$.} the metric including an observer is
\begin{equation}\label{dSobserver}
\d s^2 = -\frac{4\d x^- \d x^+}{(1- x^- x^+)^2} + \frac{\alpha^2}{(2\pi)^2}\left(\frac{1+x^- x^+}{1-x^-x^+}\right)^2\d\phi^2,
\qquad \phi\sim \phi+2\pi, \hspace{20pt} \alpha = 2\pi(1 - 4Gm).
\end{equation}
The shock wave ansatz is
\begin{equation}
\d s^2 = -\frac{4\d x^- \d x^+}{(1- x^- x^+)^2} + \frac{\alpha^2}{(2\pi)^2}\left(\frac{1+x^- x^+}{1-x^-x^+}\right)^2\d\phi^2 + 4X^+(\phi)\delta(x^-)(\d x^-)^2.
\end{equation}
Assuming a stress tensor localized along the horizon $T_{--} = T_-(\phi)\delta(x^-)$, the equation $R_{--} = 8\pi G T_{--}$ becomes
\begin{equation}\label{diffop}
2\left[-\frac{(2\pi)^2}{\alpha^2}\partial_\phi^2 -1\right]X^+(\phi) = 8\pi G T_-(\phi).
\end{equation}

Let's begin with the case of pure de Sitter space $\alpha = 2\pi$. In that case the differential operator on the LHS is $2(\ell^2-1)$, where $\ell$ is the integer angular momentum along the spatial $S^1$. For shock waves on a  black hole horizon, the analogous operator would be positive. For large $\ell$, our operator is also positive, but there are two exceptions for small $\ell$. For $\ell = 0$, the operator is negative. Note that $T_-(\phi) \ge 0$ by the the averaged null-energy condition, so this means the average over $\phi$ of $X^+(\phi)$ will be negative, corresponding to a time advance.

There is another exception at $\ell = 1$, where the differential operator vanishes. This is because the corresponding shock wave is actually pure gauge -- it glues together the two halves of de Sitter space with the action of an isometry in between (we will discuss the specific isometry in more detail below). Correspondingly, the $\ell = 1$ component of $T_-(\phi)$ is required to vanish by the gravitational constraint associated to this isometry. This makes the equation mathematically consistent, but it sounds a little puzzling. A baseball flying across the horizon in a particular direction would have nonzero $\ell = 1$ component of $T_-(\phi)$. Does this mean that observers in de Sitter cannot throw baseballs? 

Obviously not, but to resolve the puzzle, we have to take into account the recoil of whoever threw the baseball. In our context, this means keeping track of the gravitational field of the observer, which breaks the isometry and removes the gravitational constraint. In terms of the shock wave solution, for $\alpha < 2\pi$ the differential operator in (\ref{diffop}) no longer has any zero modes, so we can solve the equation for any $T_-(\phi)$. In particular, for the case of a single baseball aimed at $\phi = 0$ 
\begin{equation}
T_{-} = -p_-\frac{2\pi}{\alpha}\delta(\phi), \hspace{20pt} p_- < 0,
\end{equation}
the solution is
\begin{equation}
X^+(\phi) = (-p_-)h(\phi), \hspace{20pt} h(\phi) = -\frac{2\pi G}{\sin(\alpha/2)}\cos\!\left[\frac{\alpha}{2\pi}(\pi-|\phi|)\right],
\qquad -\pi \le \phi \le \pi.\label{shocksolution}
\end{equation}
\begin{figure}[t]
\centering
\includegraphics[width=.4\linewidth]{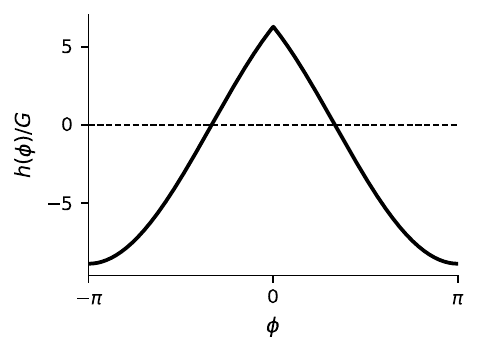}
\caption{Shown is a plot of $h(\phi)/G$ for the case $\alpha = 3\pi/2$. Note that for $\alpha < \pi$, $h(\phi)$ would be negative everywhere. In higher dimensions, the shock wave profile is positive (and divergent) at small impact parameter, although its average value is negative.}\label{hplot}
\end{figure}See figure \ref{hplot} for a plot. To understand what happened to the $\ell = 1$ zero modes, we can take the limit $G m \ll 1$, meaning $\alpha \to 2\pi$. In this limit, 
\begin{equation}\label{recoillight}
    h(\phi) =\frac{1}{2m}\cos(\phi) -2G(\pi-|\phi|)\sin|\phi|+O(G^2m).
\end{equation}
The first term is in the $\ell = 1$ subspace, and in appendix \ref{app:shock_waves_in_higher_dimensions} we show that this precisely captures the recoil of the observer. Our conclusion is that the shock wave incorporates both recoil and gravitational backreaction.

\subsection{Eikonal OTOC in static patch dS JT gravity}\label{sec:OTOCJT}
In the previous section, we saw that the shock wave in de Sitter has the opposite sign compared to black holes (at least if we average over impact parameter). One could model this in JT gravity by artificially flipping a sign in the shock wave action. Somewhat remarkably, there is an interesting system where this is actually the right answer for the OTOC, namely the static patch dS JT gravity theory discussed in appendix \ref{app:static_patch_jt_gravity}. This system does not contain the observer recoil effect associated to the $\ell = 1$ modes or the positive $\ell \ge 2$ modes, but it is a good model of the $\ell = 0$ mode of the gravitational shock wave in higher dimensions. The formulas are simple, and there is a clear comparison to the standard black hole case, so we will study this model as preparation before returning to dS$_3$ in the next section.

\subsubsection{Ordinary JT}
Let's start with the eikonal formula for the OTOC in ordinary JT gravity, see (6.56) of \cite{Maldacena:2016upp}. After adapting to our notation, this is
\begin{equation}\label{eqn:JTOTOC}
\begin{aligned}
\mathcal{F}_{\rm JT}({\color{red}\theta_1},{\color{blue}\theta_2},{\color{red}\theta_3},{\color{blue}\theta_4})&=
\int_{-\infty}^\infty \d X^+ \d X^- e^{-2\i C X^+ X^-}
\left[\frac{1}{2\sin\frac{\theta_3-\theta_1}{2} - \i e^{-\i(\theta_1+\theta_3)/2}X^-}\right]^{2\Delta}
\\
&\hspace{.7cm}\times
\left[\frac{1}{2\sin\frac{\theta_4-\theta_2}{2}-\i e^{\i(\theta_2+\theta_4)/2}X^+}\right]^{2\Delta}.
\end{aligned}
\end{equation}
The integral over $X^\pm$ is the eikonal approximation to the full path integral of gravity (see appendix C of \cite{Stanford:2021bhl}). The $X^\pm$ variables parametrize the strength of shock waves on the two horizons. The gravitational action is the $e^{-2\i C X^+ X^-}$ term, and the two expressions in brackets are the $WW$ and $VV$ two point functions on this background. The formulas are a bit simpler if we specialize to the configuration
\begin{align}
\mathcal{F}_{\rm JT}(t) &\equiv \mathcal{F}_{\rm JT}(-\pi + \i t,-\tfrac{\pi}{2},\i t,\tfrac{\pi}{2})\\ 
\label{orig}
&=
\int_{-\infty}^{\infty}\d X^-\d X^+\,e^{-2\i C X^+X^-}
\left[\frac{1}{2+e^t X^-}\right]^{2\Delta}
\left[\frac{1}{2-\i X^+}\right]^{2\Delta}.
\end{align}
The integrand has a branch point singularity along the real $X^-$ axis, corresponding to the lightcone singularity in the $VV$ two point function that would be present for a sufficiently strong {\bf negative} shock wave. Of course, negative shocks are part of the off-shell gravity integral. However, in order for JT gravity to preserve the causal separation between the two sides of the black hole, they must exactly cancel out of the final answer. We can see how this works by doing the $X^+$ integral first. This integral is proportional to $\Theta(CX^-)$, so if $C > 0$ as in the case of ordinary JT gravity, the $X^-$ integral gets restricted to positive shocks. 

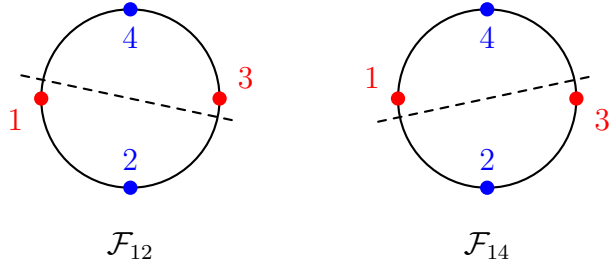
\begin{figure}[t]
\centering
\begin{tikzpicture}[scale=1.18, line cap=round, line join=round, baseline=(current bounding box.center)]
\begin{scope}[xshift=-2cm]
\draw[thick] (0,0) circle (1);
\draw[thick, dashed] (168:1.25)--(-12:1.25);
\fill[red] (180:1) circle (0.08);
\fill[red] (0:1) circle (0.08);
\fill[blue] (-90:1) circle (0.08);
\fill[blue] (90:1) circle (0.08);
\node at (0,-1.65) {$\mathcal{F}_{12}$};
\node[below] at (90:.92) {${\color{blue}4}$};
\node[below left] at (180:1.08) {${\color{red}1}$};
\node[above right] at (0:1.08) {${\color{red}3}$};
\node[above] at (-90:.92) {${\color{blue}2}$};
\end{scope}
\begin{scope}[xshift=2cm]
\draw[thick] (0,0) circle (1);
\draw[thick, dashed] (-168:1.25)--(12:1.25);
\fill[red] (180:1) circle (0.08);
\fill[red] (0:1) circle (0.08);
\fill[blue] (-90:1) circle (0.08);
\fill[blue] (90:1) circle (0.08);
\node[below] at (90:.92) {${\color{blue}4}$};
\node[above left] at (180:1.08) {${\color{red}1}$};
\node[below right] at (0:1.08) {${\color{red}3}$};
\node[above] at (-90:.92) {${\color{blue}2}$};
\node at (0,-1.65) {$\mathcal{F}_{14}$};
\end{scope}
\end{tikzpicture}
\caption{There are different perspectives on the choice that leads to $\mathcal{F}_{12}$ and $\mathcal{F}_{14}$. A mathematical perspective is simply that these are different lateral Borel resummations of the same asymptotic series. A physical interpretation is that they differ in which cut of the Euclidean geometry (shown dashed above) is continued forwards into Lorentzian time to define the shock waves. The operator ordering of $1$ and $3$ relative to this cut is reversed, so in the case $C < 0$ where shock waves allow causal separation between the points, the functions are different.}\label{figf12f14}
\end{figure}

If we rotate the contour $X^\pm \to X^\pm e^{\pm \i \gamma}$, this becomes
\begin{equation}\label{gamma}
\mathcal{F}_{\rm JT}(t)
=
\int_{-\infty}^{\infty}\d X^+\d X^-\,e^{-2\i C X^+X^-}
\left[\frac{1}{2+e^t X^- e^{-\i\gamma}}\right]^{2\Delta}
\left[\frac{1}{2-\i X^+e^{+\i\gamma}}\right]^{2\Delta}.
\end{equation}
Recall that $X^\pm$ are shock wave modes in a Lorentzian continuation of the geometry. This contour rotation changes the time slice of the Euclidean geometry that gets continued forward into Lorentzian time (see figure \ref{figf12f14}). We label the corresponding functions $\mathcal{F}_{12}$ and $\mathcal{F}_{14}$. 

\subsubsection{Static patch dS JT coupled to a 2d BCFT}\label{sec:wrongsign-JT}
Let's now discuss the case $C < 0$. The original integral (\ref{orig}) is then undefined because $\Theta(C X^-)$ imposes $X^- < 0$, meaning a negative shock wave, so we encounter the branch point. More generally, the result of (\ref{gamma}) now depends on the sign of $\gamma$ --- there are two different OTOC formulas. These two functions can be extended to all $\theta_i$ by analytic continuation.

For certain configurations of the points on the Euclidean circle, one of the two answers seems more reasonable than the other. In particular, consider a one-sided configuration $\{{\color{red}\theta_1},{\color{blue}\theta_2},{\color{red}\theta_3},{\color{blue}\theta_4}\}  = \{-3\epsilon+\i t,-\epsilon,\epsilon+\i t,3\epsilon\}$. Then the operators are all part of the same Lorentzian geometry. To use the $X^\pm$ contour corresponding to real shock waves on this Lorentzian geometry, we set $\gamma = 0$:
\begin{equation}\label{plot1}
\mathcal{F}_{12}(t)
=
\int_{-\infty}^{\infty}\d X^+\d X^-\,e^{-2\i C X^+X^-}
\left[\frac{1}{2\sin(2\epsilon)-\i e^{t+\i\epsilon}X^-}\right]^{2\Delta}
\left[\frac{1}{2\sin(2\epsilon)-\i e^{\i\epsilon}X^+}\right]^{2\Delta}
\end{equation}
The other OTOC can be obtained from $\gamma = -\pi/2$:
\begin{align}\label{plot2}
\mathcal{F}_{14}(t)
&=
\int_{-\infty}^{\infty}\d X^+\d X^-\,e^{-2\i C X^+X^-}
\left[\frac{1}{2\sin(2\epsilon)+ e^{t+\i\epsilon}X^-}\right]^{2\Delta}
\left[\frac{1}{2\sin(2\epsilon)- e^{\i\epsilon}X^+}\right]^{2\Delta}.
\end{align}
Despite the fact that these functions have the same $1/C$ perturbation theory, they are dramatically different, as shown in  figure \ref{fig:jt-onesided-contours}.
\begingroup
\renewcommand{\theHfigure}{jtcontours.\arabic{figure}}
\begin{figure}[H]
\centering
\vspace{-0.25em}
\begin{minipage}[t]{0.48\textwidth}
\centering
\begin{tikzpicture}[scale=1.18, baseline=(current bounding box.center)]
\def\epsang{7}
\path[use as bounding box] (-1.25,-1.25) rectangle (1.25,1.25);
\draw[thick] (0,0) circle (1);
\draw[thick, dashed] (180:1.25)--(0:1.25);
\fill[red] (-3*\epsang:1) circle (0.075);
\fill[blue] (-\epsang:1) circle (0.075);
\fill[red] (\epsang:1) circle (0.075);
\fill[blue] (3*\epsang:1) circle (0.075);
\node at (3.5*\epsang:1.3) {${\color{blue}4}$};
\node at (-3.5*\epsang:1.3) {${\color{red}1}$};
\node at (1.2*\epsang:1.3) {${\color{red}3}$};
\node at (-1.2*\epsang:1.3) {${\color{blue}2}$};
\end{tikzpicture}
\vspace{0.25em}
\includegraphics[width=\linewidth,height=0.55\linewidth]{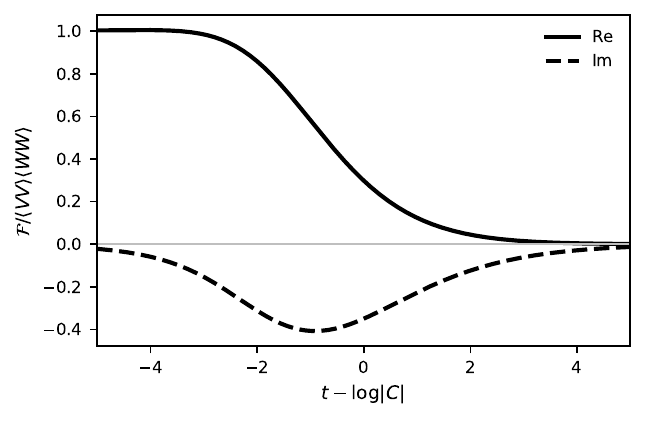}
\end{minipage}
\hfill
\begin{minipage}[t]{0.48\textwidth}
\centering
\begin{tikzpicture}[scale=1.18, baseline=(current bounding box.center)]
\def\epsang{7}
\path[use as bounding box] (-1.25,-1.25) rectangle (1.25,1.25);
\draw[thick] (0,0) circle (1);
\draw[thick, dashed, line cap=round, line join=round, rounded corners=2pt] (90:1.25) -- (90:1.1) arc[start angle=90, end angle=2*\epsang, radius=1.1] -- (2*\epsang:.9) arc[start angle=2*\epsang, end angle=90, radius=.9] -- (-90:.9) arc[start angle=-90, end angle=-2*\epsang, radius=.9] -- (-2*\epsang:1.1) arc[start angle=-2*\epsang, end angle=-90, radius=1.1] -- (-90:1.25);
\fill[red] (-3*\epsang:1) circle (0.075);
\fill[blue] (-\epsang:1) circle (0.075);
\fill[red] (\epsang:1) circle (0.075);
\fill[blue] (3*\epsang:1) circle (0.075);
\node at (3.5*\epsang:1.3) {${\color{blue}4}$};
\node at (-3.5*\epsang:1.3) {${\color{red}1}$};
\node at (1.2*\epsang:1.3) {${\color{red}3}$};
\node at (-1.2*\epsang:1.3) {${\color{blue}2}$};
\end{tikzpicture}
\vspace{0.25em}
\includegraphics[width=\linewidth,height=0.55\linewidth]{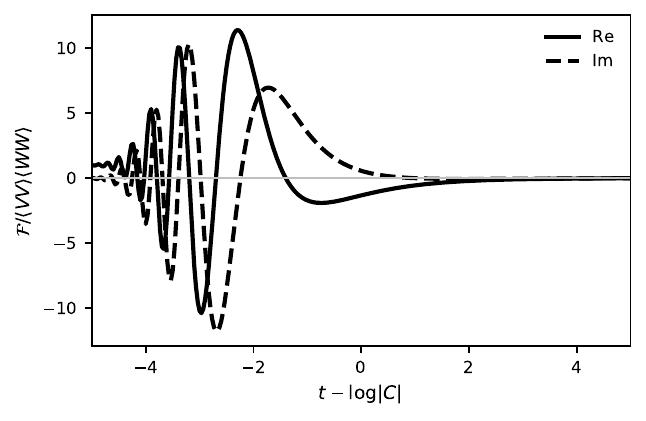}
\end{minipage}
\caption{Comparison of (\ref{plot1}) (left) and (\ref{plot2}) (right) for the case $C < 0$, $\Delta=1/2$ and $\epsilon=1/10$. Note that if we continue $\theta_1$ the long way around the circle until it is close to $\theta_4$, the behavior of the functions will be switched.}
\label{fig:jt-onesided-contours}
\end{figure}
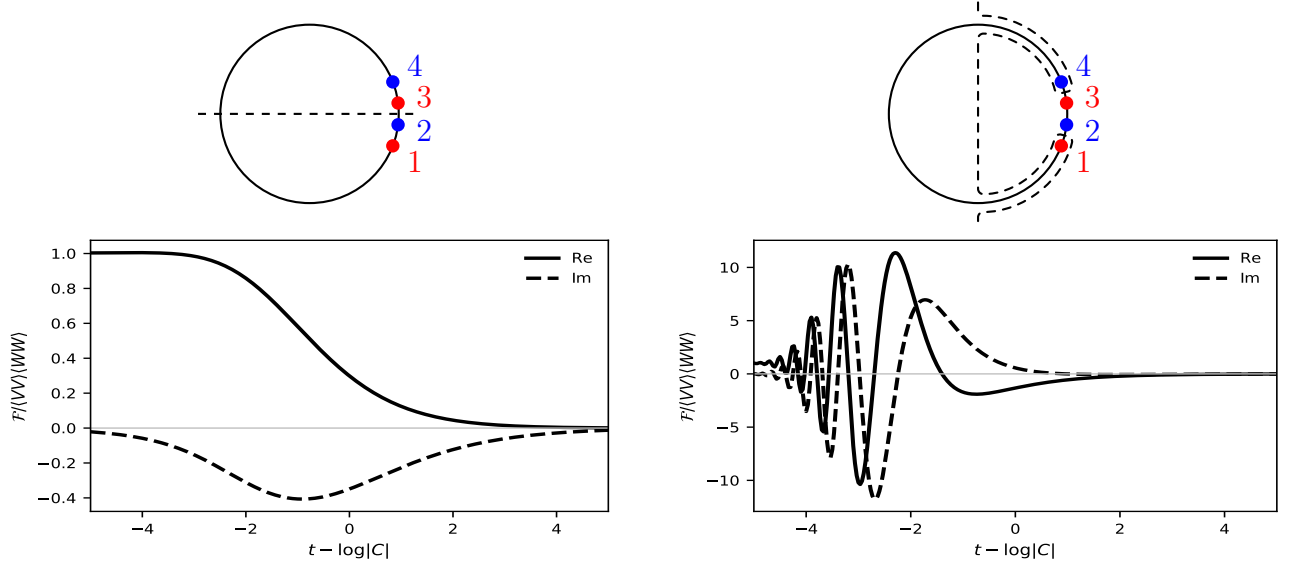
\endgroup

The exact answer for the $X^\pm$ integral can be written in terms of the confluent hypergeometric function $U(a,1,x)=\Gamma(a)^{-1}\int_0^{\infty}\d s e^{-s x} {s^{a-1}\over (1+s)^a}$ \cite{Maldacena:2016upp}:
\be\label{eqn:JTU}
{\mathcal{F}_{\rm JT}({\color{red}\theta_1},{\color{blue}\theta_2},{\color{red}\theta_3},{\color{blue}\theta_4})\over \langle WW\rangle\langle VV\rangle}
=
{U(2\Delta,1,{1\over z})\over z^{2\Delta}},
\qquad
z=-{\i\over 8 C}
{e^{\i(\theta_2+\theta_4-\theta_1-\theta_3)/2}
\over
\sin{\theta_{31}\over 2}\sin{\theta_{42}\over 2}}.
\ee
This function has a branch point at $z=0$, and the different integration contours for $X^{\pm}$ can lead to different sheets. In the above example of $\{{\color{red}\theta_1},{\color{blue}\theta_2},{\color{red}\theta_3},{\color{blue}\theta_4}\}  = \{-3\epsilon+\i t,-\epsilon,\epsilon+\i t,3\epsilon\}$ with $C$ negative, $z\propto -{\i\over C} e^t$ is along the positive imaginary axis.  $\mathcal{F}_{12}$ is given by $U$ with the principal prescription, and $\mathcal{F}_{14}$ is obtained by continuing clockwise around the branch point.

 \subsection{Eikonal formulas in dS\texorpdfstring{$_3$}{3}}
 There are various perspectives one can use to derive the eikonal formula for high-energy gravitational scattering \cite{tHooft:1987vrq,Amati:1987uf,Verlinde:1991iu,Kabat:1992tb,tHooft:1990fkf}. Perhaps the least idiosyncratic one is as follows. We approximate the gravitational path integral by an integral over gravitational shock wave modes, with a Gaussian approximation for their action. In three dimensions, these modes are functions $X^\pm(\phi)$ defined on the de Sitter horizon, so the integral is a functional integral. The action is
 \begin{equation}
\log(e^{-I}) = \i \int \frac{\alpha}{2\pi}\d\phi \left[-X^+(\phi) \mathcal{D} X^-(\phi) + X^+(\phi) T_+(\phi) + X^-(\phi) T_-(\phi)\right]
 \end{equation}
 where $\mathcal{D}$ is proportional to the operator that appears in the shock wave equation (\ref{diffop}). For fixed $T_\pm(\phi)$, one can integrate out $X^\pm(\phi)$ to get
 \begin{align}
\log(e^{-I}) &= \i \int \left(\frac{\alpha}{2\pi}\right)^2\d\phi \d\phi'\ T_+(\phi) T_-(\phi') h(\phi-\phi')\\
&=\i p_+ p_- h(\phi_1 - \phi_2)\label{insert}
 \end{align}
 where we have used that $\frac{\alpha}{2\pi}h$ is the inverse to $\mathcal{D}$. In the last line, we substituted in the answer for the case corresponding to particles with momenum $p_+$ and $p_-$ at impact parameters $\phi_1,\phi_2$. This is sometimes called the 't Hooft S-matrix for gravitational scattering, and it captures the gravitational part of the calculation. What remains is to unzip the $\langle VV\rangle$ and $\langle WW\rangle$ correlators in a basis with definite values of $T_\pm(\phi)$ and then zip them back up again with this S-matrix inserted \cite{kitaev2014hidden,Shenker:2014cwa}.

 In this computation, we will encounter two different $\mathbb{Z}_2$ choices, leading to four different eikonal OTOC formulas:
 \begin{equation}
\mathcal{F}^{t>0}_{12}, \hspace{20pt} \mathcal{F}^{t<0}_{12}, \hspace{20pt} \mathcal{F}^{t>0}_{14},\hspace{20pt} \mathcal{F}^{t<0}_{14}.
 \end{equation}
 One of the choices is whether the $W$ operators are at positive Lorentzian time (relative to the $V$ operators) or negative. This determines whether we should decompose the $W$ operators in the $p_-$ basis or the $p_+$ basis. Then $\mathcal{F}^{t>0}$ captures all $(G e^t)^n$ effects, and $\mathcal{F}^{t<0}$ captures all $(G e^{-t})^n$ effects. To get a good approximation for all $t$, one should sum over both, subtracting the disconnected term that would otherwise be counted twice:
 \begin{align}\label{summing}
\mathcal{F}_{12} &= \mathcal{F}_{12}^{t>0} + \mathcal{F}_{12}^{t < 0} - \text{disconnected}\\
\mathcal{F}_{14} &= \mathcal{F}_{14}^{t>0} + \mathcal{F}_{14}^{t < 0} - \text{disconnected}.
 \end{align}
The remaining choice is special to the case of wrong-sign shocks and has to do with the contour choice explained in the JT example above.

\subsubsection{The OTOCs in the time domain}
We will set up the computation of $\mathcal{F}_{12}^{t<0}$ and later determine the three others from this one. The basic step is to rewrite the $W$ and $V$ two point functions as Klein-Gordon inner products of wave functions on the appropriate horizons. For this case, the $\langle WW\rangle$ correlator should be expanded in terms of states with definite $p_+$, which can be done by inserting a resolution of the identity along the $x^- = 0$ horizon. Similarly, the $\langle VV\rangle$ correlator should be expanded in terms of definite $p_-$, by integrating along the $x^+ = 0$ horizon:
\begin{align}
   \langle W(\theta_4) W(\theta_2)\rangle \hspace{5pt}&= \hspace{8pt}\begin{tikzpicture}[scale=1, line cap=round, line join=round, >=stealth, baseline=(current bounding box.center)]
    \coordinate (LB) at (-1.15,-1.15);
    \coordinate (LT) at (-1.15,1.15);
    \coordinate (RB) at (1.15,-1.15);
    \coordinate (RT) at (1.15,1.15);
    \def\R{1.15}
    \def\zx{-0.55}
    \def\zy{-0.55}
    \def\thetaL{220}
    \def\thetaR{45}
    \coordinate (Lop) at ({\R*cos(\thetaL)+\zx*\R*sin(\thetaL)}, {\zy*\R*sin(\thetaL)});
    \coordinate (Rop) at ({\R*cos(\thetaR)+\zx*\R*sin(\thetaR)}, {\zy*\R*sin(\thetaR)});
    \coordinate (Hop) at (0.4,0.4);
    \fill[gray!6, domain=180:360, samples=80, variable=\t] plot ({\R*cos(\t)+\zx*\R*sin(\t)}, {\zy*\R*sin(\t)}) -- cycle;
    \fill[gray!20] (LB) -- (LT) -- (RT) -- (RB) -- cycle;
    \draw[gray, thick, dotted, domain=180:300, samples=80, variable=\t] plot ({\R*cos(\t)+\zx*\R*sin(\t)}, {\zy*\R*sin(\t)});
    \draw[thick] (LB) -- (LT) -- (RT) -- (RB) -- cycle;
    \draw[gray!55, thick] (LB) -- (RT);
    \fill[gray!6, domain=0:180, samples=80, variable=\t] plot ({\R*cos(\t)+\zx*\R*sin(\t)}, {\zy*\R*sin(\t)}) -- cycle;
    \draw[black, thick, domain=-60:180, samples=80, variable=\t] plot ({\R*cos(\t)+\zx*\R*sin(\t)}, {\zy*\R*sin(\t)});
    \draw[blue!50, thick, dotted] (Lop) .. controls (-0.05,0.27) and (0.10,0.54) .. (Hop);
    \draw[blue, thick] (Hop) .. controls (0.48,0.08) and (0.21,-0.07) .. (Rop);
    \fill[blue!50] (Lop) circle (0.06) node[above left, xshift=2pt] {${\theta_4}$};
    \fill[blue] (Rop) circle (0.06) node[below right] {${ \theta_2}$};
    \fill[black] (Hop) circle (0.06);
    \draw[thick, black] (-1.15,0) -- (1.15,0);
\end{tikzpicture}
, \hspace{40pt} \langle V(\theta_3)V(\theta_1)\rangle \hspace{5pt} = \hspace{0pt} 
\begin{tikzpicture}[scale=1, line cap=round, line join=round, >=stealth, baseline=(current bounding box.center)]
    \coordinate (LB) at (-1.15,-1.15);
    \coordinate (LT) at (-1.15,1.15);
    \coordinate (RB) at (1.15,-1.15);
    \coordinate (RT) at (1.15,1.15);
    \def\R{1.15}
    \def\zx{-0.55}
    \def\zy{-0.55}
    \def\thetaL{-70}
    \def\thetaR{135}
    \coordinate (Lop) at ({\R*cos(\thetaL)+\zx*\R*sin(\thetaL)}, {\zy*\R*sin(\thetaL)});
    \coordinate (Rop) at ({\R*cos(\thetaR)+\zx*\R*sin(\thetaR)}, {\zy*\R*sin(\thetaR)});
    \coordinate (Hop) at (-0.6,0.6);
    \fill[gray!6, domain=180:360, samples=80, variable=\t] plot ({\R*cos(\t)+\zx*\R*sin(\t)}, {\zy*\R*sin(\t)}) -- cycle;
    \fill[gray!20] (LB) -- (LT) -- (RT) -- (RB) -- cycle;
    \draw[gray, thick, dotted, domain=180:300, samples=80, variable=\t] plot ({\R*cos(\t)+\zx*\R*sin(\t)}, {\zy*\R*sin(\t)});
    \draw[thick] (LB) -- (LT) -- (RT) -- (RB) -- cycle;
    \draw[gray!55, thick] (LT) -- (RB);
    \fill[gray!6, domain=0:180, samples=80, variable=\t] plot ({\R*cos(\t)+\zx*\R*sin(\t)}, {\zy*\R*sin(\t)}) -- cycle;
    \draw[black, thick, domain=-60:180, samples=80, variable=\t] plot ({\R*cos(\t)+\zx*\R*sin(\t)}, {\zy*\R*sin(\t)});
    \draw[red!50, thick, dotted] (Lop) .. controls (0.23,0.35) and (0.20,0.46) .. (Hop);
    \draw[red, thick] (Hop) .. controls (-0.64,0.05) and (-0.52,-0.10) .. (Rop);
    \fill[red!50] (Lop) circle (0.06) node[above left, xshift=2pt] {${\theta_3}$};
    \fill[red] (Rop) circle (0.06) node[left] {${ \theta_1}$};
    \fill[black] (Hop) circle (0.06);
    \draw[thick, black] (-1.15,0) -- (1.15,0);
\end{tikzpicture}
\end{align}
The picture on the left corresponds to the equation
\begin{align}  \langle W(\theta_4) W(\theta_2)\rangle&=2\i \int_{-\infty}^\infty \d x^+ \int_0^{2\pi}\d\phi \ \langle W(\theta_4)W(x^+,\phi)\rangle\partial_+ \langle W(x^+,\phi)W(\theta_2)\rangle.
\end{align}
To separate out the components with definite $p_+$, one can insert a Fourier representation
\begin{align}
\int \d x^+ e^{-\i x^+p_+} \langle W(x^+,\phi)W(\theta_2)\rangle &= \int \d x^+ e^{-\i x^+p_+}\frac{1}{4\pi^2}\int_0^\infty \d s \frac{\cos(\nu s)}{\cosh(s)-x^+ e^{\i\theta_2}}\\
   % &= \theta(-p_+)(-2\pi\i)\int_0^\infty \frac{\d s}{4\pi^2} \cos(\nu s)e^{-\i\theta_2 -\i e^{-\i\theta_2}p_+\cosh(s)}\\
    &=\theta(-p_+)\frac{-\i e^{-\i\theta_2}}{2\pi}K_{\i\nu}(\i e^{-\i\theta_2}p_+).
\end{align}
This leads to 
\begin{equation}
     \langle W(\theta_4) W(\theta_2)\rangle = -\frac{e^{-\i(\theta_2+\theta_4)}}{\pi\alpha}\int_{-\infty}^0 p_+\d p_+\int_{-\pi}^\pi \frac{\d\phi}{2\pi}K_{\i\nu}(-\i e^{-\i\theta_4}p_+)K_{\i\nu}(\i e^{-\i\theta_2}p_+).
\end{equation}
In this formula, we snuck in a small generalization to make it apply to dS$_3$ with an observer present (\ref{dSobserver}) -- for this geometry, the free field two point function along the worldline of the observer is just $(2\pi/\alpha)$ times the one for pure de Sitter.\footnote{This can be derived by studying the two-point function on $S^3/\mathbb{Z}_k$ with $k = 2\pi/\alpha \in \mathbb{Z}$, where one can use the method of images, and then analytically continue to general $\alpha$.}

Combining with a similar expression for $\langle VV\rangle$ and inserting the S-matrix (\ref{insert}), we get
\begin{align}
\mathcal{F}^{t<0}_{12}(\theta_1,\theta_2,\theta_3,\theta_4|m)
&=
\begin{tikzpicture}[scale=1, baseline=(current bounding box.center)]
\draw[dotted, thick] (180:1.25)--(0:1.25);
\draw[thick] (0,0) circle (1);
\draw[blue] (160:1) -- (-20:1);
\draw[red] (200:1) -- (200:.15);
\draw[red] (20:1) -- (20:.15);
\fill[blue] (160:1) circle (0.08) node[left] {$\theta_4$};
\fill[blue] (-20:1) circle (0.08) node[right] {$\theta_2$};
\fill[red] (20:1) circle (0.08) node[right] {$\theta_3$};
\fill[red] (200:1) circle (0.08) node[left] {$\theta_1$};
\node at (-35:.7) {${\color{blue}\nu}$};
\node at (33:.7) {${\color{red}\nu'}$};
\end{tikzpicture}
\notag\\
&=\frac{e^{-\i(\theta_2+\theta_4)}}{\pi\alpha}
\int_{-\infty}^{0}p_+\,\d p_+\,
K_{\i\nu}(-\i e^{-\i\theta_4}p_+)
K_{\i\nu}(\i e^{-\i\theta_2}p_+)
\notag\\
&\hspace{.6cm}\times
\frac{e^{\i(\theta_1+\theta_3)}}{\pi\alpha}
\int_{-\infty}^{0}p_-\,\d p_-\,
K_{\i\nu'}(\i e^{\i\theta_3}p_-)
K_{\i\nu'}(-\i e^{\i\theta_1}p_-)
\notag\\
&\hspace{.6cm}\times
\int_{-\pi}^{\pi}\frac{\d\phi}{2\pi}
\exp\!\left\{ \i p_+p_-h(\phi)\right\}.
\label{eq:F12-time}
\end{align}
The remaining OTOCs can be obtained from this one. To compute $\mathcal{F}_{12}^{t>0}$, we should switch to decomposing $\langle WW\rangle$ in terms of $p_-$ and $\langle VV\rangle$ in terms of $p_+$. This can be accomplished by interchanging $\theta_1\leftrightarrow \theta_2$, $\theta_4\leftrightarrow \theta_3$ and $\nu \leftrightarrow \nu'$. However, now the scattering event takes place on a backwards-going Lorentzian sheet, so the sign in the 't Hooft action is reversed. Equivalently, the OTOC becomes an inner product $\langle \text{in}|\text{out}\rangle$ instead of $\langle \text{out}|\text{in}\rangle$, so we should conjugate the S-matrix. This can be accomplished by flipping $h \to -h$. All together,
\begin{align}
\mathcal{F}^{t>0}_{12}(\theta_1,\theta_2,\theta_3,\theta_4|m)
&= \mathcal{F}^{t<0}_{12}(\theta_2,\theta_1,\theta_4,\theta_3|m)\Big|_{h\to -h, \nu \leftrightarrow \nu'}\notag\\
&=\frac{e^{-\i(\theta_1+\theta_3)}}{\pi\alpha}
\int_{-\infty}^{0}p_+\,\d p_+\,
K_{\i\nu'}(-\i e^{-\i\theta_3}p_+)
K_{\i\nu'}(\i e^{-\i\theta_1}p_+)
\notag\\
&\hspace{.6cm}\times
\frac{e^{\i(\theta_2+\theta_4)}}{\pi\alpha}
\int_{-\infty}^{0}p_-\,\d p_-\,
K_{\i\nu}(\i e^{i\theta_4}p_-)
K_{\i\nu}(-\i e^{i\theta_2}p_-)
\notag\\
&\hspace{.6cm}\times
\int_{-\pi}^{\pi}\frac{\d\phi}{2\pi}
\exp\!\left\{ -\i p_+p_-h(\phi)\right\}.
\label{eq:F12-time-pos}
\end{align}
We can also obtain the $\mathcal{F}_{14}$ functions from the $\mathcal{F}_{12}$ functions, by rotating the configuration:
\begin{align}
\mathcal{F}^{t<0}_{14}(\theta_1,\theta_2,\theta_3,\theta_4|m)
&=
\begin{tikzpicture}[scale=1, baseline=(current bounding box.center)]
\draw[dotted, thick] (90:1.25)--(-90:1.25);
\draw[thick] (0,0) circle (1);
\draw[blue] (160:1) -- (-20:1);
\draw[red] (200:1) -- (200:.15);
\draw[red] (20:1) -- (20:.15);
\fill[blue] (160:1) circle (0.08) node[left] {$\theta_4$};
\fill[blue] (-20:1) circle (0.08) node[right] {$\theta_2$};
\fill[red] (20:1) circle (0.08) node[right] {$\theta_3$};
\fill[red] (200:1) circle (0.08) node[left] {$\theta_1$};
\node at (-35:.7) {${\color{blue}\nu}$};
\node at (33:.7) {${\color{red}\nu'}$};
\end{tikzpicture} = \mathcal{F}_{12}^{t>0}(\theta_4-\tfrac{3\pi}{2},\theta_1+\tfrac{\pi}{2},\theta_2+\tfrac{\pi}{2},\theta_3+\tfrac{\pi}{2})\Big|_{\nu\leftrightarrow \nu'}
\end{align}
and similarly for $\mathcal{F}_{14}^{t>0}$. After summing over the $t < 0$ and $t > 0$ functions as in (\ref{summing}),
\begin{equation}
\mathcal{F}_{14}(\theta_1,\theta_2,\theta_3,\theta_4) = \mathcal{F}_{12}(\theta_4-2\pi,\theta_1,\theta_2,\theta_3)\Big|_{\nu\leftrightarrow \nu'},
\end{equation}
where we used translation invariance to shift all of the coordinates by $\pi/2$.\footnote{Under what circumstances do the $p_\pm$ integrals converge? Using that $K_{\i\nu}(x)\sim (\text{subleading})\cdot e^{-x}$ for large $x$, we find that the integrals for $\mathcal{F}_{12}$ converge iff
\be
\text{Re}(-\i e^{-\i\theta_4} + \i e^{-\i \theta_2}) < 0 \hspace{20pt} \text{and} \hspace{20pt} \text{Re}(\i e^{\i\theta_3} - \i e^{\i \theta_1}) < 0.
\ee
However, we can expand this range by rotating the contour for the $p_\pm$ variables. As we do this, it is important to keep $p_+p_-$ real, since the shock wave profile can take both signs. This means that we can rotate by opposite phase. One can check that this is enough to cover any OTOC configuration of points.} So a cyclic permutation on the circle interchanges these two functions.

\subsubsection{Perturbative expansion}
If Newton's constant $G$ is small and the mass of the observer $m$ is large, the OTOC reduces approximately to the factorized product of two point functions. One can get a perturbative series of corrections to this by expanding the $e^{\pm \i p_+ p_- h(\phi)}$ factor and then doing the $p_-,p_+$ integrals. For this computation, it isn't necessary to distinguish between $\mathcal{F}_{12}$ and $\mathcal{F}_{14}$, because they are equal in perturbation theory. But we do have to distinguish between $\mathcal{F}^{t<0}$ and $\mathcal{F}^{t>0}$. One finds
\begin{align}
\mathcal{F}^{t\lessgtr 0}(\theta_1,\theta_2,\theta_3,\theta_4)
&\sim
\sum_{n=0}^{\infty}\Upsilon^{n}_{\nu}(\theta_{42})
\Upsilon^{n}_{\nu'}(\theta_{31})\frac{e^{\pm\frac{\i n}{2}(\pi + \theta_2+\theta_4 - \theta_1 - \theta_3)}}{n!}\int_{-\pi}^{\pi}\frac{\d\phi}{2\pi}h(\phi)^n.
\end{align}
Here, the $\pm$ refers to the $\lessgtr$, and the $\Upsilon_\nu^n$ functions are essentially moments of $(-p_\pm)$\footnote{Here is a closed form for $\Upsilon^n_\nu(\theta)$: \[-\frac{2^{n-1}}{\pi\alpha}
\Gamma\!\left(1+\frac{n}{2}\right)
e^{\i(1+\frac{n}{2})(\pi-\theta)}\Bigg[
e^{\nu(\pi-\theta)}
\Gamma\!\left(1+\frac{n}{2}-\i\nu\right)\Gamma(\i\nu)
{}_2F_1\!\left(1+\frac{n}{2}-\i\nu,1+\frac{n}{2};1-\i\nu;e^{-2\i\theta}\right) + (\nu\to-\nu)
\Bigg]\]}
\begin{align}
\Upsilon^{n}_{\nu}(\theta)
&\equiv
-\frac{1}{\pi\alpha}
\int_0^\infty y^{1+n}\d y
K_{\i\nu}(-\i e^{\i\theta/2}y)
K_{\i\nu}(\i e^{-\i\theta/2}y)
.
\end{align}
For the first few orders
\begin{align}
\int_{-\pi}^{\pi}\frac{\d\phi}{2\pi}h(\phi)&=-\frac{4\pi G}{\alpha} = -2G + O(G^2 m),\\
\int_{-\pi}^{\pi}\frac{\d\phi}{2\pi}h(\phi)^2&=
\frac{2\pi^2G^2[\alpha+\sin(\alpha)]}{\alpha\sin^2(\alpha/2)}  = \frac{1}{8m^2} - \frac{G}{2m} + \frac{2\pi^2-6}{3}G^2 + O(G^3 m).
\end{align}

If the two small parameters $G$ and $1/m$ scale the same way, then this is an expansion in $G e^t$, containing all positive integer powers. However, it is also interesting to study the ``recoil'' limit \cite{Kolchmeyer:2024fly}, where $G\to 0$ with $1/m$ finite. Then $h(\phi)$ becomes proportional to $\cos(\phi)$, so its odd moments vanish. This means that in the recoil limit only the even powers appear, and we get a series in $e^{2t}/m^2$, as in \cite{Kolchmeyer:2024fly}.\footnote{This property is also true in higher dimensions, see appendix~\ref{app:recoil in higher d}.} One can understand the reason that the pure recoil effect starts only at quadratic order $1/m^2$ by noting that the observer follows a geodesic without the recoil, so the change in its action vanishes under linear perturbation. However, with the order $G$ backreaction which breaks the rotation symmetry (see \eqref{recoillight}), we can have a $G/m$ piece. 

To write a more explicit formula, we can use the fact that $\Upsilon^n_\nu$ simplifies if $\theta = \pi$ so the operator insertions are antipodal. Taking the configuration where both pairs of operators are antipodal
\begin{equation}
\theta_1=-\frac{\pi}{2},\hspace{20pt}\theta_2=-\i t,\hspace{20pt}\theta_3=\frac{\pi}{2},\hspace{20pt}\theta_4=\pi-\i t
\end{equation}
and summing over the $t>0$ and $t<0$ pieces as in (\ref{summing}), one finds
\begin{equation}\label{pertseries}
\begin{aligned}
\frac{\mathcal{F}(t)}{\mathcal{F}_0}\sim 1&+\frac{4\pi}{\alpha}\left[\frac{(1+4\nu^2)\pi\tanh(\pi\nu)}{16\nu}\right]\left[\frac{(1+4\nu'^2)\pi\tanh(\pi\nu')}{16\nu'}\right]G(e^t + e^{-t})\\
&+\frac{\pi^2(\alpha+\sin\alpha)}{\alpha\sin^2(\alpha/2)}\left[\frac{2(1+\nu^2)}{3}\right]\left[\frac{2(1+\nu'^2)}{3}\right]G^2(e^{2t}+ e^{-2t})+\dots
\end{aligned}
\end{equation}
The displayed terms are positive for real $t$, and indeed one can check that the entire perturbative series is positive. This is to be contrasted with the more familiar case of black holes with right-sign shock waves, where this would have been an alternating series.

\subsubsection{The energy domain}\label{sec:eikonalenergy}
We would now like to go back to the full eikonal formula and use an inverse Laplace transform to compute the OTOC in the energy domain. It is convenient to define
\begin{equation}\label{eqn:FtobF}
 \mathcal{F}(\omega_i|m)= \frac{1}{Z_{\text{sphere}}(m)}\bm{\mathcal{F}}(m_i).
\end{equation}
where the frequencies $\omega_i$ are 
\begin{align}\label{omegainverse}
\omega_i &= m_i - m_{i+1},\hspace{20pt}
\sum_i\omega_i=0.%, \hspace{20pt} m = \tfrac{1}{4}\textstyle\sum_i m_i \\ m_i &= m + s_i, \hspace{20pt} s_i = \tfrac{1}{4}(2\omega_i + \omega_{i+1}-\omega_{i-1}).
\end{align}
Substituting into (\ref{FFrel}), one finds that the ``regularized OTOC'' configuration for $\mathcal{F}(\theta_i|m)$ is simply the Fourier transform of $\mathcal{F}(\omega_i|m)$:
\begin{align}
\delta({\textstyle\sum_i} \omega_i)
\mathcal{F}(\omega_i|m) = \int \frac{\d^4 t_i}{(2\pi)^4}e^{-\i\sum_i\omega_i t_i}\mathcal{F}({\color{red}-\tfrac{\pi}{2}+\i t_1},{\color{blue}\i t_2}, {\color{red}\tfrac{\pi}{2} + \i t_3},{\color{blue}\pi+\i t_4}|m).
\end{align}
All of the Fourier integrals can be carried out using
\begin{align}
\int_{-\infty}^\infty \d t e^{\i\omega t} e^t K_{\i\nu}(x e^t ) &= 2^{\i\omega-1}x^{-1-\i\omega}\Gamma\left(\frac{1 + \i\omega\pm \i \nu}{2}\right), \hspace{20pt} \text{Re}(x) > 0,
\end{align}
and we find
\begin{equation}
\begin{aligned}
\delta({\textstyle\sum_i} \omega_i)\mathcal{F}_{12}^{t<0}(\omega_i|m) = &\frac{1}{(2\pi)^4}\frac{e^{-\frac{\pi}{2}(\omega_1+\omega_3)}}{(4\pi\alpha)^2}{\textstyle
\Gamma\!\left(\frac{1+\i\omega_1\pm\i\nu}{2}\right)
\Gamma\!\left(\frac{1+\i\omega_3\pm\i\nu}{2}\right)\Gamma\!\left(\frac{1-\i\omega_2\pm\i\nu'}{2}\right)
\Gamma\!\left(\frac{1-\i\omega_4\pm\i\nu'}{2}\right)} \\
&\times \int_{-\infty}^0\frac{\d p_+}{p_+}\left(\frac{-p_+}{2}\right)^{-\i (\omega_1+\omega_3)}
\int_{-\infty}^0\frac{\d p_-}{p_-}\left(\frac{-p_-}{2}\right)^{\i (\omega_2+\omega_4)}
\int_{-\pi}^\pi \frac{\d\phi}{2\pi}e^{\i p_+p_-h(\phi)}.
\end{aligned}
\end{equation}
The integral over $p_+,p_-$ makes sense if $h$ is real (or in the upper half-plane) and we get
\begin{equation}\label{Fener1}
\begin{aligned}
\mathcal{F}_{12}^{t<0}(\omega_i|m) = &\frac{1}{(2\pi)^3}\frac{1}{(4\pi\alpha)^2}
{\textstyle\Gamma\!\left(\frac{1+\i\omega_1\pm\i\nu}{2}\right)
\Gamma\!\left(\frac{1+\i\omega_3\pm\i\nu}{2}\right)\Gamma\!\left(\frac{1-\i\omega_2\pm\i\nu'}{2}\right)
\Gamma\!\left(\frac{1-\i\omega_4\pm\i\nu'}{2}\right)} \\
&\times \Gamma\left(-\i(\omega_1+\omega_3)+ 0^+\right)\int_{-\pi}^\pi \frac{\d\phi}{2\pi}\left(4h(\phi)+\i 0^+\right)^{\i(\omega_1+\omega_3)}.
\end{aligned}
\end{equation}
Similarly, for the other three cases
\begin{equation}\label{Fener2}
\begin{aligned}
\mathcal{F}_{12}^{t>0}(\omega_i|m) = &\frac{1}{(2\pi)^3}\frac{1}{(4\pi\alpha)^2}
{\textstyle\Gamma\!\left(\frac{1-\i\omega_1\pm\i\nu}{2}\right)
\Gamma\!\left(\frac{1-\i\omega_3\pm\i\nu}{2}\right)\Gamma\!\left(\frac{1+\i\omega_2\pm\i\nu'}{2}\right)
\Gamma\!\left(\frac{1+\i\omega_4\pm\i\nu'}{2}\right)} \\
&\times \Gamma\left(-\i(\omega_2+\omega_4)+ 0^+\right)\int_{-\pi}^\pi \frac{\d\phi}{2\pi}\left(4h(\phi)-\i 0^+\right)^{\i(\omega_2+\omega_4)}.
\end{aligned}
\end{equation}
\begin{equation}\label{Fener3}
\begin{aligned}
\mathcal{F}_{14}^{t<0}(\omega_i|m) = &\frac{1}{(2\pi)^3}\frac{1}{(4\pi\alpha)^2}
{\textstyle\Gamma\!\left(\frac{1+\i\omega_1\pm\i\nu}{2}\right)
\Gamma\!\left(\frac{1+\i\omega_3\pm\i\nu}{2}\right)\Gamma\!\left(\frac{1-\i\omega_2\pm\i\nu'}{2}\right)
\Gamma\!\left(\frac{1-\i\omega_4\pm\i\nu'}{2}\right)} \\
&\times \Gamma\left(-\i(\omega_1+\omega_3)+ 0^+\right)\int_{-\pi}^\pi \frac{\d\phi}{2\pi}\left(4h(\phi)-\i 0^+\right)^{\i(\omega_1+\omega_3)}.
\end{aligned}
\end{equation}
\begin{equation}\label{Fener4}
\begin{aligned}
\mathcal{F}_{14}^{t>0}(\omega_i|m) = &\frac{1}{(2\pi)^3}\frac{1}{(4\pi\alpha)^2}
{\textstyle\Gamma\!\left(\frac{1-\i\omega_1\pm\i\nu}{2}\right)
\Gamma\!\left(\frac{1-\i\omega_3\pm\i\nu}{2}\right)\Gamma\!\left(\frac{1+\i\omega_2\pm\i\nu'}{2}\right)
\Gamma\!\left(\frac{1+\i\omega_4\pm\i\nu'}{2}\right)} \\
&\times \Gamma\left(-\i(\omega_2+\omega_4)+ 0^+\right)\int_{-\pi}^\pi \frac{\d\phi}{2\pi}\left(4h(\phi)+\i 0^+\right)^{\i(\omega_2+\omega_4)}.
\end{aligned}
\end{equation}

In these formulas, the only difference between $\mathcal{F}_{12}$ and $\mathcal{F}_{14}$ is the factor $(h \pm \i 0^+)$. This makes it clear that in the standard case of positive $h > 0$ shocks, the two functions would be equal. We can also see that the expressions are equal perturbatively: the $G$ perturbation theory is obtained from the residues of the poles of the $\Gamma(-\i(\omega_1 + \omega_3))$ or $\Gamma(-\i(\omega_2 + \omega_4))$ functions, and at these residues the power of $(h\pm \i 0^+)$ is a positive integer, so the $\i 0^+$ doesn't matter.

\subsection{Recoil OTOC as a tetrahedron graph on $S^d$}\label{sec:recoil}

In this section, we discuss a different approach to compute the OTOC in energy space, valid in the regime where the masses $m_i, \nu,\nu' \gg 1$ and yet $Gm_i, G\nu,G \nu' \rightarrow 0$. As a consequence, there is no gravitational backreaction in this limit. Nonetheless, the recoil effect of the observer is still present in this limit and leads to interesting features in the OTOC \cite{Kolchmeyer:2024fly}. For now, we will explore this approach independently from the previous sections. The connection of this approach with the eikonal approach will be explored in Section \ref{sec:compare} and further in Section \ref{sec:connect}. 

We will be working in $d$-dimensional Euclidean de Sitter space, i.e. $S^d$. It is convenient to use the embedding space coordinates
\begin{equation}
	X^1,\,..., X^{d+1}, \quad\quad (X^1)^2 + ... + (X^{d+1})^2 = 1\,.
\end{equation}
Since the masses are large, we can approximate the observer and the particles as massive worldlines on $S^d$. This motivates considering the following action
\begin{equation}\label{tetraaction}
	I_{\textrm{tetra}} = \sum_{i<j} \mathbf{m}_{ij} d_{ij} \,, \quad\quad d_{ij} \equiv \arccos \left(\vec{X}_i \cdot \vec{X}_j\right)\,.
\end{equation}
Here $d_{ij}$ is the (shortest) geodesic distance between two points along the observer worldline with embedding space coordinates $\vec{X}_i$ and $\vec{X}_j$, and the matrix $\mathbf{m}$ encodes the masses associated with the edges.  More explicitly, using the convention of Section \ref{sec:otocsetup}, we have
\begin{equation}\label{massmat}
	\mathbf{m} = \begin{pmatrix}
		0 & m_2 & \nu' & m_1 \\
		m_2 & 0 & m_3 & \nu \\
		\nu' & m_3 & 0 & m_4 \\
		m_1 & \nu & m_4 & 0 
	\end{pmatrix}\,.
\end{equation}
We can extremize $I_{\textrm{tetra}}$ over the locations of $\vec{X}_1,...,\vec{X}_4$ to find the saddle point configuration.  We anticipate a saddle point with the topology of a tetrahedron in $S^d$, where the two particle worldlines lie on \emph{opposite} sides of the observer worldline,  see Figure \ref{fig:opposite} (a).  This explains the subscript ``tetra" in the action. As we show in Appendix \ref{app:tetra}, the six geodesic edges in fact lie on a single totally geodesic $S^2$ inside $S^d$, which explains why we can represent the configuration with a two dimensional figure.
\newcommand{\omegaTetraPanel}{%
\begin{tikzpicture}[scale=\tetfigscale, line cap=round, line join=round]
\path[use as bounding box] (-1.55,-1.55) rectangle (1.55,1.55);
\pgfmathdeclarefunction{tetarea}{3}{%
  \pgfmathparse{0.25*sqrt(max(0,(##1+##2-##3)*(##1+##3-##2)*(##2+##3-##1)*(##1+##2+##3)))}%
}
\pgfmathdeclarefunction{tetcos}{6}{%
  \pgfmathparse{((##1*##1+##2*##2-##3*##3)*(##1*##1+##4*##4-##5*##5)
    -2*##1*##1*(##2*##2+##4*##4-##6*##6))/(16*tetarea(##1,##2,##3)*tetarea(##1,##4,##5))}%
}

\pgfmathsetmacro{\mtwelve}{\tetfigmtwo}
\pgfmathsetmacro{\mthirteen}{\tetfignup}
\pgfmathsetmacro{\mfourteen}{\tetfigmone}
\pgfmathsetmacro{\mtwentythree}{\tetfigmthree}
\pgfmathsetmacro{\mtwentyfour}{\tetfignu}
\pgfmathsetmacro{\mthirtyfour}{\tetfigmfour}

\pgfmathsetmacro{\ctwelve}{min(0.999999,max(-0.999999,tetcos(\mtwelve,\mthirteen,\mfourteen,\mtwentythree,\mtwentyfour,\mthirtyfour)))}
\pgfmathsetmacro{\cthirteen}{min(0.999999,max(-0.999999,tetcos(\mthirteen,\mtwelve,\mfourteen,\mtwentythree,\mthirtyfour,\mtwentyfour)))}
\pgfmathsetmacro{\cfourteen}{min(0.999999,max(-0.999999,tetcos(\mfourteen,\mtwelve,\mthirteen,\mtwentyfour,\mthirtyfour,\mtwentythree)))}
\pgfmathsetmacro{\ctwentythree}{min(0.999999,max(-0.999999,tetcos(\mtwentythree,\mtwelve,\mtwentyfour,\mthirteen,\mthirtyfour,\mfourteen)))}
\pgfmathsetmacro{\ctwentyfour}{min(0.999999,max(-0.999999,tetcos(\mtwentyfour,\mtwelve,\mtwentythree,\mfourteen,\mthirtyfour,\mthirteen)))}
\pgfmathsetmacro{\cthirtyfour}{min(0.999999,max(-0.999999,tetcos(\mthirtyfour,\mthirteen,\mtwentythree,\mfourteen,\mtwentyfour,\mtwelve)))}

\pgfmathsetmacro{\sTwelve}{sqrt(max(0,1-\ctwelve*\ctwelve))}
\pgfmathsetmacro{\YoneX}{1}
\pgfmathsetmacro{\YoneY}{0}
\pgfmathsetmacro{\YoneZ}{0}
\pgfmathsetmacro{\YtwoX}{\ctwelve}
\pgfmathsetmacro{\YtwoY}{\sTwelve}
\pgfmathsetmacro{\YtwoZ}{0}
\pgfmathsetmacro{\YthreeX}{\cthirteen}
\pgfmathsetmacro{\YthreeY}{(\ctwentythree-\ctwelve*\cthirteen)/\sTwelve}
\pgfmathsetmacro{\YthreeZ}{sqrt(max(0,1-\YthreeX*\YthreeX-\YthreeY*\YthreeY))}
\pgfmathsetmacro{\YfourX}{\cfourteen}
\pgfmathsetmacro{\YfourY}{(\ctwentyfour-\ctwelve*\cfourteen)/\sTwelve}
\pgfmathsetmacro{\YfourZ}{(\cthirtyfour-\YthreeX*\YfourX-\YthreeY*\YfourY)/\YthreeZ}

\pgfmathsetmacro{\xone}{\YoneX}
\pgfmathsetmacro{\yone}{\YoneY}
\pgfmathsetmacro{\zone}{\YoneZ}
\pgfmathsetmacro{\xtwo}{\YtwoX}
\pgfmathsetmacro{\ytwo}{\YtwoY}
\pgfmathsetmacro{\ztwo}{\YtwoZ}
\pgfmathsetmacro{\xthree}{\YthreeX}
\pgfmathsetmacro{\ythree}{\YthreeY}
\pgfmathsetmacro{\zthree}{\YthreeZ}
\pgfmathsetmacro{\xfour}{\YfourX}
\pgfmathsetmacro{\yfour}{\YfourY}
\pgfmathsetmacro{\zfour}{\YfourZ}

% Optional rotations of the computed configuration, applied in the order x, y, z.
\pgfmathsetmacro{\xoneold}{\xone}
\pgfmathsetmacro{\yoneold}{\yone}
\pgfmathsetmacro{\zoneold}{\zone}
\pgfmathsetmacro{\xonea}{\xoneold}
\pgfmathsetmacro{\yonea}{\yoneold*cos(\figxrot)-\zoneold*sin(\figxrot)}
\pgfmathsetmacro{\zonea}{\yoneold*sin(\figxrot)+\zoneold*cos(\figxrot)}
\pgfmathsetmacro{\xoneb}{\xonea*cos(\figyrot)+\zonea*sin(\figyrot)}
\pgfmathsetmacro{\yoneb}{\yonea}
\pgfmathsetmacro{\zoneb}{-\xonea*sin(\figyrot)+\zonea*cos(\figyrot)}
\pgfmathsetmacro{\xone}{\xoneb*cos(\figzrot)-\yoneb*sin(\figzrot)}
\pgfmathsetmacro{\yone}{\xoneb*sin(\figzrot)+\yoneb*cos(\figzrot)}
\pgfmathsetmacro{\zone}{\zoneb}

\pgfmathsetmacro{\xtwoold}{\xtwo}
\pgfmathsetmacro{\ytwoold}{\ytwo}
\pgfmathsetmacro{\ztwoold}{\ztwo}
\pgfmathsetmacro{\xtwoa}{\xtwoold}
\pgfmathsetmacro{\ytwoa}{\ytwoold*cos(\figxrot)-\ztwoold*sin(\figxrot)}
\pgfmathsetmacro{\ztwoa}{\ytwoold*sin(\figxrot)+\ztwoold*cos(\figxrot)}
\pgfmathsetmacro{\xtwob}{\xtwoa*cos(\figyrot)+\ztwoa*sin(\figyrot)}
\pgfmathsetmacro{\ytwob}{\ytwoa}
\pgfmathsetmacro{\ztwob}{-\xtwoa*sin(\figyrot)+\ztwoa*cos(\figyrot)}
\pgfmathsetmacro{\xtwo}{\xtwob*cos(\figzrot)-\ytwob*sin(\figzrot)}
\pgfmathsetmacro{\ytwo}{\xtwob*sin(\figzrot)+\ytwob*cos(\figzrot)}
\pgfmathsetmacro{\ztwo}{\ztwob}

\pgfmathsetmacro{\xthreeold}{\xthree}
\pgfmathsetmacro{\ythreeold}{\ythree}
\pgfmathsetmacro{\zthreeold}{\zthree}
\pgfmathsetmacro{\xthreea}{\xthreeold}
\pgfmathsetmacro{\ythreea}{\ythreeold*cos(\figxrot)-\zthreeold*sin(\figxrot)}
\pgfmathsetmacro{\zthreea}{\ythreeold*sin(\figxrot)+\zthreeold*cos(\figxrot)}
\pgfmathsetmacro{\xthreeb}{\xthreea*cos(\figyrot)+\zthreea*sin(\figyrot)}
\pgfmathsetmacro{\ythreeb}{\ythreea}
\pgfmathsetmacro{\zthreeb}{-\xthreea*sin(\figyrot)+\zthreea*cos(\figyrot)}
\pgfmathsetmacro{\xthree}{\xthreeb*cos(\figzrot)-\ythreeb*sin(\figzrot)}
\pgfmathsetmacro{\ythree}{\xthreeb*sin(\figzrot)+\ythreeb*cos(\figzrot)}
\pgfmathsetmacro{\zthree}{\zthreeb}

\pgfmathsetmacro{\xfourold}{\xfour}
\pgfmathsetmacro{\yfourold}{\yfour}
\pgfmathsetmacro{\zfourold}{\zfour}
\pgfmathsetmacro{\xfoura}{\xfourold}
\pgfmathsetmacro{\yfoura}{\yfourold*cos(\figxrot)-\zfourold*sin(\figxrot)}
\pgfmathsetmacro{\zfoura}{\yfourold*sin(\figxrot)+\zfourold*cos(\figxrot)}
\pgfmathsetmacro{\xfourb}{\xfoura*cos(\figyrot)+\zfoura*sin(\figyrot)}
\pgfmathsetmacro{\yfourb}{\yfoura}
\pgfmathsetmacro{\zfourb}{-\xfoura*sin(\figyrot)+\zfoura*cos(\figyrot)}
\pgfmathsetmacro{\xfour}{\xfourb*cos(\figzrot)-\yfourb*sin(\figzrot)}
\pgfmathsetmacro{\yfour}{\xfourb*sin(\figzrot)+\yfourb*cos(\figzrot)}
\pgfmathsetmacro{\zfour}{\zfourb}

\pgfmathsetmacro{\vx}{cos(\viewel)*cos(\viewaz)}
\pgfmathsetmacro{\vy}{cos(\viewel)*sin(\viewaz)}
\pgfmathsetmacro{\vz}{sin(\viewel)}

\ifnum\tetfigtwentyfourvertical=1
  \pgfmathsetmacro{\alignx}{\xfour-\xtwo}
  \pgfmathsetmacro{\aligny}{\yfour-\ytwo}
  \pgfmathsetmacro{\alignz}{\zfour-\ztwo}
  \pgfmathsetmacro{\aligndepth}{\alignx*\vx+\aligny*\vy+\alignz*\vz}
  \pgfmathsetmacro{\pyx}{\alignx-\aligndepth*\vx}
  \pgfmathsetmacro{\pyy}{\aligny-\aligndepth*\vy}
  \pgfmathsetmacro{\pyz}{\alignz-\aligndepth*\vz}
\else
  \pgfmathsetmacro{\updepth}{\vz}
  \pgfmathsetmacro{\pyx}{-\updepth*\vx}
  \pgfmathsetmacro{\pyy}{-\updepth*\vy}
  \pgfmathsetmacro{\pyz}{1-\updepth*\vz}
\fi
\pgfmathsetmacro{\pynorm}{sqrt(\pyx*\pyx+\pyy*\pyy+\pyz*\pyz)}
\pgfmathsetmacro{\eyx}{\pyx/\pynorm}
\pgfmathsetmacro{\eyy}{\pyy/\pynorm}
\pgfmathsetmacro{\eyz}{\pyz/\pynorm}

\pgfmathsetmacro{\exx}{\eyy*\vz-\eyz*\vy}
\pgfmathsetmacro{\exy}{\eyz*\vx-\eyx*\vz}
\pgfmathsetmacro{\exz}{\eyx*\vy-\eyy*\vx}

\newcommand{\DrawGeoSegment}[3]{%
  \pgfmathsetmacro{\midtime}{0.5*(##1+##2)}%
  \pgfmathsetmacro{\middepth}{\depthA*cos(\midtime)+\depthN*sin(\midtime)}%
  \ifdim \middepth pt > 0pt
    \draw[thick,##3] plot[domain=##1:##2, samples=60, variable=\t]
      ({\projAx*cos(\t)+\projNx*sin(\t)},
       {\projAy*cos(\t)+\projNy*sin(\t)});
  \else
    \draw[thick,##3!50,dashed] plot[domain=##1:##2, samples=60, variable=\t]
      ({\projAx*cos(\t)+\projNx*sin(\t)},
       {\projAy*cos(\t)+\projNy*sin(\t)});
  \fi
}

\newcommand{\TryRoot}[1]{%
  \pgfmathsetmacro{\candidate}{##1}%
  \ifdim \candidate pt > 0.001pt
    \ifdim \candidate pt < \edgeangle pt
      \edef\rootangle{\candidate}%
      \def\hasroot{1}%
    \fi
  \fi
}

\newcommand{\DrawGeoEdge}[7]{%
  \pgfmathsetmacro{\cosedge}{##1*##4+##2*##5+##3*##6}%
  \pgfmathsetmacro{\edgeangle}{acos(\cosedge)}%
  \pgfmathsetmacro{\sinedge}{sqrt(1-\cosedge*\cosedge)}%
  \pgfmathsetmacro{\nxedge}{(##4-\cosedge*##1)/\sinedge}%
  \pgfmathsetmacro{\nyedge}{(##5-\cosedge*##2)/\sinedge}%
  \pgfmathsetmacro{\nzedge}{(##6-\cosedge*##3)/\sinedge}%
  \pgfmathsetmacro{\projAx}{\sphereR*(\exx*##1+\exy*##2+\exz*##3)}%
  \pgfmathsetmacro{\projAy}{\sphereR*(\eyx*##1+\eyy*##2+\eyz*##3)}%
  \pgfmathsetmacro{\projNx}{\sphereR*(\exx*\nxedge+\exy*\nyedge+\exz*\nzedge)}%
  \pgfmathsetmacro{\projNy}{\sphereR*(\eyx*\nxedge+\eyy*\nyedge+\eyz*\nzedge)}%
  \pgfmathsetmacro{\depthA}{\vx*##1+\vy*##2+\vz*##3}%
  \pgfmathsetmacro{\depthN}{\vx*\nxedge+\vy*\nyedge+\vz*\nzedge}%
  \pgfmathsetmacro{\rootzero}{atan2(-\depthA,\depthN)}%
  \def\hasroot{0}%
  \TryRoot{\rootzero}%
  \TryRoot{\rootzero+180}%
  \TryRoot{\rootzero-180}%
  \TryRoot{\rootzero+360}%
  \TryRoot{\rootzero-360}%
  \ifnum\hasroot=1
    \DrawGeoSegment{0}{\rootangle}{##7}%
    \DrawGeoSegment{\rootangle}{\edgeangle}{##7}%
  \else
    \DrawGeoSegment{0}{\edgeangle}{##7}%
  \fi
}

\newcommand{\PlaceVertex}[6]{%
  \pgfmathsetmacro{\px}{\sphereR*(\exx*##1+\exy*##2+\exz*##3)}%
  \pgfmathsetmacro{\py}{\sphereR*(\eyx*##1+\eyy*##2+\eyz*##3)}%
  \pgfmathsetmacro{\vdepth}{\vx*##1+\vy*##2+\vz*##3}%
  \ifdim \vdepth pt > 0pt
    \fill[##5] (\px,\py) circle (0.04) node[##6,text=##5] {$##4$};%
  \else
    \fill[##5!50] (\px,\py) circle (0.04) node[##6,text=##5!50] {$##4$};%
  \fi
}

\shade[inner color=gray!0, outer color=gray!20] (0,0) circle (\sphereR);
\draw[gray!45] (0,0) circle (\sphereR);

\DrawGeoEdge{\xone}{\yone}{\zone}{\xtwo}{\ytwo}{\ztwo}{black}
\DrawGeoEdge{\xthree}{\ythree}{\zthree}{\xfour}{\yfour}{\zfour}{black}
\DrawGeoEdge{\xone}{\yone}{\zone}{\xthree}{\ythree}{\zthree}{red}
\DrawGeoEdge{\xtwo}{\ytwo}{\ztwo}{\xfour}{\yfour}{\zfour}{blue}
\DrawGeoEdge{\xone}{\yone}{\zone}{\xfour}{\yfour}{\zfour}{black}
\DrawGeoEdge{\xtwo}{\ytwo}{\ztwo}{\xthree}{\ythree}{\zthree}{black}

\PlaceVertex{\xone}{\yone}{\zone}{1}{red}{tetfigone}
\PlaceVertex{\xtwo}{\ytwo}{\ztwo}{2}{blue}{tetfigtwo}
\PlaceVertex{\xthree}{\ythree}{\zthree}{3}{red}{tetfigthree}
\PlaceVertex{\xfour}{\yfour}{\zfour}{4}{blue}{tetfigfour}
\end{tikzpicture}
}%

\newcommand{\sameSidePanel}{%
\begin{tikzpicture}[scale=\samefigscale, line cap=round, line join=round]
\path[use as bounding box] (-1.55,-1.55) rectangle (1.55,1.55);
\pgfmathsetmacro{\sameP}{(4*(\samefigm)^2-(\samefignu)^2)*((\samefignu)^2-(\samefigomega)^2)}
\pgfmathsetmacro{\ctwelve}{min(0.999999,max(-0.999999,((\samefignu)^4+2*\samefigm*\samefigomega*(4*(\samefigm)^2-2*(\samefignu)^2+(\samefigomega)^2)-4*(\samefigm)^2*(\samefigomega)^2)/\sameP))}
\pgfmathsetmacro{\cfourteen}{min(0.999999,max(-0.999999,((\samefignu)^4-2*\samefigm*\samefigomega*(4*(\samefigm)^2-2*(\samefignu)^2+(\samefigomega)^2)-4*(\samefigm)^2*(\samefigomega)^2)/\sameP))}
\pgfmathsetmacro{\cthirteen}{min(0.999999,max(-0.999999,(3*(\samefignu)^4-(\samefignu)^2*(\samefigomega)^2-4*(\samefigm)^2*((\samefignu)^2+(\samefigomega)^2))/\sameP))}
\pgfmathsetmacro{\sTwelve}{sqrt(max(0,1-\ctwelve*\ctwelve))}
\pgfmathsetmacro{\xone}{1}
\pgfmathsetmacro{\yone}{0}
\pgfmathsetmacro{\zone}{0}
\pgfmathsetmacro{\xtwo}{\ctwelve}
\pgfmathsetmacro{\ytwo}{\sTwelve}
\pgfmathsetmacro{\ztwo}{0}
\pgfmathsetmacro{\xthree}{\cthirteen}
\pgfmathsetmacro{\ythree}{(\cfourteen-\ctwelve*\cthirteen)/\sTwelve}
\pgfmathsetmacro{\zthree}{sqrt(max(0,1-\xthree*\xthree-\ythree*\ythree))}
\pgfmathsetmacro{\xfour}{1-\xtwo+\xthree}
\pgfmathsetmacro{\yfour}{-\ytwo+\ythree}
\pgfmathsetmacro{\zfour}{\zthree}

\pgfmathsetmacro{\xoneold}{\xone}
\pgfmathsetmacro{\yoneold}{\yone}
\pgfmathsetmacro{\zoneold}{\zone}
\pgfmathsetmacro{\xonea}{\xoneold}
\pgfmathsetmacro{\yonea}{\yoneold*cos(\figxrot)-\zoneold*sin(\figxrot)}
\pgfmathsetmacro{\zonea}{\yoneold*sin(\figxrot)+\zoneold*cos(\figxrot)}
\pgfmathsetmacro{\xoneb}{\xonea*cos(\figyrot)+\zonea*sin(\figyrot)}
\pgfmathsetmacro{\yoneb}{\yonea}
\pgfmathsetmacro{\zoneb}{-\xonea*sin(\figyrot)+\zonea*cos(\figyrot)}
\pgfmathsetmacro{\xone}{\xoneb*cos(\figzrot)-\yoneb*sin(\figzrot)}
\pgfmathsetmacro{\yone}{\xoneb*sin(\figzrot)+\yoneb*cos(\figzrot)}
\pgfmathsetmacro{\zone}{\zoneb}

\pgfmathsetmacro{\xtwoold}{\xtwo}
\pgfmathsetmacro{\ytwoold}{\ytwo}
\pgfmathsetmacro{\ztwoold}{\ztwo}
\pgfmathsetmacro{\xtwoa}{\xtwoold}
\pgfmathsetmacro{\ytwoa}{\ytwoold*cos(\figxrot)-\ztwoold*sin(\figxrot)}
\pgfmathsetmacro{\ztwoa}{\ytwoold*sin(\figxrot)+\ztwoold*cos(\figxrot)}
\pgfmathsetmacro{\xtwob}{\xtwoa*cos(\figyrot)+\ztwoa*sin(\figyrot)}
\pgfmathsetmacro{\ytwob}{\ytwoa}
\pgfmathsetmacro{\ztwob}{-\xtwoa*sin(\figyrot)+\ztwoa*cos(\figyrot)}
\pgfmathsetmacro{\xtwo}{\xtwob*cos(\figzrot)-\ytwob*sin(\figzrot)}
\pgfmathsetmacro{\ytwo}{\xtwob*sin(\figzrot)+\ytwob*cos(\figzrot)}
\pgfmathsetmacro{\ztwo}{\ztwob}

\pgfmathsetmacro{\xthreeold}{\xthree}
\pgfmathsetmacro{\ythreeold}{\ythree}
\pgfmathsetmacro{\zthreeold}{\zthree}
\pgfmathsetmacro{\xthreea}{\xthreeold}
\pgfmathsetmacro{\ythreea}{\ythreeold*cos(\figxrot)-\zthreeold*sin(\figxrot)}
\pgfmathsetmacro{\zthreea}{\ythreeold*sin(\figxrot)+\zthreeold*cos(\figxrot)}
\pgfmathsetmacro{\xthreeb}{\xthreea*cos(\figyrot)+\zthreea*sin(\figyrot)}
\pgfmathsetmacro{\ythreeb}{\ythreea}
\pgfmathsetmacro{\zthreeb}{-\xthreea*sin(\figyrot)+\zthreea*cos(\figyrot)}
\pgfmathsetmacro{\xthree}{\xthreeb*cos(\figzrot)-\ythreeb*sin(\figzrot)}
\pgfmathsetmacro{\ythree}{\xthreeb*sin(\figzrot)+\ythreeb*cos(\figzrot)}
\pgfmathsetmacro{\zthree}{\zthreeb}

\pgfmathsetmacro{\xfourold}{\xfour}
\pgfmathsetmacro{\yfourold}{\yfour}
\pgfmathsetmacro{\zfourold}{\zfour}
\pgfmathsetmacro{\xfoura}{\xfourold}
\pgfmathsetmacro{\yfoura}{\yfourold*cos(\figxrot)-\zfourold*sin(\figxrot)}
\pgfmathsetmacro{\zfoura}{\yfourold*sin(\figxrot)+\zfourold*cos(\figxrot)}
\pgfmathsetmacro{\xfourb}{\xfoura*cos(\figyrot)+\zfoura*sin(\figyrot)}
\pgfmathsetmacro{\yfourb}{\yfoura}
\pgfmathsetmacro{\zfourb}{-\xfoura*sin(\figyrot)+\zfoura*cos(\figyrot)}
\pgfmathsetmacro{\xfour}{\xfourb*cos(\figzrot)-\yfourb*sin(\figzrot)}
\pgfmathsetmacro{\yfour}{\xfourb*sin(\figzrot)+\yfourb*cos(\figzrot)}
\pgfmathsetmacro{\zfour}{\zfourb}

\pgfmathsetmacro{\vx}{cos(\viewel)*cos(\viewaz)}
\pgfmathsetmacro{\vy}{cos(\viewel)*sin(\viewaz)}
\pgfmathsetmacro{\vz}{sin(\viewel)}

\ifnum\samefigtwentyfourvertical=1
  \pgfmathsetmacro{\alignx}{\xfour-\xtwo}
  \pgfmathsetmacro{\aligny}{\yfour-\ytwo}
  \pgfmathsetmacro{\alignz}{\zfour-\ztwo}
  \pgfmathsetmacro{\aligndepth}{\alignx*\vx+\aligny*\vy+\alignz*\vz}
  \pgfmathsetmacro{\pyx}{\alignx-\aligndepth*\vx}
  \pgfmathsetmacro{\pyy}{\aligny-\aligndepth*\vy}
  \pgfmathsetmacro{\pyz}{\alignz-\aligndepth*\vz}
\else
  \pgfmathsetmacro{\updepth}{\vz}
  \pgfmathsetmacro{\pyx}{-\updepth*\vx}
  \pgfmathsetmacro{\pyy}{-\updepth*\vy}
  \pgfmathsetmacro{\pyz}{1-\updepth*\vz}
\fi
\pgfmathsetmacro{\pynorm}{sqrt(\pyx*\pyx+\pyy*\pyy+\pyz*\pyz)}
\pgfmathsetmacro{\eyx}{\pyx/\pynorm}
\pgfmathsetmacro{\eyy}{\pyy/\pynorm}
\pgfmathsetmacro{\eyz}{\pyz/\pynorm}

\pgfmathsetmacro{\exx}{\eyy*\vz-\eyz*\vy}
\pgfmathsetmacro{\exy}{\eyz*\vx-\eyx*\vz}
\pgfmathsetmacro{\exz}{\eyx*\vy-\eyy*\vx}

\newcommand{\DrawSameSegment}[3]{%
  \pgfmathsetmacro{\midtime}{0.5*(##1+##2)}%
  \pgfmathsetmacro{\middepth}{\depthA*cos(\midtime)+\depthN*sin(\midtime)}%
  \ifdim \middepth pt > 0pt
    \draw[thick,##3] plot[domain=##1:##2, samples=60, variable=\t]
      ({\projAx*cos(\t)+\projNx*sin(\t)},
       {\projAy*cos(\t)+\projNy*sin(\t)});
  \else
    \draw[thick,##3!50,dashed] plot[domain=##1:##2, samples=60, variable=\t]
      ({\projAx*cos(\t)+\projNx*sin(\t)},
       {\projAy*cos(\t)+\projNy*sin(\t)});
  \fi
}

\newcommand{\TrySameRoot}[1]{%
  \pgfmathsetmacro{\candidate}{##1}%
  \ifdim \candidate pt > \arcstart pt
    \ifdim \candidate pt < \arcend pt
      \ifnum\nroots=0
        \edef\rootone{\candidate}%
        \def\nroots{1}%
      \else
        \ifnum\nroots=1
          \ifdim \candidate pt < \rootone pt
            \edef\roottwo{\rootone}%
            \edef\rootone{\candidate}%
          \else
            \edef\roottwo{\candidate}%
          \fi
          \def\nroots{2}%
        \fi
      \fi
    \fi
  \fi
}

\newcommand{\DrawSameEdge}[8]{%
  \pgfmathsetmacro{\cosedge}{##1*##4+##2*##5+##3*##6}%
  \pgfmathsetmacro{\edgeangle}{acos(max(-0.999999,min(0.999999,\cosedge)))}%
  \pgfmathsetmacro{\sinedge}{sqrt(max(0.000001,1-\cosedge*\cosedge))}%
  \pgfmathsetmacro{\nxedge}{(##4-\cosedge*##1)/\sinedge}%
  \pgfmathsetmacro{\nyedge}{(##5-\cosedge*##2)/\sinedge}%
  \pgfmathsetmacro{\nzedge}{(##6-\cosedge*##3)/\sinedge}%
  \pgfmathsetmacro{\projAx}{\sphereR*(\exx*##1+\exy*##2+\exz*##3)}%
  \pgfmathsetmacro{\projAy}{\sphereR*(\eyx*##1+\eyy*##2+\eyz*##3)}%
  \pgfmathsetmacro{\projNx}{\sphereR*(\exx*\nxedge+\exy*\nyedge+\exz*\nzedge)}%
  \pgfmathsetmacro{\projNy}{\sphereR*(\eyx*\nxedge+\eyy*\nyedge+\eyz*\nzedge)}%
  \pgfmathsetmacro{\depthA}{\vx*##1+\vy*##2+\vz*##3}%
  \pgfmathsetmacro{\depthN}{\vx*\nxedge+\vy*\nyedge+\vz*\nzedge}%
  \ifnum##8=1
    \pgfmathsetmacro{\arcstart}{\edgeangle}%
    \pgfmathsetmacro{\arcend}{360}%
  \else
    \pgfmathsetmacro{\arcstart}{0}%
    \pgfmathsetmacro{\arcend}{\edgeangle}%
  \fi
  \pgfmathsetmacro{\rootzero}{atan2(-\depthA,\depthN)}%
  \def\nroots{0}%
  \TrySameRoot{\rootzero}%
  \TrySameRoot{\rootzero+180}%
  \TrySameRoot{\rootzero-180}%
  \TrySameRoot{\rootzero+360}%
  \TrySameRoot{\rootzero-360}%
  \ifnum\nroots=0
    \DrawSameSegment{\arcstart}{\arcend}{##7}%
  \else
    \ifnum\nroots=1
      \DrawSameSegment{\arcstart}{\rootone}{##7}%
      \DrawSameSegment{\rootone}{\arcend}{##7}%
    \else
      \DrawSameSegment{\arcstart}{\rootone}{##7}%
      \DrawSameSegment{\rootone}{\roottwo}{##7}%
      \DrawSameSegment{\roottwo}{\arcend}{##7}%
    \fi
  \fi
}%

\newcommand{\PlaceSameVertex}[6]{%
  \pgfmathsetmacro{\px}{\sphereR*(\exx*##1+\exy*##2+\exz*##3)}%
  \pgfmathsetmacro{\py}{\sphereR*(\eyx*##1+\eyy*##2+\eyz*##3)}%
  \pgfmathsetmacro{\vdepth}{\vx*##1+\vy*##2+\vz*##3}%
  \ifdim \vdepth pt > 0pt
    \fill[##5] (\px,\py) circle (0.04) node[##6,text=##5] {$##4$};%
  \else
    \fill[##5!50] (\px,\py) circle (0.04) node[##6,text=##5!50] {$##4$};%
  \fi
}

\shade[inner color=gray!0, outer color=gray!20] (0,0) circle (\sphereR);
\draw[gray!45] (0,0) circle (\sphereR);
\DrawSameEdge{\xone}{\yone}{\zone}{\xtwo}{\ytwo}{\ztwo}{black}{0}
\DrawSameEdge{\xthree}{\ythree}{\zthree}{\xfour}{\yfour}{\zfour}{black}{0}
\DrawSameEdge{\xone}{\yone}{\zone}{\xfour}{\yfour}{\zfour}{black}{0}
\DrawSameEdge{\xtwo}{\ytwo}{\ztwo}{\xthree}{\ythree}{\zthree}{black}{0}
\DrawSameEdge{\xone}{\yone}{\zone}{\xthree}{\ythree}{\zthree}{red}{1}
\DrawSameEdge{\xtwo}{\ytwo}{\ztwo}{\xfour}{\yfour}{\zfour}{blue}{1}
\PlaceSameVertex{\xone}{\yone}{\zone}{3}{red}{samefigone}
\PlaceSameVertex{\xtwo}{\ytwo}{\ztwo}{2}{blue}{samefigtwo}
\PlaceSameVertex{\xthree}{\ythree}{\zthree}{1}{red}{samefigthree}
\PlaceSameVertex{\xfour}{\yfour}{\zfour}{4}{blue}{samefigfour}
\end{tikzpicture}
}%

\begin{figure}[h!]
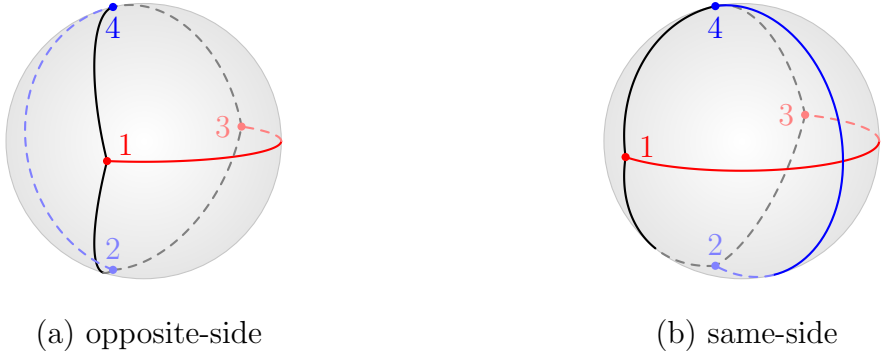

\centering
\begin{minipage}{0.25\textwidth}
\centering
\begingroup
\def\viewaz{100}
\def\viewel{-30}
\def\figxrot{165.522487814}
\def\figyrot{14.963217433}
\def\figzrot{90}
\def\tetfigtwentyfourvertical{1}
\def\tetfigscale{1.35}
\def\sphereR{1.35}
\def\tetfigmone{4}
\def\tetfigmtwo{4}
\def\tetfigmthree{4}
\def\tetfigmfour{4}
\def\tetfignu{2}
\def\tetfignup{2}
\tikzset{
  tetfigone/.style={right,xshift=0pt,yshift=6pt},
  tetfigtwo/.style={above},
  tetfigthree/.style={left},
  tetfigfour/.style={below}
}
\resizebox{\linewidth}{!}{\omegaTetraPanel}
\endgroup
\par\vspace{2pt}{(a) opposite-side}
\end{minipage}\hspace{0.2\textwidth}
\begin{minipage}{0.25\textwidth}
\centering
\begingroup
\def\viewaz{0}
\def\viewel{0}
\def\figxrot{40}
\def\figyrot{-150}
\def\figzrot{0}
\def\samefigtwentyfourvertical{1}
\def\samefigm{4}
\def\samefignu{2}
\def\samefigomega{0}
\def\samefigscale{1.35}
\def\sphereR{1.35}
\tikzset{%I swap one and three here
  samefigone/.style={left},
  samefigtwo/.style={above},
  samefigthree/.style={right,xshift=1pt,yshift=4pt},
  samefigfour/.style={below}
}
\resizebox{\linewidth}{!}{\sameSidePanel}
\endgroup
\par\vspace{2pt}{(b) same-side}
\end{minipage}
\caption{In the large mass yet non-backreacting limit, the OTOC can be captured by geodesic graphs on $S^d$, where (a) the two particle edges are on the opposite side of the observer worldline, forming a tetrahedron shape; and (b) the particle worldlines are on the same side of the observer worldline.}
\label{fig:opposite}
\end{figure}

In Appendix \ref{app:tetra}, we explain the detailed steps to find the saddle point configuration explicitly. The action of the saddle point is  presented in (\ref{allds}) and (\ref{Itetrasaddle}). 
 
 As can be seen from Figure \ref{fig:opposite} (a), since the two particle edges lie on opposite sides of the observer worldline,  the worldline trajectory follows a ``zig-zag" pattern due to the pulls from the particles. This leads to two edges $d_{{\color{red}{13}}}, d_{{\color{blue}{24}}} < \pi$, consistent with our assumption that the edges correspond to the shortest geodesics between points. On the other hand, one can also consider a different configuration where the two worldlines are on the \emph{same} side of the observer worldline, see Figure \ref{fig:opposite} (b). To balance with the forces from the particles, the observer worldline must be displaced in a direction opposite to the particles, leading to two particle edges with lengths greater than $\pi$. The {\color{red}{red}} and {\color{blue}{blue}} worldlines cross in this configuration, but they simply pass through each other in the approximation that they correspond to free fields in the bulk. Further details of this configuration is studied also in Appendix \ref{app:sameside}. Combining the  opposite-side tetrahedron configuration with the same-side configuration, whose action we denote as $I_{\textrm{same}}$,  the pure recoil OTOC in the energy space is then given by
 \begin{equation}
 \bm{\mathcal{F}}_{\rm recoil} \approx e^{- I_{\textrm{tetra}}} + e^{- I_{\textrm{same}}}\,.
 \end{equation}
 Since the same-side configuration is further suppressed compared to the opposite-side solution, unless otherwise noted,  below we will be mostly focusing on the opposite-side configuration.

We also note in passing that even though here we focus on the $G\rightarrow 0$ limit, in the case of dS$_3$ it is possible to extend the construction into the fully backreacted regime. The relevant geometries are spherical tetrahedrons and we plan to report on their properties elsewhere \cite{chen_hartman_stanford_tang_wip}.

\subsubsection{Special configuration with $m_1= m_3,\, m_2 = m_4$}\label{sec:special}

Instead of studying the tetrahedron configuration with the most general masses, we will be considering situations where the six masses satisfy
\begin{equation}
	m_1 = m_3 = m + \frac{\omega}{2}, \qquad m_2 = m_4 = m - \frac{\omega}{2}, \qquad \nu = \nu'\,,
\label{eq: special mass conf1}
\end{equation}
or, using the definition of the frequencies $\omega_i$ in (\ref{omegainverse}), 
\begin{equation}
	\omega_1 = \omega_3 = -\omega_2 = -\omega_4= \omega\,. 
\end{equation}
In this case, using (\ref{allds}) and (\ref{Itetrasaddle}), the saddle point action reduces to
\begin{equation}\label{Itetraomega}
	I_{\textrm{tetra}} =
	2\left(m-\frac{\omega}{2}\right)d_{12}
	+2\nu d_{13}
	+2\left(m+\frac{\omega}{2}\right)d_{14}\,,
\end{equation}
where
\begin{equation}\label{dijs}
\begin{aligned}
d_{12}=d_{34}
={}&\arccos\!\left[
\frac{4m^2\omega^2-\nu^4-2m\omega(4m^2-2\nu^2+\omega^2)}
{(4m^2-\nu^2)(\nu^2-\omega^2)}
\right]\,,\\
d_{14}=d_{23}
={}&\arccos\!\left[
\frac{4m^2\omega^2-\nu^4+2m\omega(4m^2-2\nu^2+\omega^2)}
{(4m^2-\nu^2)(\nu^2-\omega^2)}
\right]\,,\\
d_{13}=d_{24}
={}&\arccos\!\left[
\frac{3\nu^4-\nu^2\omega^2-4m^2(\nu^2+\omega^2)}
{(4m^2-\nu^2)(\nu^2-\omega^2)}
\right]\,.
\end{aligned}
\end{equation}
To understand these expressions, it is useful to examine some special limits. First, as a simple check, if we set $\omega = 0$ and expand the action in the regime $\nu\ll m$, we get
\begin{equation}\label{Itetrasmallnu}
	I_{\textrm{tetra}} \approx 2\pi m + 2\pi \nu - \frac{\nu^2}{m} + \mathcal{O}(\nu^3)\,.
\end{equation}
The $2\pi m$  term comes from the observer worldline wrapping a great circle, while the $2\pi \nu -\nu^2/m = 2\times  \nu (\pi - \nu/(2m))$ comes from the two particle edges, each of length slightly less than $\pi$. Instead, if we set $\nu=m$ so that all six edges have the same mass, all the $d_{ij}$ in (\ref{dijs}) become the same as $\arccos (-1/3)$, which corresponds to the edge length of a regular tetrahedron in $S^d$.

\begin{figure}[h!]
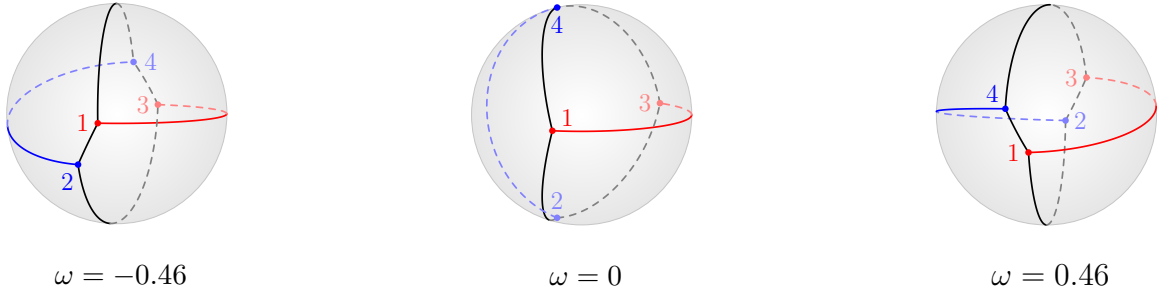

\centering
\def\omegapanelwidth{0.2\textwidth}
\def\omegasidegap{0.04\textwidth}
\hspace*{\omegasidegap}%
\begin{minipage}{\omegapanelwidth}
\centering
\begingroup
\def\viewaz{0}
\def\viewel{0}
\def\figxrot{-118}
\def\figyrot{5}
\def\figzrot{-10}
\def\tetfigtwentyfourvertical{0}
\def\tetfigscale{1.35}
\def\sphereR{1.35}
\def\tetfigmone{3.77}
\def\tetfigmtwo{4.23}
\def\tetfigmthree{3.77}
\def\tetfigmfour{4.23}
\def\tetfignu{2}
\def\tetfignup{2}
\tikzset{
  tetfigone/.style={left},
  tetfigtwo/.style={below,xshift=-5pt,yshift=0pt},
  tetfigthree/.style={left},
  tetfigfour/.style={right,xshift=1pt,yshift=0pt}
}
\resizebox{\linewidth}{!}{\omegaTetraPanel}
\endgroup
\par\vspace{2pt}{\small $\omega=-0.46$}
\end{minipage}\hfill
\begin{minipage}{\omegapanelwidth}
\centering
\begingroup
\def\viewaz{100}
\def\viewel{-30}
\def\figxrot{165.522487814}
\def\figyrot{14.963217433}
\def\figzrot{90}
\def\tetfigtwentyfourvertical{1}
\def\tetfigscale{1.35}
\def\sphereR{1.35}
\def\tetfigmone{4}
\def\tetfigmtwo{4}
\def\tetfigmthree{4}
\def\tetfigmfour{4}
\def\tetfignu{2}
\def\tetfignup{2}
\tikzset{
  tetfigone/.style={right,xshift=0pt,yshift=6pt},
  tetfigtwo/.style={above},
  tetfigthree/.style={left},
  tetfigfour/.style={below}
}
\resizebox{\linewidth}{!}{\omegaTetraPanel}
\endgroup
\par\vspace{2pt}{\small $\omega=0$}
\end{minipage}\hfill
\begin{minipage}{\omegapanelwidth}
\centering
\begingroup
\def\viewaz{0}
\def\viewel{0}
\def\figxrot{-90}
\def\figyrot{20}
\def\figzrot{-10}
\def\tetfigtwentyfourvertical{0}
\def\tetfigscale{1.35}
\def\sphereR{1.35}
\def\tetfigmone{4.23}
\def\tetfigmtwo{3.77}
\def\tetfigmthree{4.23}
\def\tetfigmfour{3.77}
\def\tetfignu{2}
\def\tetfignup{2}
\tikzset{
  tetfigone/.style={left,xshift=0pt,yshift=0pt},
  tetfigtwo/.style={right},
  tetfigthree/.style={left},
  tetfigfour/.style={left,xshift=1pt,yshift=7pt}
}
\resizebox{\linewidth}{!}{\omegaTetraPanel}
\endgroup
\par\vspace{2pt}{$\omega= 0.46$}
\end{minipage}\hspace*{\omegasidegap}
\caption{Some example tetrahedron configurations with different $\omega$. We've chosen $m = 4$ and $\nu=2$ so $\omega = \pm 0.46$ are close to $\pm \nu^2/(2m) =\pm 0.5$ where pairs of points collide. }
\label{fig:omega}
\end{figure}

Now, let's hold fixed $m,\nu$ and look at how (\ref{dijs}) vary with respect to $\omega$. If we start from $\omega=0$ and gradually increase it, the first special place that we encounter is  
\begin{equation}
 \omega\rightarrow \frac{\nu^2}{2m}\,, \quad \textrm{where} \quad	d_{14}, d_{23} \rightarrow 0, \quad d_{12}, d_{34}, d_{13}, d_{24} \rightarrow \pi\,.
\end{equation} 
On the other hand, if we move in the negative direction, we have
\begin{equation}
 \omega\rightarrow -\frac{\nu^2}{2m}\,, \quad \textrm{where} \quad	d_{12}, d_{34} \rightarrow 0, \quad d_{14}, d_{23}, d_{13}, d_{24} \rightarrow \pi\,.
\end{equation} 
Therefore, $\omega = \pm \nu^2/(2m)$, which are branch points of the action (\ref{Itetraomega}), correspond to places where pairs of points collide in the Euclidean tetrahedron. Continuing beyond these points, the tetrahedron saddle becomes Lorentzian. In Appendix~\ref{app: lorenzian tetra}, we explicitly show what Lorentzian configurations the tetrahedron become. Coming back to the Euclidean configurations with $\omega$ in between $\pm \nu^2/(2m)$, we show some examples in Figure \ref{fig:omega}.

\begin{figure}[h!]
\centering
\begin{tikzpicture}[scale=2]
\draw[->] (-2,0) -- (2,0);
\draw[->] (0,-1.5) -- (0,1.5);
\draw[decorate, decoration={zigzag, segment length=5pt, amplitude=1pt}] (0.2,0) -- (1.9,0);
\draw[decorate, decoration={zigzag, segment length=5pt, amplitude=1pt}] (-0.2,0) -- (-1.9,0);
\draw[decorate, decoration={zigzag, segment length=5pt, amplitude=1pt}] (0.7,0.0) -- (1.8,0.0);
\draw[decorate, decoration={zigzag, segment length=5pt, amplitude=1pt}] (-0.7,0.0) -- (-1.8,0.0);
\draw[decorate, decoration={zigzag, segment length=5pt, amplitude=1pt}] (0,1.) -- (0,1.45);
\draw[decorate, decoration={zigzag, segment length=5pt, amplitude=1pt}] (0,-1.) -- (0,-1.45);
\node[circle, fill=black, inner sep=1.5pt] at (0.2,0) {};
\node[circle, fill=black, inner sep=1.5pt] at (-0.2,0) {};
\node[circle, fill=black, inner sep=1.5pt] at (0,1.) {};
\node[circle, fill=black, inner sep=1.5pt] at (0,-1.) {};
\node[font=\small, inner sep=0pt] at (0.7,0.0) {$\times$};
\node[font=\small, inner sep=0pt] at (-0.7,0.0) {$\times$};
\node[font=\small] at (0.3,-0.3) {$\frac{\nu^2}{2m}$};
\node[font=\small] at (-0.3,-0.3) {$-\frac{\nu^2}{2m}$};
\node[font=\small] at (0.7,-0.3) {$\nu$};
\node[font=\small] at (-0.8,-0.3) {$-\nu$};
\node[font=\fontsize{9}{8}\selectfont, anchor=west] at (0.05,1.) {$\i\sqrt{4m^2-2\nu^2}$};
\node[font=\fontsize{9}{8}\selectfont, anchor=west] at (0.05,-1.) {$-\i\sqrt{4m^2-2\nu^2}$};
\node[anchor=north east, font=\small] (omegalabel) at (2,1.5) {$\omega$};
\draw (1.65,1.5) -- (1.65,1.2) -- (2,1.2);
\end{tikzpicture}
\caption{We sketch the analytic structure of the function $I_{\rm tetra}(\omega)$. The branch cuts associated with the dots are two-sheeted, while the branch cuts associated with the crosses are infinite-sheeted.}
\label{fig:omega_branch_cuts}
\end{figure}
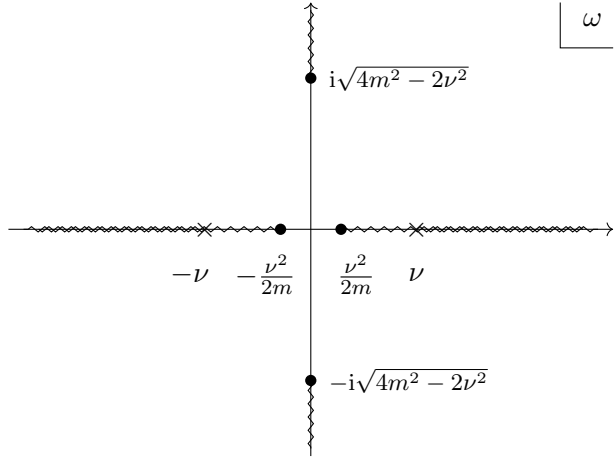

Apart from the branch points $\omega = \pm \nu^2/(2m)$, the semiclassical action $I_{\textrm{tetra}}(\omega)$ for fixed $m,\nu$ also has other branch points in the complex plane of $\omega$, which are $\omega = \pm \nu$ and $\omega = \pm \i \sqrt{4m^2-2\nu^2}$.  The $\omega = \pm \nu$ branch points can be attributed to the various $\Gamma((1\pm \i\omega_i\pm \i\nu)/2)$ factors in the eikonal OTOC. On the other hand, the $\omega = \pm \i \sqrt{4m^2-2\nu^2}$ branch points correspond to saddle point configurations where the length of the particle edges goes to zero. 
We summarize the analytic structure of $I_{\textrm{tetra}}(\omega)$ in Figure \ref{fig:omega_branch_cuts}.

\subsubsection{Comparing with the eikonal OTOCs in the pure recoil limit}\label{sec:compare}

The pure recoil OTOC that we worked out the previous section and the eikonal OTOC in Section \ref{sec:eikonalenergy} have an overlapping regime of validity when
\begin{equation}\label{recoillimit}
m \gg \omega , \nu \gg 1\,, \quad \textrm{while} \quad	Gm, G\omega, G\nu \rightarrow 0\,. 
\end{equation}
Therefore, it is natural to compare the two computations in this limit.\footnote{We will continue to focus on the special case where $m_1 = m_3 = m+ \omega/2,\,m_2 = m_4 = m - \omega/2$ as it simplifies the expressions and already captures the main features.} Of course, an immediate obstruction for making such a comparison is the apparent existence of two different OTOCs $\mathcal{F}_{12}$ and $\mathcal{F}_{14}$, whose differences persist in the limit (\ref{recoillimit}) as we check in Appendix \ref{app:recoileikonal}.  It is therefore not immediately clear which OTOC we should compare with, if either. However, we will put this issue aside temporarily and first work out what various expressions become in this limit.

\paragraph{Opposite-side saddle}
As a first step, we would like to find out what the tetrahedron (opposite-side) saddle becomes in the limit $m\gg \omega,\nu\gg 1$. The branch cuts starting from $\pm\nu^2/(2m)$ as seen in Figure \ref{fig:omega_branch_cuts} join at the origin in the large $m$ limit, and we get two different answers depending on whether $\omega$ is in the upper or lower half plane.
When $\textrm{Im}(\omega) > 0$, the tetrahedron action (\ref{Itetraomega}) gives
\begin{equation}\label{Itetraupper}
	I_{\textrm{tetra}}^{\rm upper} \approx{} 2\pi m + 2\pi \nu  -\i\varphi_+(\omega)\,,
\end{equation}
in which we defined
\begin{equation}\label{phipm}
	\varphi_{\pm}(\omega) \equiv
	2\omega+2\omega\log(\mp\i  4m\omega)
	+2\left[
	(\nu-\omega)\log(\nu-\omega)
	-(\nu+\omega)\log(\nu+\omega)
	\right]\,,
\end{equation}
where depending on the $\pm$ sign, the branch cut of $\log(\omega)$ is chosen to be along the negative/positive imaginary axis   such that $\varphi_\pm (\omega)$ is analytic in the upper/lower half plane.  
%We notice that expression (\ref{phipm}) also correctly reflects the branch points at $\omega = \pm \nu$ and can be analytically continued. 
Note that $\varphi_\pm$ are not real functions when $\omega$ takes real values. If $\textrm{Im}(\omega) < 0$, we instead arrive at
\begin{equation}\label{Itetralower}
	I_{\textrm{tetra}}^{\rm lower} \approx{} 2\pi m + 2\pi \nu  +\i\varphi_-(\omega)\,.
\end{equation}

Now, we would like to compare (\ref{Itetraupper}) and (\ref{Itetralower}) with the eikonal OTOCs in the limit (\ref{recoillimit}). Each individual eikonal OTOC in Section \ref{sec:eikonalenergy} involves an integral over $\phi$. As we explain in Appendix \ref{app:recoileikonal}, in the semiclassical and pure recoil limit, the integral receives contributions from two saddle points at $\phi = 0$ and $\phi=\pi$.\footnote{Away from the pure recoil limit, since $h(\phi)$ is non-analytic at $\phi=0$, it becomes an end point contribution rather than a saddle point.} These two saddle points should be compared with the same-side and opposite-side saddles, respectively. Below, we will use subscript $\phi=0$ and $\phi=\pi$ to distinguish these two contributions. Focusing on the $\phi=\pi$ contribution, as shown in Appendix \ref{app:recoileikonal}, we have
\begin{equation}\label{eikonalrecoilF}
\begin{aligned}
-\log\bm{\mathcal{F}}_{12,\phi= \pi}^{t<0}
\sim{}&
2\pi m+2\pi\nu+2\pi\omega+\i\varphi_+(\omega)\,,\\
-\log\bm{\mathcal{F}}_{12,\phi= \pi}^{t>0}
\sim{}&
2\pi m+2\pi\nu+2\pi\omega-\i\varphi_-(\omega)\,,\\
-\log\bm{\mathcal{F}}_{14,\phi= \pi}^{t<0}
\sim{}&
2\pi m+2\pi\nu-2\pi\omega+\i\varphi_+(\omega)\,,\\
-\log\bm{\mathcal{F}}_{14,\phi= \pi}^{t>0}
\sim{}&
2\pi m+2\pi\nu-2\pi\omega-\i\varphi_-(\omega)\,.
\end{aligned}
\end{equation}
In particular, the $t>0$ functions have  branch cuts along the positive imaginary axis, coming from the semiclassical approximation of the $\Gamma(-\i (\omega_2 + \omega_4)) = \Gamma(2\i \omega)$ factor, while the $t<0$ functions have branch cuts along the negative imaginary axis due to the $\Gamma(-2\i \omega)$ factor.

Comparing (\ref{eikonalrecoilF}) with (\ref{Itetraupper}) and (\ref{Itetralower}), one finds that the upper/lower half planes correspond to $t>0$ and $t<0$, respectively, while $\bm{\mathcal{F}}_{12}$ and $\bm{\mathcal{F}}_{14}$ can be identified with whether $\omega$ is in the left/right half planes. In other words, $I_{\textrm{tetra}}$ in each quadrant has a unique identification with one of the eikonal OTOCs computed in Section \ref{sec:eikonalenergy},\footnote{For example, to see that $I_{\textrm{tetra}}^{\rm upper}
=-\log\bm{\mathcal{F}}_{14,\phi= \pi}^{t>0}$, one uses $\varphi_+(\omega) = \varphi_- (\omega) - 2\pi \i \omega$ when $\textrm{Re}(\omega), \textrm{Im}(\omega) > 0$. } 
\begin{equation}\label{quadrants}
\begin{aligned}
I_{\textrm{tetra}}^{\rm upper}
={}&-\log\bm{\mathcal{F}}_{14,\phi= \pi}^{t>0}\,,
\quad \textrm{Re}(\omega)>0,\quad \textrm{Im}(\omega)>0\,,\\
I_{\textrm{tetra}}^{\rm upper}
={}&-\log\bm{\mathcal{F}}_{12,\phi= \pi}^{t>0}\,,
\quad \textrm{Re}(\omega)<0,\quad \textrm{Im}(\omega)>0\,,\\
I_{\textrm{tetra}}^{\rm lower}
={}&-\log\bm{\mathcal{F}}_{14,\phi= \pi}^{t<0}\,,
\quad \textrm{Re}(\omega)>0,\quad \textrm{Im}(\omega)<0\,,\\
I_{\textrm{tetra}}^{\rm lower}
={}&-\log\bm{\mathcal{F}}_{12,\phi= \pi}^{t<0}\,,
\quad \textrm{Re}(\omega)<0,\quad \textrm{Im}(\omega)<0\,.
\end{aligned}
\end{equation}
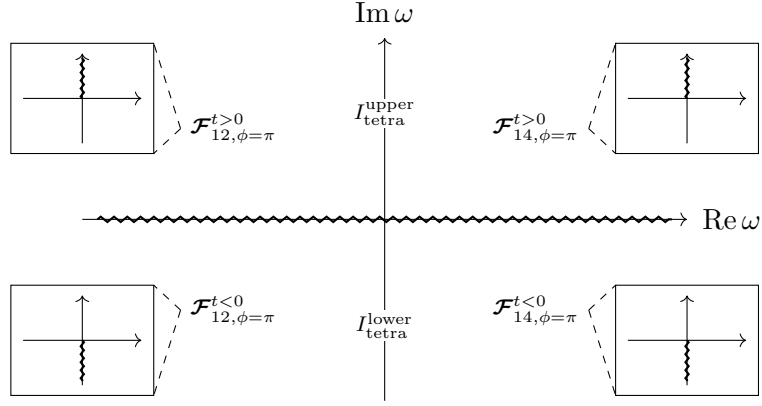
\begin{figure}[t!]
\centering
\begin{tikzpicture}[scale=2]
\draw[->] (-2,0) -- (2,0);
\draw[->] (0,-1.2) -- (0,1.2);
\node[anchor=west, font=\small] at (2.03,0) {$\mathrm{Re}\,\omega$};
\node[anchor=south, font=\small] at (0,1.23) {$\mathrm{Im}\,\omega$};
\draw[decorate, thick, decoration={zigzag, segment length=5pt, amplitude=1pt}] (-1.9,0) -- (1.9,0);
\node[font=\scriptsize, fill=white, inner sep=1pt] at (0,0.7) {$I_{\textrm{tetra}}^{\rm upper}$};
\node[font=\scriptsize, fill=white, inner sep=1pt] at (0,-0.7) {$I_{\textrm{tetra}}^{\rm lower}$};
\node[font=\scriptsize, align=center] (Ffourteentp) at (1,0.6) {$\bm{\mathcal{F}}_{14,\phi= \pi}^{t>0}$};
\node[font=\scriptsize, align=center] (Fonetwotp) at (-1,0.6) {$\bm{\mathcal{F}}_{12,\phi= \pi}^{t>0}$};
\node[font=\scriptsize, align=center] (Ffourteenlt) at (1,-0.6) {$\bm{\mathcal{F}}_{14,\phi= \pi}^{t<0}$};
\node[font=\scriptsize, align=center] (Fonetwolt) at (-1,-0.6) {$\bm{\mathcal{F}}_{12,\phi= \pi}^{t<0}$};
\begin{scope}[shift={(-2.,0.8)}, scale=0.27]
	\fill[white] (-1.75,-1.35) rectangle (1.75,1.35);
	\draw[thin] (-1.75,-1.35) rectangle (1.75,1.35);
	\coordinate (FonetwotpBoxUR) at (1.75,1.35);
	\coordinate (FonetwotpBoxLR) at (1.75,-1.35);
	\draw[->, thin] (-1.45,0) -- (1.45,0);
	\draw[->, thin] (0,-1.1) -- (0,1.1);
	\draw[decorate, thick, decoration={zigzag, segment length=3pt, amplitude=0.6pt}] (0,0) -- (0,1);
	\end{scope}
	\draw[dashed] (-1.35,0.6) -- (FonetwotpBoxUR);
	\draw[dashed] (-1.35,0.6) -- (FonetwotpBoxLR);
		\begin{scope}[shift={(2.,0.8)}, scale=0.27]
	\fill[white] (-1.75,-1.35) rectangle (1.75,1.35);
	\draw[thin] (-1.75,-1.35) rectangle (1.75,1.35);
	\coordinate (FfourteentpBoxUL) at (-1.75,1.35);
	\coordinate (FfourteentpBoxLL) at (-1.75,-1.35);
	\draw[->, thin] (-1.45,0) -- (1.45,0);
	\draw[->, thin] (0,-1.1) -- (0,1.1);
	\draw[decorate, thick, decoration={zigzag, segment length=3pt, amplitude=0.6pt}] (0,0) -- (0,1);
	\end{scope}
		\draw[dashed] (1.35,0.6) -- (FfourteentpBoxUL);
		\draw[dashed] (1.35,0.6) -- (FfourteentpBoxLL);
	\begin{scope}[shift={(-2.,-0.8)}, scale=0.27]
	\fill[white] (-1.75,-1.35) rectangle (1.75,1.35);
	\draw[thin] (-1.75,-1.35) rectangle (1.75,1.35);
	\coordinate (FonetwoltBoxUR) at (1.75,1.35);
	\coordinate (FonetwoltBoxLR) at (1.75,-1.35);
	\draw[->, thin] (-1.45,0) -- (1.45,0);
	\draw[->, thin] (0,-1.1) -- (0,1.1);
	\draw[decorate, thick, decoration={zigzag, segment length=3pt, amplitude=0.6pt}] (0,0) -- (0,-1);
	\end{scope}
		\draw[dashed] (-1.35,-0.6) -- (FonetwoltBoxUR);
		\draw[dashed] (-1.35,-0.6) -- (FonetwoltBoxLR);
	\begin{scope}[shift={(2.,-0.8)}, scale=0.27]
	\fill[white] (-1.75,-1.35) rectangle (1.75,1.35);
	\draw[thin] (-1.75,-1.35) rectangle (1.75,1.35);
	\coordinate (FfourteenltBoxUL) at (-1.75,1.35);
	\coordinate (FfourteenltBoxLL) at (-1.75,-1.35);
	\draw[->, thin] (-1.45,0) -- (1.45,0);
	\draw[->, thin] (0,-1.1) -- (0,1.1);
	\draw[decorate, thick, decoration={zigzag, segment length=3pt, amplitude=0.6pt}] (0,0) -- (0,-1);
	\end{scope}
		\draw[dashed] (1.35,-0.6) -- (FfourteenltBoxUL);
		\draw[dashed] (1.35,-0.6) -- (FfourteenltBoxLL);
\end{tikzpicture}
\caption{In the limit $m\rightarrow \infty$, the semiclassical action $I_{\textrm{tetra}}$ has a branch cut on the real axis, while being analytic in both half planes. In each quadrant, it matches with a different eikonal OTOCs. The boxed insets attached to the four eikonal OTOCs show their respective analytic structures in the semiclassical approximation. We've omitted the branch cuts starting from $\pm \nu$ in the figure as they are common for all the functions.}
\label{fig:identification}
\end{figure}

We can summarize the identifications (\ref{quadrants}) with Figure \ref{fig:identification}, where we also sketch the analytic structure of individual eikonal OTOCs. Let's focus on the upper half plane, since the statement in the lower half plane would be analogous. In the semiclassical limit, both $\log\bm{\mathcal{F}}_{12,\phi= \pi}^{t>0}$ and $\log\bm{\mathcal{F}}_{14,\phi= \pi}^{t>0}$ have cuts along the positive imaginary axis. What we learn from Figure \ref{fig:identification} is that their semiclassical approximation are in fact related by going under the cut into the second sheet, and this analytic continuation is unified by a single semiclassical tetrahedron answer $I_{\rm tetra}^{\rm upper}$. 
An interesting feature is that on the principal sheet, $I_{\rm tetra}$ always picks out the \emph{larger} contribution (smaller action) among $\bm{\mathcal{F}}_{12,\phi= \pi}$ and $\bm{\mathcal{F}}_{14,\phi= \pi}$. For instance, along the real axis approached from the upper half plane, we have
\begin{equation}\label{maxoftwo}
\begin{aligned}
	\log |\bm{\mathcal{F}}_{12,\phi= \pi}^{t>0}| &  =  -2\pi m - 2\pi \nu - \pi \omega - 2\pi \omega \Theta(\omega) \,,\\
	\log |\bm{\mathcal{F}}_{14,\phi= \pi}^{t>0}| &  =  -2\pi m - 2\pi \nu + \pi \omega + 2\pi \omega  \Theta (-\omega) \,,\\
	-\textrm{Re}(I_{\rm tetra}^{\rm upper}) & = \textrm{max}\left\{ \log |\bm{\mathcal{F}}_{12,\phi= \pi}^{t>0}| , \log |\bm{\mathcal{F}}_{14,\phi= \pi}^{t>0}| \right\} \\
    & = -2\pi m - 2\pi \nu + \pi |\omega|\,.
\end{aligned}
\end{equation}
Similar statement holds also for the lower half plane. 

Away from the semiclassical limit, $\bm{\mathcal{F}}_{12,\phi= \pi}^{t>0}$ and $\bm{\mathcal{F}}_{14,\phi= \pi}^{t>0}$ are meromorphic, where the cuts are replaced by series of poles along the positive imaginary axis. Therefore,  there is no clear sense that they can be continued into one another. Nonetheless, $I_{\rm tetra}^{\rm upper}$ provides a semiclassical approximation that interpolates between the two different eikonal OTOCs.

We emphasize that away from the $m\rightarrow \infty$ limit, the branch cuts from $\omega = \pm\nu^2/(2m)$ do not cover the entire real axis, and we can go freely between the upper and lower half plane through the gap around $\omega=0$. Therefore, $I_{\rm tetra}$ also plays the role of uniting the $t>0$ and $t<0$ functions. In summary, instead of finding that $I_{\rm tetra}$ matches with a particular eikonal OTOC, what we are seeing is that a single function $I_{\rm tetra}$ interpolates smoothly between the semiclassical and pure recoil limits of the eikonal OTOCs in (\ref{Fener1}) - (\ref{Fener4}).

\paragraph{Same-side saddle}

The analysis of the same-side saddle is simpler. The reason is that in the eikonal computation, the same-side saddle corresponds to the contribution near $\phi=0$, at which in the pure recoil limit, $h(0) = 1/(2m) > 0$. Therefore, we have a shockwave scattering similar to that of AdS spacetime and the distinction between $\bm{\mathcal{F}}_{12}$ and $\bm{\mathcal{F}}_{14}$ disappears.

 In Appendix \ref{app:sameside}, we worked out what the semiclassical action $I_{\rm same}$ of the same-side configuration is. It shares the same analytic structure as $I_{\rm tetra}$, and in the limit $m\rightarrow \infty$ it has a branch cut on the entire real axis. One finds
 \begin{equation}
 	I_{\rm same}^{\rm upper} \approx 
 	2\pi m+2\pi\nu+\i\varphi_+(\omega)\,,
 	\end{equation}
  \begin{equation}
 	I_{\rm same}^{\rm lower} \approx 
 	2\pi m+2\pi\nu-\i\varphi_-(\omega)\,.
 \end{equation}
 The relevant limits of the eikonal OTOC formulas are also worked out in Appendix \ref{app:recoileikonal}. This agrees with the $\phi=0$ contributions to the eikonal OTOCs,
 \begin{equation}\label{samesideidentify}
 \begin{aligned}
 I_{\rm same}^{\rm upper}
 ={}&-\log\bm{\mathcal{F}}_{12,\phi=0}^{t<0}
 =-\log\bm{\mathcal{F}}_{14,\phi=0}^{t<0}\,,\\
 I_{\rm same}^{\rm lower}
 ={}&-\log\bm{\mathcal{F}}_{12,\phi=0}^{t>0}
 =-\log\bm{\mathcal{F}}_{14,\phi=0}^{t>0}\,.
 \end{aligned}
 \end{equation}
In particular, notice the difference from the opposite-side configuration that here the $t>0$ and $t<0$ OTOCs are associated with lower and upper half planes instead. This makes sense since the $t<0$ function is analytic in the upper half plane, so it can match with $I_{\rm same}^{\rm upper}$ on the entire upper half plane instead of a single quadrant. 

~

 To summarize the main point, what we have found in this section is that the Euclidean tetrahedron saddle point provides a \emph{seemingly} unambiguous answer that interpolates different eikonal OTOC answers. This leads to the hope that perhaps one can follow the Euclidean approach and define an unique answer for the OTOC.  To really achieve this, however, one would need to go beyond the saddle point approximation and define an integration contour non-perturbatively. As we will soon see, one encounters the same difficulties of choosing integration contours, similar to the discussion in Section \ref{sec:wrongsign-JT}.

 \subsection{Connecting the two approaches in the large $\nu'$ limit}\label{sec:connect}
 To clarify the relationship between the eikonal formulas and the tetrahedron saddle point, we will study a model that is intermediate between the two, where $m,{\color{red}\nu'} \gg 1$ but ${\color{blue}\nu}$ is held fixed. We will continue to work in the recoil limit, where $Gm_i,G\nu,G\nu' \to 0$. In this limit, the observer and the ${\color{red}\nu'}$ particle are backreacted into a ``theta graph'' on the sphere, and the OTOC is proportional to the $\langle {\color{blue}WW}\rangle$ two point function on this background. We will choose to study an antipodal configuration labeled by a complex parameter $t$:
\begin{equation}
\begin{tikzpicture}[scale=1, line cap=round, line join=round, baseline=(current bounding box.center)]
  \def\sphereR{1.5}
  \def\axisangle{30}
  \def\axisdepth{0.52}
  \pgfmathsetmacro{\equatordepth}{sqrt(1-\axisdepth*\axisdepth)}
  \coordinate (N) at ({\sphereR*\axisdepth*cos(\axisangle)},
    {\sphereR*\axisdepth*sin(\axisangle)});
  \coordinate (S) at ({-\sphereR*\axisdepth*cos(\axisangle)},
    {-\sphereR*\axisdepth*sin(\axisangle)});

  \shade[inner color=gray!0, outer color=gray!20] (0,0) circle (\sphereR);

  \drawSphereMeridian{210}{thick, red}
  \drawSphereMeridian{105}{thick, black}
  \drawSphereMeridian{315}{thick, black}

  %\draw[thick] (0,0) circle (\sphereR);
  \setSpherePoint{105}{130}{blueA}
  \setSpherePoint{315}{50}{blueB}
  \fill[blue] (blueA) circle (0.05) node[right] {$4$};
  \fill[blue!50] (blueB) circle (0.05) node[right] {$2$};
  \fill[red!50] (N) circle (0.05) node[left] {$3$};
  \fill[red] (S) circle (0.05) node[left] {$1$};
\end{tikzpicture} \hspace{40pt} 
\begin{tikzpicture}[scale=1.2,baseline=(current bounding box.center)]
\draw[thick] (0,0) circle (1);
\draw[red] (-90:1) -- (90:1);
\fill[red] (-90:1) circle (0.08) node[below] {$\theta_1 = -\frac{\pi}{2}$};;
\fill[red] (90:1) circle (0.08) node[above] {$\theta_3 = \frac{\pi}{2}$};
\fill[blue] (10:1) circle (0.08) node[right] {$\theta_2 = \i t$};
\fill[blue] (190:1) circle (0.08) node[left] {$\theta_4 = \pi + \i t$};
\end{tikzpicture}
\end{equation}

The ${\color{red} \nu'}$ particle induces a kink in the observer's trajectory, with some angle $\gamma$ that will be determined below. Then the distance between points ${\color{blue}2}$ and ${\color{blue}4}$ satisfies
\begin{align}
\cos(d_{24}) &= \sin(\theta_2)\sin(\theta_4) + \cos(\gamma)\cos(\theta_2)\cos(\theta_4).
\end{align}
Plugging in $\theta_2 = \i t$ and $\theta_4 = \pi + \i t$ and simplifying, we find
\begin{equation}
\cos(\tfrac{d_{24}}{2}) = \sin(\tfrac{\gamma}{2})\cosh(t).
\end{equation}
The heavy-light OTOC is proportional to the correlator of the light operators on the heavy background. Normalizing by the disconnected answer $\mathcal{F}_{0}$,
\begin{align}\label{gammaexp}
\frac{\mathcal{F}(t,\gamma)}{\mathcal{F}_0} &= \frac{G_\nu(d_{24})}{G_{\nu}(\pi)} = \frac{\sinh[\nu(\pi - d_{24})]}{\nu\sin(d_{24})}=
 {}_2F_1\!\left(
1-i\nu,1+i\nu;\tfrac{3}{2};
\sin^2(\tfrac{\gamma}{2})\cosh^2(t)
\right).
\end{align}
This function has branch-point singularities at $\cosh(t) = \pm1/\sin(\gamma/2)$, when the points ${\color{blue}2}$ and ${\color{blue} 4}$ become null separated.

This answer is a function of the kink angle $\gamma$, and to complete the calculation we need to determine it. We will do this in the limit $\nu'/m \ll 1$, keeping both $\nu',m \gg 1$. If the $\nu'$ particle is emitted perpendicular to the observer's worldline, momentum conservation determines $\gamma = \nu' / m$. If, on the other hand, the $\nu'$ particle is emitted at angle $\theta$ away from perpendicular, then $\gamma = \cos(\theta) \nu'/m$. For the antipodal configuration that we are considering, the length of the $\nu'$ geodesic is independent of the angle, so the probability measure just comes from the volume element $\cos(\theta)\d\theta$. So $\gamma$ does not take a fixed value, but is instead a random variable with distribution
\begin{equation}
\mathbb{E}_\gamma f(\gamma) = \frac{1}{2}\int_{-\frac{\pi}{2}}^{\frac{\pi}{2}} \cos(\theta)\d\theta f(\cos(\theta)\nu'/m),
\end{equation}
where the $1/2$ is for normalization. Applying this expectation value to (\ref{gammaexp}) gives
\begin{equation}\label{4f3largenu}
\frac{\mathcal{F}(t)}{\mathcal{F}_0}= \mathbb{E}_\gamma \frac{\mathcal{F}(t,\gamma)}{\mathcal{F}_0} = {}_3F_2\!\left(
\begin{matrix}
1,1-i\nu,1+i\nu\\
\frac32,\frac32
\end{matrix}
;q^2
\right), \hspace{20pt} q = \frac{\nu'}{2m}\cosh(t).
\end{equation}
Note that for small $\nu'/m$ and sufficiently small $t$, this formula has a convergent power series.\footnote{If we further take $\nu$ large, this leads to an entire function
\begin{equation}
\frac{\mathcal{F}(t)}{\mathcal{F}_0}= {}_1F_2(1;\tfrac{3}{2},\tfrac{3}{2};\hat{q}^2),\hspace{20pt} \hat{q} = \frac{\nu\nu'}{2m}\cosh(t).
\end{equation} Now the branch cuts are replaced by ``ridges'' where the function grows double-exponentially in $t$.} The power series breaks down at branch points $q = \pm 1$ where the two $W$ operators become lightlike separated from each other. The behavior of the function in the $t$ plane is shown below:
\begin{equation}
\begin{tikzpicture}[scale=1.25, line cap=round, line join=round,baseline=(current bounding box.center)]
  \def\xmax{2}
  \def\ymax{2}
  \def\halfpi{1.35}
  \def\branchy{1.15}
  \draw[-, gray!70] (-\xmax,0) -- (\xmax,0);
  \draw[-, gray!70] (0,-\ymax) -- (0,\ymax);
  \draw[black, thick,decorate,decoration={zigzag,segment length=5pt,amplitude=1pt}] (0,\branchy) -- (0,\ymax);
   \draw[black, thick,decorate,decoration={zigzag,segment length=5pt,amplitude=1pt}] (.7*\xmax,\branchy) -- (.7*\xmax,\ymax);
   \draw[black,thick,decorate,decoration={zigzag,segment length=5pt,amplitude=1pt}] (.7*\xmax,-\branchy) -- (.7*\xmax,-\ymax);
   \draw[black, thick,decorate,decoration={zigzag,segment length=5pt,amplitude=1pt}] (-.7*\xmax,-\branchy) -- (-.7*\xmax,-\ymax);
   \draw[black,thick,decorate,decoration={zigzag,segment length=5pt,amplitude=1pt}] (-.7*\xmax,\branchy) -- (-.7*\xmax,\ymax);
  \draw[black, thick,decorate,decoration={zigzag,segment length=5pt,amplitude=1pt}] (0,-\branchy) -- (0,-\ymax);
  \fill[black] (0,\branchy) circle (0.035);
  \fill[black] (0,-\branchy) circle (0.035);
  \fill[black] (.7*\xmax,\branchy) circle (0.035);
  \fill[black] (.7*\xmax,-\branchy) circle (0.035);
  \fill[black] (-.7*\xmax,\branchy) circle (0.035);
  \fill[black] (-.7*\xmax,-\branchy) circle (0.035);

  \draw[thick] (1.8,2.02) -- (1.8,1.7) -- (2.2,1.7);
  \node at (2.01,1.86) {$\i t$};
  \node at (-.5*\halfpi,1.5*\ymax/2) {$\mathcal{F}_{12}^{t>0}$};
  \node at (.5*\halfpi,1.5*\ymax/2) {$\mathcal{F}_{14}^{t>0}$};
  \node at (-.5*\halfpi,-1.5*\ymax/2) {$\mathcal{F}_{12}^{t<0}$};
  \node at (.5*\halfpi,-1.5*\ymax/2) {$\mathcal{F}_{14}^{t<0}$};
    \draw[black, thin] (.7*\xmax,-0.06) -- (.7*\xmax,0.06);
  \draw[black, thin] (-.7*\xmax,-0.06) -- (-.7*\xmax,0.06);
  \node[below] at (.7*\xmax,-0.06) {$\pi$};
  \node[below] at (-.7*\xmax,-0.06) {$-\pi$};
\end{tikzpicture}
\end{equation}

On this plot, we also indicated where in the $t$ plane the function approximates the recoil limit of the eikonal $\mathcal{F}_{12}$ and $\mathcal{F}_{14}$ functions. To establish this, one can simply compute the eikonal formulas by taking the $G\to 0$ and large $\nu'$ limit of the perturbative series (\ref{pertseries}). We will not show the calculation here, but in the antipodal configuration, those limits turn that asymptotic series into one with a finite radius of convergence. Moreover, it exactly matches (\ref{4f3largenu}) but with $q = \frac{\nu'}{2m} \frac{e^{\pm t}}{2}$, depending on whether we are studying $\mathcal{F}^{t>0}$ or $\mathcal{F}^{t<0}$. In particular, because the perturbative series for $\mathcal{F}_{12}$ and $\mathcal{F}_{14}$ are the same, they give the same function within the radius of convergence of the series. Continuing past the branch points where the series breaks down, we get either $\mathcal{F}_{12}$ or $\mathcal{F}_{14}$, depending on the side of the cut. So, in this limit, $\mathcal{F}_{12}$ and $\mathcal{F}_{14}$ are related to each other by analytic continuation through the ``small time'' region where the perturbative series converges.

Let's now discuss the Fourier transform\footnote{The factor of two in the exponent is to agree with the $\omega$ defined in (\ref{eq: special mass conf1}). Because we remain in the antipodal configuration, this is like taking the full Fourier transform and then integrating over the $\omega_i$ variables orthogonal to $\omega$.}
\begin{equation}
\mathcal{F}(\omega) = \int \d t e^{2\i\omega t} \mathcal{F}(t).
\end{equation}
This integral can be considered as a model of the full gravity path integral that computes $\bm{\mathcal{F}}(m_i)$, similar to the discussion in section \ref{sec:wrongsign-JT}. As there, the problem is that the integral is ambiguous: there is a singularity along the real $t$ contour. We can, for example, define
\begin{align}\label{F12F14}
    \mathcal{F}_{12}(\omega) &= \int_{\mathbb{R} + \i \epsilon} \d t e^{2\i\omega t} \mathcal{F}(t)\\
     \mathcal{F}_{14}(\omega) &= \int_{\mathbb{R} -\i\epsilon}\d t e^{2\i\omega t} \mathcal{F}(t).
\end{align}

\begin{figure}
    \begin{center}
        \includegraphics[scale = 1.1]{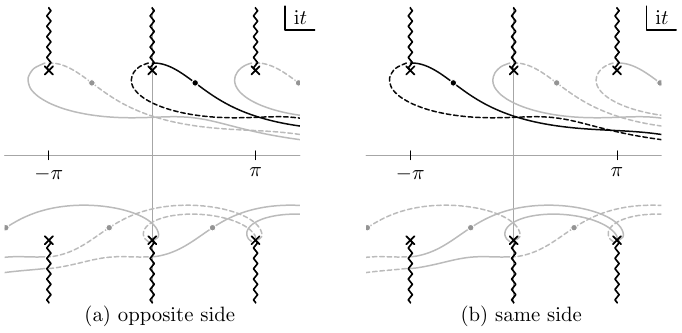}
        \caption{Saddles (dots) and steepest-ascent curves for the semiclassical approximation to $\mathcal{F}(\omega)$, with $\omega/\nu = 0.5 + 0.1\i$ and $\gamma = 0.1$. A saddle contributes if its steepest-ascent curve intersects the defining contour. The main tetrahedron saddle discussed in the previous section is the black dot in panel (a). It contributes to $\mathcal{F}_{14}(\omega)$, for which the defining contour is a little to the right of the imaginary axis, on the first sheet. It does not contribute to $\mathcal{F}_{12}(\omega)$, for which the defining contour is a little to the left of the imaginary axis. The situation would be reversed if $\text{Re}(\omega) < 0$. The ``same side'' saddle discussed in the previous section is shown black in panel (b) and it contributes to both contours.}\label{contourfig}
    \end{center}
\end{figure}
Let's now connect these formulas to the saddle point discussion in the last section. There, we found an ``opposite side'' tetrahedron saddle point that gave the approximate answer for $\mathcal{F}_{14}$ for $\text{Re}(\omega) > 0$ and the approximate answer for $\mathcal{F}_{12}$ for $\text{Re}(\omega) < 0$. We can understand this by looking at the saddle point analysis for the integral
\begin{equation}
\int \d t \exp\left[{\nu\left(2\i\frac{\omega}{\nu}t \pm (\pi - d_{24}(t))\right)}\right].
\end{equation}
Here the $+$ sign is for the ``opposite side'' saddle points as defined in the previous section, and the $-$ sign is for the ``same side.'' In figure \ref{contourfig} we plot the saddle points and steepest-ascent contours. The main ``opposite side'' saddle contributes to only one or the other of $\mathcal{F}_{12}$ and $\mathcal{F}_{14}$, depending on the sign of $\text{Re}(\omega)$. The ``same side'' saddle contributes to both. These correctly reproduce the match of the $\mathcal{F}_{12}$ and $\mathcal{F}_{14}$ with the ``opposite side" and ``same side" saddles in various quadrants, see (\ref{quadrants}) and (\ref{samesideidentify}).

The discussion here also makes clear the difficulty of promoting the semiclassical tetrahedron saddle into a non-perturbatively unambiguous answer for the OTOC. Despite the tetrahedron saddle being Euclidean (for small $\omega$), in order to compute the energy basis OTOC, one would like to integrate the separations between operators along the observer worldline along  imaginary/Lorentzian contours. We do not understand these contours precisely, but we expect that the $t$ integral in (\ref{F12F14}) to play a part.  Again, one would have to confront the choice of which side of the branch cuts to perform the integral, and choosing one-side or the other would break the cyclicity property.

Of course, given that our understanding of integration contour in the gravitational path integral is quite limited, we are not ruling out the possibility that there is in fact a contour that does preserve the cyclicity property and gives an answer that approximates $\mathcal{F}_{12}(\omega)$ and $\mathcal{F}_{14}(\omega)$ in their respective quadrants. As discussed around (\ref{maxoftwo}), since the tetrahedron saddle always approximates the greater of the two, this hypothetical scenario would give an OTOC that is well approximated by the sum (or average) of $\mathcal{F}_{12}$ and $\mathcal{F}_{14}$. It would be interesting to pursue this possibility, though as we will show in Section \ref{sec:chaosbound}, this ultimately does not lead to an answer that is consistent with the conjecture (\ref{conj:static-patch-holography}) despite being cyclic.

\section{A problem for static patch holography}\label{sec:chaosbound}
In this section, we will point out a contradiction between the first perturbative correction to the OTOC and the version of static patch holography conjecture that we formulated in the introduction (\ref{eq:necklace}). The essential point, which was noticed in \cite{Narovlansky:2025tpb}, is that the first perturbative correction violates the chaos bound -- not the bound on the Lyapunov exponent (although that is also violated in the recoil limit \cite{Kolchmeyer:2024fly}), but the more basic bound that the growing corrections should lead the OTOC to decrease rather than increase. 

\subsection{A contradiction}
The chaos bound argument \cite{Maldacena:2015waa} is based on the fact that if a function $f$ is analytic and bounded in a half-infinite strip and satisfies $|f| \le 1$ on its boundary, then $|f| \le 1$ everywhere inside too. This is a version of the maximum modulus principle known as the Phragm\'en--Lindel\"of principle. We will apply this to the function
\begin{align}
f(\i t) = c\cdot \tr(\sigma {\color{blue} W} e^{(\frac{\pi}{2}-\i t)M}\sigma {\color{red} V} e^{(\frac{\pi}{2}+\i t)M}\sigma {\color{blue} W} e^{(\frac{\pi}{2}-\i t)M}\sigma {\color{red} V} e^{(\frac{\pi}{2}+\i t)M}), \hspace{20pt} \sigma = e^{-\delta^2(m-M)^2}.
\end{align}
Here $M$ is the operator that represents the observer mass, $m$ is the target value of the observer mass, and $\delta\sim 1/\sqrt{m}$ is a small parameter. The constant $c$ is chosen so that for small $t\approx t_0$, the function approaches one. The time $t_0$ is chosen large enough to suppress correlations between $V,W$ but small compared to the scrambling time, see \cite{Maldacena:2015waa}. We will assume that for nonzero $G$, this function is analytic in the strip $-\frac{\pi}{2} < \text{Im}(t) < \frac{\pi}{2}$, $\text{Re}(t) > t_0$. This seems reasonable because if the trace is over a finite number $e^{S_{\rm dS}}$ of states as suggested by (\ref{eqn:sphere}) then $f$ is actually entire.

To derive the contradiction, the first step is to assume the static patch holography conjecture 
\begin{equation}\label{falsified}
    \bm{\mathcal{F}}(m_1,m_2,m_3,m_4) \stackrel{?}{=} \tr\!\left({\color{blue} W}\Pi_{m_4}{\color{red} V}\Pi_{m_3}
{\color{blue} W}\Pi_{m_2}{\color{red} V}\Pi_{m_1}\right).
\end{equation}
and then compare $f(\i t)$ to a special case of the OTOC we computed above, namely
\begin{align}
\mathcal{F}(\i t) &\equiv \mathcal{F}(-\tfrac{\pi}{2},\i t,\tfrac{\pi}{2}, \pi + \i t|m)\\
&\propto \int \d^4 m_i \delta(m - \bar{m})e^{\i t(m_1 -m_2 + m_3 - m_4) + 2\pi \bar{m}}\bm{\mathcal{F}}(m_i), \hspace{20pt} \bar{m} \equiv \tfrac{1}{4}{\textstyle \sum_i}m_i.\label{F(t)}
\end{align}
Apart from normalization, the only difference is that in one case we have an insertion of $\delta(m - \bar{m})$ and in the other case we have an insertion of 
\begin{equation}\label{imprecision}
\prod_{i =1}^4 \sigma(m_i) = e^{-4\delta^2(m-\overline{m})^2-O(\delta^2\omega^2)},
\end{equation}
where $\omega$ here stands for a difference $m_i - m_j$. Comparing the RHS of this expression to $\delta(m - \bar{m})$, there are two types of error. One is due to the nonzero variance $\frac{1}{\delta^2}$ in the resulting distribution for $\bar{m}$. This leads to corrections of order $\frac{1}{\delta^2} \partial_m^2\mathcal{F}(\i t)\sim\frac{1}{\delta^2m^2}$. The second type of error is due to the $\delta^2\omega^2$ factor in the exponent. Away from a singularity of the function $\mathcal{F}$, the typical $\omega$ will also be of order one. So, choosing $\delta\sim \frac{1}{\sqrt{m}}$, both errors are small, of order $\frac{1}{m}$. Our perturbative computation of the OTOC implies that near the center of the strip, $\mathcal{F}(\i t)$ exceeds its early time value. Choosing $t$ to be large enough (but still small compared to the scrambling time, so the first perturbative correction is accurate), it will exceed the early time value by an amount bigger than errors just discussed. So the first perturbative term in our OTOC computation, together with the conjecture (\ref{falsified}), implies that $|f(t)|$ exceeds one inside the strip.

To show that this is a contradiction, we just need to show that $|f|\le 1$ along the edges of the strip, using the Cauchy-Schwarz argument from \cite{Maldacena:2015waa}. We will repeat this argument explicitly, in order to emphasize the two different Cauchy-Schwarz cuts that are used. It is useful to resolve the argument $\i t$ into its real and imaginary parts, and write
\begin{align}
f(\tau+\i s)&= c\cdot \tr\!\left(\sigma {\color{blue} W} e^{(\frac{\pi}{2}-\tau-\i s)M}\sigma {\color{red} V} e^{(\frac{\pi}{2}+\tau+\i s)M}\sigma {\color{blue} W} e^{(\frac{\pi}{2}-\tau-\i s)M}\sigma {\color{red} V} e^{(\frac{\pi}{2}+\tau+\i s)M}\right).
\end{align}
Defining
\begin{equation}
A_{12}=e^{\alpha M}\sqrt{\sigma} {\color{blue} W} e^{(\frac{\pi}{2}-\tau-\i s)M}\sigma {\color{red} V} e^{(\frac{\pi}{2}+\tau+\i s)M}\sqrt{\sigma} e^{-\alpha M},
\end{equation}
we can use 
\begin{align}
|f(\tau+\i s)|&=c\left|\tr(A_{12}^2)\right|\le c\tr(A_{12}A_{12}^\dagger)\notag\\
&=c\tr\!\left({\color{blue} W}\sigma e^{2\alpha M} {\color{blue} W} \sigma e^{(\frac{\pi}{2}-\tau-\i s)M}{\color{red} V} \sigma e^{(\pi + 2\tau -2\alpha)M}{\color{red} V} \sigma e^{(\frac{\pi}{2}-\tau + \i s)M}\right)\notag\\
&\approx \frac{G_\nu(2\alpha)G_{\nu'}(\pi + 2\tau - 2\alpha)}{G_\nu(\pi)G_{\nu'}(\pi)}.
\end{align}
In the last step, we assumed that time-ordered correlators approximately factorize.\footnote{We believe this is a safe assumption of the bulk theory, based on the idea of replacing the $WW$ operators by an OPE. The factorized answer is the contribution of the identity, and the corrections involve propagators of physical fields, which will be small for small $G$ and large $m$ and should not grow exponentially in time.} This gives a family of bounds parametrized by $\alpha$, and it can be optimized by minimizing with respect to $\alpha$ over the region where $2\alpha,\pi + 2\tau - 2\alpha$ are both positive. Alternatively, if we define
\begin{equation}
A_{14}=e^{\alpha M}\sqrt{\sigma} {\color{red} V} e^{(\frac{\pi}{2}+\tau+\i s)M}\sigma {\color{blue} W} e^{(\frac{\pi}{2}-\tau-\i s)M}\sqrt{\sigma} e^{-\alpha M},
\end{equation}
then one can use cyclicity of the trace to write
\begin{align}
|f(\tau+\i s)|&=c\left|\tr(A_{14}^2)\right|\le c\tr(A_{14}A_{14}^\dagger)\notag\\
&=c\tr\!\left({\color{red} V}\sigma e^{2\alpha M}{\color{red} V}\sigma e^{(\frac{\pi}{2}+\tau+\i s)M}{\color{blue} W}\sigma e^{(\pi-2\tau-2\alpha)M}{\color{blue} W}\sigma e^{(\frac{\pi}{2}+\tau-\i s)M}\right)\notag\\
&\approx \frac{G_{\nu'}(2\alpha)G_\nu(\pi-2\tau-2\alpha)}{G_\nu(\pi)G_{\nu'}(\pi)}.
\end{align}
This gives a second family of bounds, valid when $2\alpha$ and $\pi-2\tau-2\alpha$ are positive. 

The two Cauchy-Schwarz upper bounds can be combined by taking the smaller of the two. But analyticity gives a much stronger way to combine them. On the two edges of the strip, $\tau=\pm\frac{\pi}{2}$, choosing $\alpha=\frac{\pi}{2}$ in the appropriate channel gives $
|f(\pm\tfrac{\pi}{2}+\i s)|\le 1.$ The same bounds imply that $|f|$ is bounded by something (not necessarily one) everywhere inside the strip. The maximum modulus principle then implies that in fact $|f(\tau+\i s)|\le 1$ everywhere in the strip.

\subsection{\texorpdfstring{Comment on $\mathcal{F}_{12}$ and $\mathcal{F}_{14}$}{Comment on F12 and F14}}
This argument also clarifies the meaning of the two candidate OTOCs, $\mathcal{F}_{12}$ and $\mathcal{F}_{14}$. They satisfy the respective Cauchy-Schwarz bounds associated with $A_{12}$ and $A_{14}$. In each case the inequality can be proven by using $|e^{\i p_+ p_- h}|\le 1$ inside the integral formula for the eikonal OTOCs. However, $\mathcal{F}_{12}$ does not satisfy the inequality associated to $A_{14}$, and $\mathcal{F}_{14}$ does not satisfy the inequality associated to $A_{12}$. The above argument implies that no function can satisfy both inequalities if the first perturbative term has the sign we computed.  The argument also shows that the more robust problem is with positivity, not cyclicity. The average of $\mathcal{F}_{12}$ and $\mathcal{F}_{14}$ would be cyclic but would violate positivity -- this function violates both Cauchy-Schwarz inequalities above.

\subsection{Comment on Euclidean correlators}

In the argument above, we found a problem with Lorentzian correlators. This implies a problem with the Euclidean correlation functions, by a variant \cite{Birke:2002fj} of the Osterwalder-Schrader reconstruction theorem, which states that if Euclidean correlators on $S^1$ satisfy continuity, translation invariance, KMS, and reflection positivity, then they can be continued to Lorentzian correlators that satisfy unitarity and KMS. It would be interesting to find the simplest manifestation of the problem with Euclidean correlators.

\section{Discussion}\label{discussion}
The main purpose of this paper was to compute out of time order correlators (OTOCs) along an observer's worldline in de Sitter space. We worked in an eikonal approximation where the gravitational constant $G$ and the inverse observer mass $1/m$ are small, retaining the leading exponentially growing effects at each order in these small parameters. Such growing corrections can arise from two qualitatively different phenomena. First, the momentum recoil of an observer after creating a bulk particle leads to a series in $(e^t/m)^2$ \cite{Kolchmeyer:2024fly}. Second, gravitational backreaction of the emitted particle leads to a series in $G e^t$ \cite{Aalsma:2020aib}. A pleasant surprise was that both effects can be analyzed together using 't Hooft's shock wave formalism for gravitational scattering \cite{tHooft:1987vrq} -- the recoil effect is mediated by a set of would-be pure-gauge zero modes in the shock wave equation that are lifted and made physical by the existence of the observer.

Within the approximation described above, the OTOC has an asymptotic series in $G e^t$ and $e^t/m$. Schematically, 
\begin{equation}
\text{OTOC} \sim 1 + G e^t + (G^2 + G/m + 1/m^2) e^{2t} + \dots
\end{equation} 
It has been emphasized recently \cite{Chandrasekaran:2022cip,Witten:2023xze} that correlators in de Sitter can be studied in the infinite temperature tracial state. But they also make sense with Boltzmann factors inserted, and we made use of this to study OTOCs with operators spaced at arbitrary positions in imaginary time. The phase of the coefficients then depends on the particular configuration.  For the ``regularized'' configuration where the operators are equally spaced in imaginary time, the total coefficient at each order is positive. This is to be contrasted with the black hole case, where the coefficients would alternate in sign. This is a direct symptom of the ``time-advance'' nature of gravitational backreaction in de Sitter, where matter makes the Penrose diagram taller, and brings antipodal observers closer to causal contact.

Because the coefficients are all positive, the (Borel) resummation of this asymptotic series is ambiguous. To some extent, this ambiguity is not present for other configurations of points, such as the one where all operators have the same imaginary part of time. In that case, the singularity moves away from the real Borel axis, and there is a natural choice of resummation. However, this natural choice does not respect the cyclicity/KMS relation, because halfway through the continuation involved in that relation, the singularity crosses the real positive Borel axis. An average over contours that go above and below the singularity would preserve cyclicity, but would be unnatural in the sense just described. In general, we feel that the more complete perspective is that there are simply two different OTOC functions, corresponding to the different operator orderings that arise due to the causal contact mentioned above.
\begin{figure}[t]
\begin{center}
\begin{tikzpicture}[scale=1, line cap=round, line join=round, baseline=(current bounding box.center)]
  \def\R{1.35}
  \def\Ry{0.4}
  \shade[inner color=gray!0, outer color=gray!20] (0,0) circle (\R);
  \draw[thick, gray!55, dashed] (90:{\Ry} and {\R}) arc (90:270:{\Ry} and {\R});
  \draw[thick] (90:{\Ry} and {\R}) arc (90:-90:{\Ry} and {\R});
  \fill[blue!50] (140:{\Ry} and {\R}) circle (0.08) node[left] {$ W$};
  \fill[blue] (-40:{\Ry} and {\R}) circle (0.08) node[left] {${\color{blue} W}$};
  \fill[red] (20:{\Ry} and {\R}) circle (0.08) node[left] {${\color{red} V}$};
  \fill[red!50] (-160:{\Ry} and {\R}) circle (0.08) node[left] {$ V$};
\end{tikzpicture}
\hspace{18pt}
\begin{tikzpicture}[scale=1, line cap=round, line join=round, baseline=(current bounding box.center)]
  \def\R{1.35}
  \def\Ry{0.4}
  \shade[inner color=gray!0, outer color=gray!20] (0,0) circle (\R);
  \draw[thick, gray!55, dashed] (90:{\Ry} and {\R}) arc (90:270:{\Ry} and {\R});
  \draw[thick] (90:{\Ry} and {\R}) arc (90:-90:{\Ry} and {\R});
  \draw[very thick, green!60!black, opacity=0.5, dash pattern=on 8pt off 4pt] (\R,0) arc (0:180:{\R} and {0.15});
  \draw[very thick, green!60!black, dash pattern=on 8pt off 4pt] (-\R,0) arc (180:360:{\R} and {0.15});
  % \draw[very thick, green!60!black] (-\R,0) arc (180:360:{\R} and {0.15});
  \fill[blue!50] (140:{\Ry} and {\R}) circle (0.08);
  \fill[blue] (-40:{\Ry} and {\R}) circle (0.08);
  \fill[red] (20:{\Ry} and {\R}) circle (0.08);
  \fill[red!50] (-160:{\Ry} and {\R}) circle (0.08);
\end{tikzpicture}
\hspace{18pt}
\begin{tikzpicture}[scale=1, line cap=round, line join=round, baseline=(current bounding box.center)]
  \def\R{1.35}
  \def\Ry{0.4}
  \pgfmathsetmacro{\Rperp}{sqrt(\R*\R-\Ry*\Ry)}
  \shade[inner color=gray!0, outer color=gray!20] (0,0) circle (\R);
  \draw[thick, gray!55, dashed] (90:{\Ry} and {\R}) arc (90:270:{\Ry} and {\R});
   % \draw[very thick, green!60!black] (90:{\Rperp} and {\R}) arc (90:-90:{\Rperp} and {\R});
  \draw[very thick, green!60!black, opacity=0.5, dash pattern=on 8pt off 4pt] (90:{\Rperp} and {\R}) arc (90:-90:{\Rperp} and {\R});
  \draw[thick] (90:{\Ry} and {\R}) arc (90:-90:{\Ry} and {\R});
  \draw[very thick, green!60!black, dash pattern=on 8pt off 4pt] (90:{\Rperp} and {\R}) arc (90:270:{\Rperp} and {\R});
  % \draw[very thick, green!60!black] (90:{\Rperp} and {\R}) arc (90:270:{\Rperp} and {\R});
  \fill[blue!50] (140:{\Ry} and {\R}) circle (0.08);
  \fill[blue] (-40:{\Ry} and {\R}) circle (0.08);
  \fill[red] (20:{\Ry} and {\R}) circle (0.08);
  \fill[red!50] (-160:{\Ry} and {\R}) circle (0.08);
\end{tikzpicture}
\end{center}
\caption{The static patch holography conjecture requires correlation function along the observer worldline to have the cyclicity and positivity properties of a trace. Cyclicity will be true in perturbation theory, because the observer trajectory forms a circle (left). Positivity is more obviously subtle, because of the conformal mode problem in Euclidean gravity. In section \ref{sec:chaosbound}, we showed that the first perturbative correction to the OTOC is already inconsistent with analyticity together with reflection positivity across both the horizontal and vertical cuts shown.}\label{fig:discussion}
\end{figure}
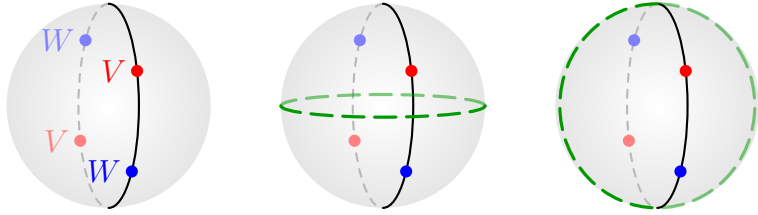

Putting aside the question of resummation, there is actually a more basic puzzle associated to the sign of the leading perturbative term (either $G e^t$ or $e^{2t}/m^2$). This causes the regularized OTOC to {\it grow} in magnitude \cite{Aalsma:2020aib,Kolchmeyer:2024fly,Narovlansky:2025tpb}, which violates the chaos bound. We used this to falsify an otherwise appealing version of the static patch holography conjecture. Roughly, this conjecture states that correlation functions along an observer's worldline can be understood as a trace. The violation of the chaos bound implies a contradiction with the positivity properties of such a trace, see figure \ref{fig:discussion}.

On the one hand, one might be tempted to ignore this problem, because it is not clear how an observer would measure the OTOCs we computed. Instead, without having a specific suggestion to offer, we remain hopeful that the distinctive ``wrong'' sign in the OTOC will eventually be understood as an essential feature of the correct fundamental description of de Sitter space.

\section*{Acknowledgements}
We are grateful to Gauri Batra, Jan Boruch, Frank Ferrari,  Tom Hartman, Xuyao Hu, Veronika Hubeny, David Kolchmeyer, Juan Maldacena, Henry Maxfield, Alexey Milekhin, Xiao-Liang Qi, Mukund Rangamani, Steve Shenker, Edward Witten, Jiuci Xu, Cynthia Yan, Wenwen Zheng for discussions. This work was supported in part by DOE grant DE-SC0026143 and by a
grant from the Simons foundation (926198, DS). YC is also supported by the Simons Foundation Empire Faculty Fellowship. ZY is supported by NSFC Grant No. 12447108, 12342501, 12475071. HT is supported by Shoucheng Zhang Graduate Fellowship. ZY gratefully acknowledges support from the Simons Center for Geometry and Physics, Stony Brook University at which some of the research for this paper was performed.

\appendix

%\newpage
\section{Static patch JT gravity}\label{app:static_patch_jt_gravity}

We now introduce a simple model for de Sitter static patch holography. It is a static patch version of de Sitter JT gravity, designed to capture the sign differences characteristic of higher-dimensional de Sitter gravity. The action is the dilaton-gravity action with $U(\phi)=-2\phi$:
\begin{equation}\label{eqn:dSJTaction}
	I=-\frac{1}{2}\int_{D} \d^2 x\sqrt{g}(\phi R-2\phi)- \frac{1-4Gm}{4G}\int_{S^1} \sqrt{h}.
\end{equation}
Here $D$ denotes the two-dimensional manifold, which has disk topology, and $S^1$ is its boundary circle. The induced boundary metric is $h$. We model the observer as a particle of proper mass $m$ living on $S^1$, and gauge the boundary diffeomorphism group.

We will study this model on its own right, but it could also be motivated as a spherical reduction of dS$_3$ adapted to the observer's worldline \cite{Svesko:2022txo,Rahman:2022jsf,Susskind:2022bia,Batra:2024qju}. In the Euclidean geometry, choose the observer to lie along a great circle of $S^3$ and reduce on the transverse angular circle. The radius of this circle is proportional to the dilaton field $r = 4G\phi$.
The result is a two-dimensional dilaton-gravity theory on a disk $D$, whose boundary $S^1$ represents the observer's great circle on $S^3$. After analytic continuation, this boundary geodesic becomes the center of the dS$_3$ static patch. The observer's backreaction creates a conical defect with a local spatial geometry:
\be
\d \rho^2+r^2 \d\vartheta^2\equiv\d \rho^2+(1-4Gm)^2\rho^2 \d\vartheta^2, \quad\vartheta\sim\vartheta+2\pi, \quad \rho\ll 1.
\ee
This fixes the normal derivative of the dilaton, $4G\partial_n\phi=1-4Gm$, which is the origin of the boundary term in \eqref{eqn:dSJTaction}.
The on-shell action of \eqref{eqn:dSJTaction} comes entirely from this boundary term and gives the entropy of the system in the presence of an observer of mass $m$: $S_m=\frac{2\pi(1-4Gm)}{4G}$.

So far, we have specified the classical action but not yet the integration contour of the field. In the standard definition of Euclidean path integral of JT gravity, the dilaton is defined to be integrated along the imaginary axis, giving a functional delta function that restricts the 2d metric to be hyperbolic. Repeating that step here, one finds the constraint $R = 2$ so the 2d metric is spherical, part of an $S^2$. What remains is to integrate over immersions of the disk $D^2$ in $S^2$.\footnote{An immersion is different from an embedding because the image is allowed to self overlap. The importance of this distinction was pointed out in \cite{Ferrari:2024ndr}. It will not matter in our one-loop computations below.} The static patch JT action reduces to that of a relativistic particle with negative mass set by the de Sitter entropy:
\be
I=-{S_m\over 2\pi} \int_{S^1} \sqrt{h}.
\ee
This is analogous to the Schwarzian action that arises in the standard JT theory. There, the action is bounded from below and the integral is convergent. Here, the action is bounded from above and the integral is not convergent. It would be interesting to try to make sense of this theory nonperturbatively. In this appendix, we will just study it in the one-loop approximation around the saddle point where the immersed disk is a hemisphere of $S^2$. Most of the fluctuations are unstable, and we will Wick-rotate them. One mode is stable, and we will not Wick-rotate it. This situation is analogous to the conformal modes in the higher dimensional sphere partition function \cite{Polchinski:1988ua}.

An immersion of the disk $D^2$ in $S^2$ can be parametrized by a locally invertible holomorphic map $\xi(z)$:\footnote{Or equivalently meromorphic functions with simple poles and nonzero derivatives on $D^2$, we thank Frank Ferrari for pointing this out to us.}
\begin{equation}\label{eqn:ccdisk}
	\d s^2={4\d\xi\d\bar\xi\over (1+\xi\bar\xi)^2}={4\partial\xi\bar\partial\bar\xi \over (1+\xi\bar\xi)^2} \d z \d \bar z\equiv e^{2\omega}\d z\d\bar z,~~~|z|\leq 1.
\end{equation}
The space of maps $\xi(z)$ is not identical to the space of constant-positive-curvature disks, for two reasons. The first is the $SO(3)$ isometry group of the round-sphere metric \eqref{eqn:ccdisk}:
\begin{equation}
	SO(3)_L:~~~\xi\rightarrow {\alpha \xi+\beta\over \bar\beta \xi-\bar\alpha},~~~\alpha,\beta \in \mathbb{C};
\end{equation}
The second is the group of $PSL(2,\mathbb R)$ conformal Killing transformations of the disk:
\begin{equation}\label{eqn:PSL2}
	PSL(2,\mathbb R)_R:~~~~z\rightarrow {\alpha' z+\beta'\over \bar\beta' z+\bar \alpha'},~~~\alpha',\beta' \in \mathbb{C}.
\end{equation}
Both transformations describe the same geometry: the first leaves $\omega$ unchanged, while the second is a coordinate transformation. We should therefore integrate only over the space of $\xi(z)$ modulo $SO(3)_L\times PSL(2,\mathbb R)_R$, with an action given by the boundary length:\footnote{The measure for $\xi(z)$ could be induced from the ultralocal measure for $\omega$; it would also include the ghosts, the Jacobian from the delta function $\delta(R-2)$, the Liouville action from gauge fixing, and the change of ultralocal measure from the metric to $\omega$.}
\begin{equation}\label{eqn:JTAction2}
	Z(m)=\int {\mathcal{D}\xi(z)\over SO(3)_L\times PSL(2,\mathbb R)_R} \exp\left({S_m\over \pi}\int_{|z|=1}|\d z| \sqrt{{\partial\xi\bar\partial\bar\xi \over (1+\xi\bar\xi)^2}}\right).
\end{equation}
%Perturbatively, we can think of such an immersion as an embedding of the disk into the round sphere. This allows us to visualize these modes as a relativistic particle moving on $S^2$ with winding number one. However, as emphasized in \cite{Ferrari:2024ndr}, the full space of immersions is much larger than the space of embeddings.

Let us now contemplate one point before moving on. From the perspective of 2d dilaton gravity, we have considered this model with fixed boundary value of $\phi_{\partial}=0$ and $\partial_n\phi_{\partial}=\frac{1}{4G}-m$, as motivated from observer in dS$_3$. Those are boundary conditions for a  microcanonical partition function in dilaton gravity. From the intrinsic dilaton-gravity point of view, however, one could instead study a canonical partition function with fixed $\phi_{\partial}=0$ and fixed total boundary proper length $\int_{S^1} \sqrt{h}=\beta$. At first sight, the classical model becomes purely topological, with two infinite families of zero modes: one family of zero modes are the immersions with fixed boundary length; and the other is a family of dilaton solutions at the particular value of $\beta=2\pi$.
To see them, consider the classical solution in round-sphere coordinates:
\be
\d s^2=\sin^2(\rho) \d\theta^2+\d\rho^2;~~~\phi=\phi_h \cos\rho.
\ee
We look for solutions satisfying the boundary conditions:
\be
\phi_{\partial}=0;~~~\int_{S^1} \sqrt{h}=\beta.
\ee
There are two classes of classical solutions satisfying the first boundary condition: either $\phi_h=0$ or $\rho={\pi \over 2}$. The case $\phi_h=0$ is analogous to the extremal black hole case: the entropy is minimized, and the dilaton is constant. Any immersion of the disk with boundary length $\beta$ is a classical solution.\footnote{Higher topologies can be suppressed by adding a topological term to the action.} For the second class of solutions, $\rho={\pi \over 2}$, the boundary length is fixed to be $\beta=2\pi$. There is a single zero-mode integral over $\phi_h$. 

At the level of classical solutions and small fluctuations, the microcanonical boundary condition removes both zero-mode problems. We will continue with those boundary conditions. Consider the perturbative expansion around the hemisphere saddle point $\xi=z$:
\begin{equation}\label{eqn:xipert}
	\xi(z)=z+\sum_{n\geq 0} \xi_n z^n.%,~~~\epsilon_n\equiv\epsilon_n^R+\i \epsilon_n^I \in \mathbb{C}.
\end{equation}
Around the classical saddle, an infinitesimal $SO(3)$ transformation takes the form
\begin{equation}\label{eqn:SU2modes}
	\delta\xi(z)= (a+\i b)+\i c z+(a-\i b) z^2,~~~a,b,c\in \mathbb{R}.
\end{equation}
while an infinitesimal $PSL(2,\mathbb R)$ transformation gives
\begin{equation}
	\delta \xi(z)= (a'+\i b')+\i c' z-(a'-\i b') z^2,~~~a',b',c'\in \mathbb{R}.
\end{equation}
As emphasized above, none of these modes is physical. Together they remove the $\xi_{0},\xi_2$ modes and the imaginary part of $\xi_1$. The real part of $\xi_1$ is physical.

Expanding the action \eqref{eqn:JTAction2} to quadratic order in $\xi_n$ gives
\begin{equation}\label{eqn:quadraticfluc}
	Z(m)\propto\int \d \xi_1\prod_{n\geq 3} \d^2\xi_n \exp\left(-{S_m\over 2}\xi_1^2+{S_m\over 4}\sum_{n\geq 3}\left((n-1)^2-1\right)|\xi_n|^2\right).
\end{equation}
Here $\xi_1$ is stable, but all of the $\xi_n$ with $n \ge 3$ are unstable and should be Wick-rotated as in the sphere partition function of gravity. The corresponding phase $\prod_{n\geq 3} (-\i)^2$ requires regularization. One way to regularize it is to count the missing negative modes in the gravitational path integral \cite{Polchinski:1988ua}. Naively, since we have removed $5$ modes in $\xi$, one might conclude that the answer should be $\i^{(5+1)}=-1$ (five zero modes plus one stable mode). This counting, however, is not correct. To determine the phase properly, one must compare with the number of modes of the local field $h$ discussed above. Since different SO$(3)$ transformations of $\xi$ lead to the same $h$, these modes should not be counted, and we get $\i^{(5+1-3)}= -\i$, reproducing the naive phase of the sphere partition function with an observer \cite{Maldacena:2024spf}. It can be understood as arising from two CKV zero modes and one stable mode describing the overall size, as in the higher-dimensional case.

As explained in \cite{Maldacena:2024spf,Chen:2025jqm}, this final $-\i$ phase can be understood as arising from the single unstable mode in the $\beta,E$ integral:
\be
\int_{\mathbb{R}}\frac{\d \delta E\d\delta \beta}{2\pi} e^{-\delta \beta \delta E}=-\i
\ee
We then compare by hand with the proper integration contour for the inverse Laplace transform:
\be
\int_{\mathbb{R}}\d \delta E \int_{\i\mathbb{R}}{\d \delta\beta \over 2\pi \i } e^{-\delta \beta \delta E}=1.
\ee
With this prescription, the final microcanonical partition function is real and positive. Note that we started from the Euclidean path integral and changed the integration contour of $\beta$ only at the end, in order to implement the inverse Laplace transform. It might be interesting to understand this more deeply within this simple model.

\subsection{Coupling to a 2d CFT}
Let's now couple the 2d de Sitter JT gravity theory to a 2d CFT, with a conformal boundary condition imposed on the $S^1$ boundary of the immersed $D^2 \subset S^2$. Since the metric \eqref{eqn:ccdisk} is already in the conformal gauge, correlators of boundary primary operators with conformal dimension $\Delta$ are directly related to the flat space disk correlators in the $z$ coordinates by a Weyl transformation: 
\be
\langle \mathcal{O}(\theta_1)\mathcal{O}(\theta_2)\rangle={e^{-\Delta \omega_1}e^{-\Delta \omega_2}\over |z_1-z_2|^{2\Delta}};~~~z=e^{\i\theta}.
\ee
We will measure the boundary length using the boundary proper time coordinate $0 < u < \beta$. The angle as a function of proper time, $\theta(u)$, can be determined from the $\xi_n$ coefficients by imposing the proper length parametrization condition
\be\label{eqn:propertime}
\d u =e^{\omega }\d\theta;~~~\theta'(u)={1+|\xi(e^{\i\theta(u)})|^2\over 2|\partial\xi(e^{\i\theta(u)})|}.
\ee
Then the two point functions in $u$ coordinates have the nice simple form completely determined by the boundary parameterization $z(u)=e^{\i\theta(u)}$:
\begin{align}\label{eqn:2ptconformal}
\langle \mathcal{O}(u_1)\mathcal{O}(u_2)\rangle&=\left({\theta_1'\theta_2'\over 4\sin^2{\theta_1(u_1)-\theta_2(u_2)\over 2}}\right)^{\Delta}.
\end{align}

\paragraph{Linear expansion} Let's now use this formula to compute the four point OTOC and TOC to leading order in the gravitational fluctuations, following a similar computation for AdS JT gravity in \cite{Maldacena:2016upp}. To proceed, we need to know $\theta(u)$ to linear order in the fluctuations $\xi_n$. The total proper length $\beta$ fluctuates, but as we saw in (\ref{eqn:quadraticfluc}), the fluctuation starts at order $\xi_n^2$. This means that one can approximate $\beta = 2\pi$ in what follows. Then writing
\begin{equation}
\theta(u)=u+\varepsilon(u)+\mathcal{O}(\xi_n^2),\qquad \varepsilon(u)=\sum_{|n|\geq 2}\varepsilon_n e^{\i n u},
\end{equation}
the expansion of (\ref{eqn:propertime}) to linear order implies simply
\begin{align}\label{varepsilonandxi}
\varepsilon_n &= {\i\over 2}\xi_{n+1},\qquad n\geq 2,\qquad \varepsilon_{-n}=\varepsilon_n^*.
\end{align}
Expanding (\ref{eqn:2ptconformal}) to linear order in the fluctuations, we get
\begin{align}
\langle \mathcal{O}(u_1)\mathcal{O}(u_2)\rangle
&={1\over \left(4\sin^2{u_{12}\over 2}\right)^\Delta}\left[1+\Delta\left(\varepsilon'(u_1)+\varepsilon'(u_2)-{\varepsilon(u_1)-\varepsilon(u_2)\over \tan{u_{12}\over 2}}\right)+\mathcal{O}(\varepsilon^2)\right].
\end{align}

\paragraph{Propagator for $\varepsilon$} To compute the four point function, we should take a product of two of these expressions, and contract the $\epsilon$ fluctuations using the propagator. To derive the propagator, we substitute (\ref{varepsilonandxi}) into (\ref{eqn:quadraticfluc}) to get the path integral weighting for $\varepsilon$ fluctuations:
\begin{equation}
\exp\left[{S_m\over 2}\int_0^{2\pi}\frac{\d u}{2\pi} \left(\dot\varepsilon^2(u)-\varepsilon^2(u)\right)\right].
\end{equation}
The corresponding propagator is
\begin{equation}
\langle \varepsilon(u_1)\varepsilon(u_2)\rangle=-{1\over S_m}\sum_{|n|\geq 2}{e^{\i n u_{12}}\over n^2-1}=-{2\over S_m}\left({1\over 2}+{\cos u_{12}\over 4}-{\pi-u_{12}\over 2}\sin u_{12}\right),\qquad 0<u_{12}<2\pi.
\end{equation}

\paragraph{OTOC and TOC} After summing over the contractions, one finds that the TOC is zero to this order. The OTOC is nonzero and has a particularly simple expression in the configuration where the pairs of operators are antipodal:
\begin{align}\label{dsJTOTOC}
\frac{\langle W(\pi+u)V(\tfrac{\pi}{2})W(u)V(-\tfrac{\pi}{2})\rangle_{c}}{\langle WW\rangle\langle VV\rangle}&=\Delta_W\Delta_V\left\langle\left(\varepsilon'(\pi+u)+\varepsilon'(u)\right)\left(\varepsilon'(\tfrac{\pi}{2})+\varepsilon'(-\tfrac{\pi}{2})\right)\right\rangle\\
&={2\pi\Delta_W\Delta_V\over S_m}\cos u,\qquad -{\pi\over2}<u<{\pi\over2}.
\end{align}

\paragraph{Eikonal resummation} After continuation $u \to \i t$, this expression includes a term of order $e^t / S_m$. Higher orders in perturbation theory will include higher powers $(e^t / S_m)^n$, and these powers can be summed using the same strategy as in \cite{Maldacena:2016upp}. That is given by integrating over the correlators on the shockwave background parametrized by the null shifts $X^{\pm}$ with the Dray t'Hooft action. As in \cite{Maldacena:2016upp}, the null shifts are given by relative $PSL(2,\mathbb R)$ transformations \eqref{eqn:PSL2} between the operators, and therefore the action is identical to the case of AdS JT gravity due to the same reparameterization form \eqref{eqn:2ptconformal}, that is given by equation (6.54) and (6.55) in \cite{Maldacena:2016upp}. The only thing we need to change is the coefficient $C$ that enters in the shockwave action. This can be matched by comparing to the tree-level answer (\ref{dsJTOTOC}), and one finds
\be
C=-{S_m\over 2\pi}.
\ee
In particular, $C$ is negative. This justifies the formulas studied in section \ref{sec:wrongsign-JT}.

Note that the above discussion can be generalized to other dilaton gravity theories. The $PSL(2,\mathbb R)$ transformations will still act the same way as in AdS JT gravity. The only input data from the gravitational theory is the value of the coupling constant $C$, which will be positive for black hole type horizons and negative for de Sitter type horizons.

\section{Pure recoil model}

\subsection{Finding the tetrahedron saddle point}\label{app:tetra}

In this appendix, we explain the detailed steps in finding the saddle point configuration of the action (\ref{tetraaction}), which we copy here:
\begin{equation}
	I_{\textrm{tetra}} = \sum_{i<j} \mathbf{m}_{ij} d_{ij} \,, \quad\quad d_{ij} \equiv \arccos \left(\vec{X}_i \cdot \vec{X}_j\right)\,.
\end{equation}
Let's consider the variation of the action with respect to $\vec{X}_i$, with $\delta\vec{X}_i$ constrained to lie in $T_{\vec X_i}S^d$. We have 
\begin{equation}
	\delta d_{ij} = - \frac{ \delta \vec{X}_i \cdot \vec{X}_j}{\sin d_{ij}} = -\delta \vec{X}_i \cdot \vec{t}_{ij}, \quad\quad \vec{t}_{ij} \equiv \frac{ \vec{X}_j - \cos d_{ij} \vec{X}_i  }{\sin d_{ij}} \,.
\end{equation}
In the second equality, we used that $\delta \vec{X}_i\cdot \vec{X}_i = 0$ to introduce $\vec{t}_{ij}$, which   is the unit tangent vector at $\vec{X}_i$ that points towards $\vec{X}_j$. Therefore, the saddle point equations can be written as
\begin{equation}\label{tetraeom}
	\sum_{j\neq i}\mathbf{m}_{ij} \vec{t}_{ij} = 0\,, \quad \forall \,i\,.
\end{equation}
These equations have the nice interpretation as implementing the local momentum conservation at each vertex.

Before we go about solving the equations (\ref{tetraeom}), we can first observe the following. Suppose we find a solution. Starting from a single vertex $i$, each of the other vertices is obtained from one of the tangent vectors by
\begin{equation}
	\vec X_j = \cos d_{ij}\vec X_i + \sin d_{ij} \vec{t}_{ij}\,.
\end{equation}
It follows that all four vertices lie in a three-dimensional linear subspace of the embedding space generated by $\vec X_i$ and a two-dimensional tangent plane spanned by $\vec{t}_{ij}, j \neq i$. The intersection of this subspace with the unit sphere is a totally geodesic $S^2$. This explains why in the main text we can represent the tetrahedra on a single $S^2$. 

\begin{figure}[H]
\centering
\begin{tikzpicture}[scale=2.0]
\coordinate (A) at (-1,0);
\coordinate (B) at (0.3,-0.6);
\coordinate (C) at (1,0);
\coordinate (D) at (0,1);
\coordinate (E1) at (0.1,0);
\coordinate (E2) at (0.25,0);
\coordinate (Fone) at (-0.5,0.2);
\coordinate (Ftwo) at (0.6,0.2);
\fill[red!12, fill opacity=0.35] (A)--(B)--(D)--cycle;
\fill[blue!12, fill opacity=0.35] (B)--(C)--(D)--cycle;
\draw (A)--(B);
\draw[dashed] (A)--(E1);
\draw[dashed] (E2)--(C);
\draw (B) -- (D);
\node at (-0.1,0.45) {$\mathbf m_{12}$};
\draw (A) -- (D);
\draw (D) -- (C);
\draw (B) -- (C);
\draw[->,  red] (Fone) -- ++(-0.4,0.25) node[left] {$\vec X_1$};
\draw[->,  blue] (Ftwo) -- ++(0.3,0.2) node[right] {$\vec X_2$};
\end{tikzpicture}
\caption{The auxiliary flat tetrahedron, where $\mathbf{m}_{ij}$ become the lengths of the edges and $\vec{X}_i$ become the normal vectors to the faces. In the illustration we've colored two faces labeled by $1$ and $2$ and labeled their normal vectors and shared edge.}
\label{fig:flattetra}
\end{figure}
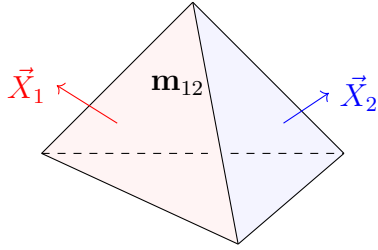

Now, let's explain how to find the general solutions of (\ref{tetraeom}). 
Rather than solving these equations directly, it is useful to introduce an auxiliary ``dual" flat tetrahedron, see Figure \ref{fig:flattetra}. By this we mean an ordinary tetrahedron in $\mathbb{R}^3$ whose faces are labeled by $1,2,3,4$, and whose edge shared by the faces $i$ and $j$ has length $\mathbf m_{ij}$. The face labeled by $i$ is a flat triangle with side lengths $\mathbf m_{ij},\mathbf m_{ik},\mathbf m_{i\ell}$. For a flat triangle, the sum of the edge lengths times the outward unit normals to the edges vanishes. So if we identify these outward unit normals with $\vec t_{ij}$, the closure equation for this face is precisely (\ref{tetraeom}). In this construction, $\vec X_i$ is the unit normal to the face labeled by $i$, since it is orthogonal to the three outward pointing normals of the edges.
With this construction, the spherical distance $d_{ij}$ is the angle between the normals to the faces $i$ and $j$ of the flat tetrahedron, as $d_{ij}=\arccos(\vec X_i\cdot \vec X_j)$ by definition. 

This reduces the problem to elementary Euclidean geometry of finding the dihedral angles given the lengths of the edges of a tetrahedron. We simply state the results here. Denote
\begin{equation}
	A(a,b,c) \equiv \frac{1}{4}\sqrt{(a+b-c)(a+c-b)(b+c-a)(a+b+c)}
\end{equation}
as the area of a flat triangle with side lengths $a,b,c$. Then for distinct $i,j,k,\ell$, we have
\begin{equation}\label{tetradgeneral}
\begin{aligned}
	d_{ij}=\arccos\left[
	\frac{\left(
	(\mathbf m_{ij}^2+\mathbf m_{ik}^2-\mathbf m_{i\ell}^2)
	(\mathbf m_{ij}^2+\mathbf m_{jk}^2-\mathbf m_{j\ell}^2)
	-2\mathbf m_{ij}^2(\mathbf m_{ik}^2+\mathbf m_{jk}^2-\mathbf m_{k\ell}^2)
	\right) }{16A(\mathbf m_{ij},\mathbf m_{ik},\mathbf m_{i\ell})
	A(\mathbf m_{ij},\mathbf m_{jk},\mathbf m_{j\ell})}\right] \,.
\end{aligned}
\end{equation}
For the mass matrix (\ref{massmat}) relevant for the OTOC computation, this gives
\begin{equation}\label{allds}
\begin{aligned}
d_{12}={}&\arccos\!\left[
\frac{(m_2^2+\nu'^2-m_1^2)(m_2^2+m_3^2-\nu^2)-2m_2^2(\nu'^2+m_3^2-m_4^2)}
{16A(m_2,\nu',m_1)A(m_2,m_3,\nu)}
\right]\,,\\
d_{13}={}&\arccos\!\left[
\frac{(\nu'^2+m_2^2-m_1^2)(\nu'^2+m_3^2-m_4^2)-2\nu'^2(m_2^2+m_3^2-\nu^2)}
{16A(\nu',m_2,m_1)A(\nu',m_3,m_4)}
\right]\,,\\
d_{14}={}&\arccos\!\left[
\frac{(m_1^2+m_2^2-\nu'^2)(m_1^2+\nu^2-m_4^2)-2m_1^2(m_2^2+\nu^2-m_3^2)}
{16A(m_1,m_2,\nu')A(m_1,\nu,m_4)}
\right]\,,\\
d_{23}={}&\arccos\!\left[
\frac{(m_3^2+m_2^2-\nu^2)(m_3^2+\nu'^2-m_4^2)-2m_3^2(m_2^2+\nu'^2-m_1^2)}
{16A(m_3,m_2,\nu)A(m_3,\nu',m_4)}
\right]\,,\\
d_{24}={}&\arccos\!\left[
\frac{(\nu^2+m_2^2-m_3^2)(\nu^2+m_1^2-m_4^2)-2\nu^2(m_2^2+m_1^2-\nu'^2)}
{16A(\nu,m_2,m_3)A(\nu,m_1,m_4)}
\right]\,,\\
d_{34}={}&\arccos\!\left[
\frac{(m_4^2+\nu'^2-m_3^2)(m_4^2+m_1^2-\nu^2)-2m_4^2(\nu'^2+m_1^2-m_2^2)}
{16A(m_4,\nu',m_3)A(m_4,m_1,\nu)}
\right]\,.
\end{aligned}
\end{equation}
The saddle value of the action is therefore
\begin{equation}\label{Itetrasaddle}
	I_{\rm tetra,\, saddle}
	= m_2d_{12}+\nu'd_{13}+m_1d_{14}
	+m_3d_{23}+\nu d_{24}+m_4d_{34}\,.
\end{equation}

\subsection{The same-side configuration}\label{app:sameside}
In writing the action (\ref{tetraaction}), we assumed that the worldlines follow the shortest geodesics on the sphere. We can also consider a different configuration where the two particle trajectories travel the longer way around the sphere. This is the same-side configuration depicted in Figure \ref{fig:opposite}.

The same-side configuration does not appear to possess a nice dual flat tetrahedron configuration, so
let's try to explicitly solve for the configuration, for the special case in Section \ref{sec:special} where we have
\begin{equation}
	m_1=m_3=m+{\omega\over2},\quad m_2=m_4=m-{\omega\over2},\quad \nu=\nu'\,.
\end{equation}
In this case, we can use the symmetry to write the distances between the points as
\begin{equation}
	a_1 = d_{12}=d_{34}\,,\quad a_2 = d_{14}=d_{23}\,,\quad  b = d_{13}=d_{24}\,.
\end{equation}
Note that here $b$ denotes the shortest geodesic distance between the endpoints of a particle trajectory, but the actual trajectories follow the longer way and have length $2\pi-b$. The action in this case, which we label by $I_{\rm same}$, is therefore
\begin{equation}
	I_{\rm same} = 2\left(m - \frac{\omega}{2}\right) a_1 + 2\left(m + \frac{\omega}{2}\right) a_2 + 2\nu \left(2\pi - b\right)\,.
\end{equation}
By varying the action  $I_{\rm same} $ with respect to $\delta \vec{X}_i$'s, we get equations
\begin{equation}
\begin{aligned}
\left(m-\frac{\omega}{2}\right)\vec t_{12}
+\left(m+\frac{\omega}{2}\right)\vec t_{14}
-\nu\vec t_{13}={}&0\,,\\
\left(m-\frac{\omega}{2}\right)\vec t_{21}
+\left(m+\frac{\omega}{2}\right)\vec t_{23}
-\nu\vec t_{24}={}&0\,,\\
\left(m+\frac{\omega}{2}\right)\vec t_{32}
+\left(m-\frac{\omega}{2}\right)\vec t_{34}
-\nu\vec t_{31}={}&0\,,\\
\left(m+\frac{\omega}{2}\right)\vec t_{41}
+\left(m-\frac{\omega}{2}\right)\vec t_{43}
-\nu\vec t_{42}={}&0\,.
\end{aligned}
\end{equation}
We can turn these equations into ones for $a_1, a_2, b$ by taking inner product with different $\vec{t}_{ij}$'s. In terms of $a_1,a_2,b$, we have
\begin{equation}
\begin{aligned}
0={}&m-\frac{\omega}{2}
+\left(m+\frac{\omega}{2}\right)
\frac{\cos b-\cos a_1\cos a_2}{\sin a_1\sin a_2}
-\nu
\frac{\cos a_2-\cos a_1\cos b}{\sin a_1\sin b}\,,\\
0={}&\left(m-\frac{\omega}{2}\right)
\frac{\cos b-\cos a_1\cos a_2}{\sin a_1\sin a_2}
+m+\frac{\omega}{2}
-\nu
\frac{\cos a_1-\cos a_2\cos b}{\sin a_2\sin b}\,,\\
0={}&\left(m-\frac{\omega}{2}\right)
\frac{\cos a_2-\cos a_1\cos b}{\sin a_1\sin b}
+\left(m+\frac{\omega}{2}\right)
\frac{\cos a_1-\cos a_2\cos b}{\sin a_2\sin b}
-\nu\,.
\end{aligned}
\end{equation}
We can solve these equations and get
\begin{equation}\label{samebranchdijs}
\begin{aligned}
\cos a_1
={}&
\frac{\nu^4+2m\omega(4m^2-2\nu^2+\omega^2)-4m^2\omega^2}
{(4m^2-\nu^2)(\nu^2-\omega^2)}\,,\\
\cos a_2
={}&
\frac{\nu^4-2m\omega(4m^2-2\nu^2+\omega^2)-4m^2\omega^2}
{(4m^2-\nu^2)(\nu^2-\omega^2)}\,,\\
\cos b
={}&
\frac{3\nu^4-\nu^2\omega^2-4m^2(\nu^2+\omega^2)}
{(4m^2-\nu^2)(\nu^2-\omega^2)}\,.
\end{aligned}
\end{equation}
As a check of these formulas, at $\omega=0$ and $\nu\ll m$ we have
\begin{equation}
	I_{\rm same}  =2\pi m+2\pi\nu+{\nu^2\over m}
	+...\,,
\end{equation}
from which we indeed see that the $\nu$ edges have lengths a bit greater than $\pi$ and this configuration is further suppressed compared to the ``opposite side" configuration (\ref{Itetrasmallnu}).

\subsection{Pure recoil limit of the eikonal OTOCs}\label{app:recoileikonal}
     In this appendix, we work out the limit of the eikonal OTOCs in the pure recoil semiclassical limit (\ref{recoillimit}), for the special case of $m_1 = m_3 = m+ \omega/2$ and $m_2 = m_4 = m- \omega/2$.\footnote{We will focus on the dS$_3$ discussion, but one can verify that for the wrong sign JT OTOCs one arrives at a similar structure as the $\phi =\pi$ case. The detailed formulas are different since in the JT discussion the observer is coupled to 2d CFTs rather than massive fields.}  In the notation of Section \ref{sec:eikonalenergy}, this means
\begin{equation}
	\omega_1=\omega_3=-\omega_2=-\omega_4=\omega\,,
\end{equation}
and we also set $\nu'=\nu$ as in Section \ref{sec:special}. Since we take $G\rightarrow 0$, we have
\begin{equation}
		\alpha \rightarrow 2\pi\,,\quad\quad h(\phi)\rightarrow {\cos\phi\over 2m}\,.
\end{equation}
The distinction between $\mathcal{F}_{12}$ and $\mathcal{F}_{14}$ in Section \ref{sec:eikonalenergy} is then entirely in the $\i 0$ prescription of the $\phi$ integral. We can perform the $\phi$ integral explicitly using
\begin{equation}\label{phiintegral}
\begin{aligned}
\int_{-\pi}^{\pi}{\d\phi\over 2\pi}
\left({2\cos\phi\over m}+\i 0^+\right)^{\i a}
={}&\left({2\over m}\right)^{\i a}
\,
{\Gamma\!\left({1+\i a\over2}\right)\over
\Gamma\!\left(1+{\i a\over2}\right)}\,{1+e^{-\pi a}\over 2\sqrt{\pi}},\\
\int_{-\pi}^{\pi}{\d\phi\over 2\pi}
\left({2\cos\phi\over m}-\i 0^+\right)^{\i a}
={}&\left({2\over m}\right)^{\i a}
\,
{\Gamma\!\left({1+\i a\over2}\right)\over
\Gamma\!\left(1+{\i a\over2}\right)}\,{1+e^{\pi a}\over 2\sqrt{\pi}}\,.
\end{aligned}
\end{equation}
Each $\phi$ integral leads to a sum of two pieces. The $``1"$ piece comes from the saddle point at $\phi=0$, while the $e^{\pm \pi a}$ piece comes from the saddle point at $\phi=\pi$. Since $\phi$ has the interpretation of the transverse separation between the two matter particles, it instructs us to separate them and compare them separately with the same-side ($\phi=0$) and opposite-side ($\phi=\pi$) configurations in Section \ref{sec:recoil}.

\subsubsection{The $\phi = \pi$ pieces}
We now focus on the $\phi = \pi$ contributions from the integrals (\ref{phiintegral}), which correspond to the $e^{\pm\pi a}$ pieces. For example, for $\mathcal{F}_{12}^{t<0}$, dropping  the delta function factor $\delta(\sum_i\omega_i)$ and power-law prefactors, we have
\begin{equation}
\begin{aligned}
\mathcal{F}_{12,\phi= \pi}^{t<0}
\sim{}&
\Gamma\!\left({1+\i\omega\pm\i\nu\over2}\right)^4  {\Gamma\!\left({1+2\i\omega\over2}\right)\over
\Gamma(1+\i\omega)}
\left({2\over m}\right)^{2\i\omega}  
\Gamma(-2\i\omega) \times e^{-2\pi\omega} \,.
\end{aligned}
\end{equation}
We apply the Stirling formulas to extract the semiclassical limit of this expression. The $\Gamma(-2\i\omega)$ factor leads to a series of poles on the negative imaginary axis, which turns into a branch cut in the semiclassical limit. As it turns out, one can utilize the $\varphi_\pm$ functions defined in (\ref{phipm}), which are analytic in the upper/lower half planes, to represent the semiclassical limit in a compact way
\begin{equation}
\begin{aligned}
-\log\bm{\mathcal{F}}_{12,\phi= \pi}^{t<0}
\sim{}&
2\pi m + 2\pi\nu + 2\pi\omega +\i\varphi_+(\omega)\,.
\end{aligned}
\end{equation}
where we introduced an additional $2\pi m$ shift to go to $\bm{\mathcal{F}}$.
Repeating the same analysis for the remaining OTOCs, we get
\begin{equation}
\begin{aligned}
-\log\bm{\mathcal{F}}_{12,\phi= \pi}^{t>0}
\sim{}&
2\pi m+2\pi\nu+2\pi\omega-\i\varphi_-(\omega)\,,\\
-\log\bm{\mathcal{F}}_{14,\phi= \pi}^{t<0}
\sim{}&
2\pi m+2\pi\nu-2\pi\omega+\i\varphi_+(\omega)\,,\\
-\log\bm{\mathcal{F}}_{14,\phi= \pi}^{t>0}
\sim{}&
2\pi m+2\pi\nu-2\pi\omega-\i\varphi_-(\omega)\,.
\end{aligned}
\end{equation}

\subsubsection{The $\phi = 0$ pieces}\label{app:phi=0}

Now let's turn to the $\phi = 0$  pieces from the integrals (\ref{phiintegral}), which correspond to the ``1"'s in the two terms on the right hand side.

Importantly, since the saddle at $\phi=0$ has $\cos\phi>0$, the $\i 0^+$ prescription does not distinguish $\mathcal{F}_{12}$ from $\mathcal{F}_{14}$ at this order. Therefore, we get
\begin{equation}
\begin{aligned}
-\log\bm{\mathcal{F}}_{12,\phi=0}^{t<0} \sim -\log\bm{\mathcal{F}}_{14,\phi=0}^{t<0}
\sim{}&
2\pi m + 2\pi\nu+\i\varphi_+(\omega)\,,\\
-\log\bm{\mathcal{F}}_{12,\phi=0}^{t>0} \sim -\log\bm{\mathcal{F}}_{14,\phi=0}^{t>0}
\sim{}&
2\pi m+2\pi\nu-\i\varphi_-(\omega)\,.\\
\end{aligned}
\end{equation}

\subsection{Lorentzian tetrahedron for eikonal OTOCs in the energy domain}
\label{app: lorenzian tetra}
In this appendix we show that the OTOC kinematics in the recoil and large observer mass limit ($m\gg\omega_i,\nu,\nu'\gg1$ while $Gm,G\omega_i, G\nu\rightarrow0$) correspond to a tetrahedron in Lorentzian dS$_3$. We focus on the special configuration \eqref{eq: special mass conf1}. Starting from \eqref{allds}, the six edge lengths satisfy (assuming $-\nu<\operatorname{Re}\omega<\nu$ so that $\omega$ does not cross the $\pm\nu$ branch cuts shown in figure \ref{fig:omega_branch_cuts}):
\begin{equation}
\cos d_{14}=\cos d_{23}=-\cos d_{12}=-\cos d_{34}=\frac{2m\omega}{\nu^2-\omega^2}+O(m^0),
\end{equation}
\begin{equation}
\cos d_{13}=\cos d_{24}=-\frac{\nu^2+\omega^2}{\nu^2-\omega^2}+O(m^{-1})
\end{equation}
Taking $\arccos$ requires a choice of branch, which depends on the analytic continuation, i.e. which  quadrant $\omega$ sits. This yields four possibilities, corresponding to $\bm{\mathcal{F}}_{12,\phi=\pi}^{t\lessgtr0}$ and $\bm{\mathcal{F}}_{14,\phi=\pi}^{t\lessgtr0}$ as in figure \ref{fig:identification}. To leading order at large $m$, one finds:
\begin{align}
\renewcommand{\arraystretch}{1.25}
\begin{array}{c|c|c|c}
 & d_{14}=d_{23} & d_{12}=d_{34} & d_{13}=d_{24} \\
\hline
\bm{\mathcal{F}}_{12,\phi=\pi}^{t>0} 
& \pi-\i(\log m+v) 
& \i(\log m+v) 
& \pi-\i u \\
\bm{\mathcal{F}}_{12,\phi=\pi}^{t<0} 
& \pi+\i(\log m+v) 
& -\i(\log m+v) 
& \pi+\i u \\
\bm{\mathcal{F}}_{14,\phi=\pi}^{t>0} 
& -\i(\log m+v) 
& \pi+\i(\log m+v) 
& \pi+\i u \\
\bm{\mathcal{F}}_{14,\phi=\pi}^{t<0} 
& \i(\log m+v) 
& \pi-\i(\log m+v) 
& \pi-\i u
\end{array}
\end{align}
\begin{equation}
\text{with}\ \ \  u:=\operatorname{arccosh}\left(\frac{\nu^2+\omega^2}{\nu^2-\omega^2}\right),\ v:=\log\left(\frac{4|\omega|}{\nu^2-\omega^2}\right)
\end{equation}
For real $\omega$, we have $u>0$ and $v\in\mathbb R$. The large logarithm $\log m\gg1$ sets the (recoil) scrambling timescale of the OTOC. In these variables, the four  choices above map to Lorentzian tetrahedra  reproducing the OTOC configurations shown below\footnote{The lines in the figures are schematic, they do not represent the actual trajectory of geodesics connecting them, since some of the geodesics live on complex geometry. }.
\begin{align}
\begin{tikzpicture}[scale=2.5, baseline=(current bounding box.center)]
% draw the gray dS_3
\coordinate (T1) at (0,0);  
\coordinate (T2) at (0,1);   
\coordinate (T3) at (1,1); 
\coordinate (T4) at (1,0); 
\draw[line width=1.5pt, draw=gray!50] (T1) -- (T2) -- (T3) -- (T4) -- cycle;
\draw[line width=1pt, draw=gray!50, dashed](0,0)--(1,1);
\draw[line width=1pt, draw=gray!50, dashed](1,0)--(0,1);
% draw the OTOC lines
\coordinate (T1p) at (0,0.1);  
\coordinate (T2p) at (0,0.9);   
\coordinate (T5) at (1,0.9); 
\coordinate (T6) at (1,0.1); 
\draw[line width=2pt, draw=black] (T1p) -- (T2p) -- (T5) -- (T6) -- cycle;
\draw[line width=1.5pt, draw=red]
(T1p)--(T5);
\draw[line width=1.5pt, draw=blue]
(T2p)--(T6);
\fill[red] (T1p) circle (0.025) node[left, font=\large] {$4$};
\fill[red] (T5) circle (0.025) node[right, font=\large] {$2$};
\fill[blue] (T2p) circle (0.025) node[left, font=\large] {$3$};
\fill[blue] (T6) circle (0.025) node[right, font=\large] {$1$};
\node[font=\Large] at (0.5,-0.3) {$\bm{\mathcal{F}}_{12,\phi=\pi}^{t>0}$};
\end{tikzpicture}\ \ 
\begin{tikzpicture}[scale=2.5, baseline=(current bounding box.center)]
% draw the gray dS_3
\coordinate (T1) at (0,0);  
\coordinate (T2) at (0,1);   
\coordinate (T3) at (1,1); 
\coordinate (T4) at (1,0); 
\draw[line width=1.5pt, draw=gray!50] (T1) -- (T2) -- (T3) -- (T4) -- cycle;
\draw[line width=1pt, draw=gray!50, dashed](0,0)--(1,1);
\draw[line width=1pt, draw=gray!50, dashed](1,0)--(0,1);
% draw the OTOC lines
\coordinate (T1p) at (0,0.1);  
\coordinate (T2p) at (0,0.9);   
\coordinate (T5) at (1,0.9); 
\coordinate (T6) at (1,0.1); 
\draw[line width=2pt, draw=black] (T1p) -- (T2p) -- (T5) -- (T6) -- cycle;
\draw[line width=1.5pt, draw=blue]
(T1p)--(T5);
\draw[line width=1.5pt, draw=red]
(T2p)--(T6);
\fill[blue] (T1p) circle (0.025) node[left, font=\large] {$3$};
\fill[blue] (T5) circle (0.025) node[right, font=\large] {$1$};
\fill[red] (T2p) circle (0.025) node[left, font=\large] {$4$};
\fill[red] (T6) circle (0.025) node[right, font=\large] {$2$};
\node[font=\Large] at (0.5,-0.3) {$\bm{\mathcal{F}}_{12,\phi=\pi}^{t<0}$};
\end{tikzpicture}\ \ 
\begin{tikzpicture}[scale=2.5, baseline=(current bounding box.center)]
% draw the gray dS_3
\coordinate (T1) at (0,0);  
\coordinate (T2) at (0,1);   
\coordinate (T3) at (1,1); 
\coordinate (T4) at (1,0); 
\draw[line width=1.5pt, draw=gray!50] (T1) -- (T2) -- (T3) -- (T4) -- cycle;
\draw[line width=1pt, draw=gray!50, dashed](0,0)--(1,1);
\draw[line width=1pt, draw=gray!50, dashed](1,0)--(0,1);
% draw the OTOC lines
\coordinate (T1p) at (0,0.1);  
\coordinate (T2p) at (0,0.9);   
\coordinate (T5) at (1,0.9); 
\coordinate (T6) at (1,0.1); 
\draw[line width=2pt, draw=black] (T1p) -- (T2p) -- (T5) -- (T6) -- cycle;
\draw[line width=1.5pt, draw=red]
(T1p)--(T5);
\draw[line width=1.5pt, draw=blue]
(T2p)--(T6);
\fill[red] (T1p) circle (0.025) node[left, font=\large] {$2$};
\fill[red] (T5) circle (0.025) node[right, font=\large] {$4$};
\fill[blue] (T2p) circle (0.025) node[left, font=\large] {$3$};
\fill[blue] (T6) circle (0.025) node[right, font=\large] {$1$};
\node[font=\Large] at (0.5,-0.3) {$\bm{\mathcal{F}}_{14,\phi=\pi}^{t>0}$};
\end{tikzpicture}\ \ 
\begin{tikzpicture}[scale=2.5, baseline=(current bounding box.center)]
% draw the gray dS_3
\coordinate (T1) at (0,0);  
\coordinate (T2) at (0,1);   
\coordinate (T3) at (1,1); 
\coordinate (T4) at (1,0); 
\draw[line width=1.5pt, draw=gray!50] (T1) -- (T2) -- (T3) -- (T4) -- cycle;
\draw[line width=1pt, draw=gray!50, dashed](0,0)--(1,1);
\draw[line width=1pt, draw=gray!50, dashed](1,0)--(0,1);
% draw the OTOC lines
\coordinate (T1p) at (0,0.1);  
\coordinate (T2p) at (0,0.9);   
\coordinate (T5) at (1,0.9); 
\coordinate (T6) at (1,0.1); 
\draw[line width=2pt, draw=black] (T1p) -- (T2p) -- (T5) -- (T6) -- cycle;
\draw[line width=1.5pt, draw=blue]
(T1p)--(T5);
\draw[line width=1.5pt, draw=red]
(T2p)--(T6);
\fill[blue] (T1p) circle (0.025) node[left, font=\large] {$3$};
\fill[blue] (T5) circle (0.025) node[right, font=\large] {$1$};
\fill[red] (T2p) circle (0.025) node[left, font=\large] {$2$};
\fill[red] (T6) circle (0.025) node[right, font=\large] {$4$};
\node[font=\Large] at (0.5,-0.3) {$\bm{\mathcal{F}}_{14,\phi=\pi}^{t<0}$};
\end{tikzpicture}
\end{align}

\section{Shock waves in higher dimensions}\label{app:shock_waves_in_higher_dimensions}
In this appendix, we discuss the shock wave solutions for higher-dimensional de Sitter space with an observer. This observer could be a horizon-free geometry, like dust or a star. For the shock wave calculation near the horizon, however, it is enough to assume that the observer is spherically symmetric and supported in a finite ball-shaped region, so that the exterior metric is de Sitter-Schwarzschild:
\begin{equation}
\d s^2 = -f(r)\d t^2 + \frac{\d r^2}{f(r)} + r^2\d\Omega_{d-2}^2,
\hspace{20pt}
f(r) = 1-r^2-\frac{16\pi G m}{(d-2)\Omega_{d-2}r^{d-3}}.
\end{equation}
Let $r_c$ be the cosmological horizon, so
\begin{equation}
f(r_c)=0,
\hspace{20pt}
f'(r_c)=-|f'(r_c)|<0.
\end{equation}
Introduce the tortoise coordinate by $\d r_* = \d r / f(r)$. Then Kruskal coordinates are
\begin{equation}
x^-x^+ = -e^{-|f'(r_c)|r_*},
\hspace{20pt}
\frac{x^+}{x^-} = -e^{|f'(r_c)|t}.
\end{equation}
The metric including a shock is
\begin{equation}
\d s^2 = -A(x^-x^+)\d x^-\d x^+ + B(x^-x^+)\d\Omega_{d-2}^2 + A(0)X^+(\Omega)\delta(x^-)(\d x^-)^2
\end{equation}
where
\begin{equation}
A(z)=-\frac{4f(r(z))}{f'(r_c)^2z},
\hspace{20pt}
B(z)=r(z)^2.
\end{equation}
The shock wave is parametrized by the function $X^+(\Omega)$, which is determined by the stress tensor $T_{--} = \delta(x^-) T_-$ by applying the ``$--$'' Einstein equation to the shock Ricci tensor in (\ref{eq:shock-ricci-component}).
The quantities that enter this equation are
\begin{equation}
A(0)
=
\lim_{z\to0}\left[-\frac{4f(r(z))}{f'(r_c)^2z}\right]
=
\frac{4r'(0)}{|f'(r_c)|},
\hspace{20pt}
B(0)=r_c^2,
\hspace{20pt}
B'(0)=2r_cr'(0).
\end{equation}
For a particle crossing the horizon at the north pole, the stress tensor is
\begin{equation}
T_{--}=-p_-\delta(x^-)\frac{\delta^{(d-2)}(\Omega)}{r_c^{d-2}},
\hspace{20pt}
p_-<0.
\end{equation}
So the shock equation is
\begin{equation}
\left[-\nabla^2_{S^{d-2}}-\mu^2\right]X^+(\Omega)
=
(-p_-)\frac{4\pi G|f'(r_c)|}{r'(0)r_c^{d-4}}\delta^{(d-2)}(\Omega),
\end{equation}
with
\begin{align}
\mu^2
&=
\frac{d-2}{2}|f'(r_c)|r_c
\notag\\
&=
(d-2)-\frac{8\pi(d-1)Gm}{\Omega_{d-2}}+O((Gm)^2).
\end{align}
On the $(d-2)$ sphere, the Laplacian has eigenvalues $\ell(\ell+d-3)$, so when the observer mass vanishes $m=0$, the $\ell=1$ modes are ``recoil'' zero modes.

The solution is proportional to the Green's function on $S^{d-2}$:
\begin{equation}\label{X+sol}
X^+(\vartheta)
=
(-p_-)\frac{4\pi G|f'(r_c)|}{r'(0)r_c^{d-4}}\times \frac{\Gamma(-j)\Gamma(j + d-3)}{(4\pi)^{\frac{d-2}{2}}\Gamma(\frac{d-2}{2})}
{}_2F_1\!\left(-j,j + d-3;\frac{d-2}{2};\cos^2\frac{\vartheta}{2}\right),
\end{equation}
where 
\begin{equation}
j(j + d-3)=\mu^2.
\end{equation}
Here are plots of normalized versions of this function in $d=4$ for two small observer masses, $Gm=1/100$ and $Gm=1/10$:
\begin{center}
\begin{minipage}[c]{0.47\textwidth}
\centering
\includegraphics[width=\linewidth]{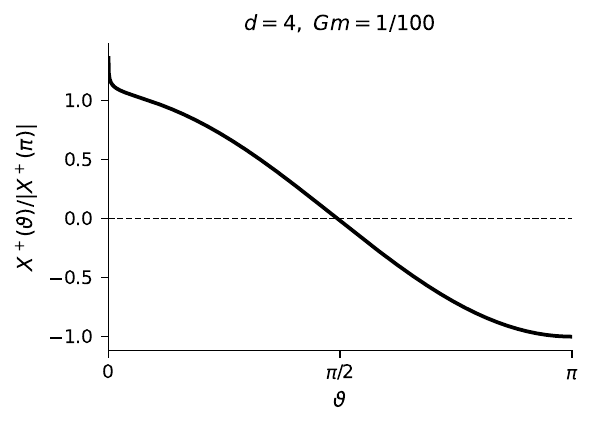}
\end{minipage}
\hfill
\begin{minipage}[c]{0.47\textwidth}
\centering
\includegraphics[width=\linewidth]{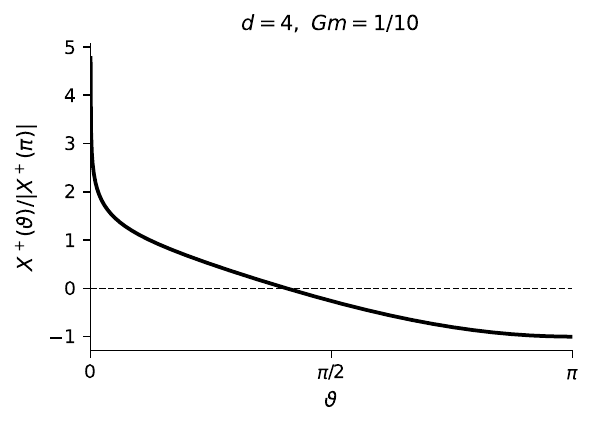}
\end{minipage}
\end{center}

\subsection{Recoil limit}

On the $(d-2)$ sphere, the Laplacian has eigenvalues $\ell(\ell+d-3)$, so when the observer mass vanishes $m=0$, the $\ell=1$ modes are ``recoil'' zero modes. For small but nonzero $m$, we find that the shock wave profile (\ref{X+sol}) becomes
\begin{align}
X^+(\vartheta)
&=
-\frac{p_-}{r'(0)m}\cos\vartheta+O(m^0).
\end{align}
We would like to explain this in terms of the recoil of a non-backreacting observer in pure de Sitter space. For pure de Sitter, $m=0$, so $r_c=1$, $|f'(r_c)|=2$, and we can choose
\begin{equation}
r_*=\frac{1}{2}\log\frac{1+r}{1-r},
\hspace{20pt}
x^-x^+=-e^{-2r_*}=-\frac{1-r}{1+r},
\hspace{20pt}
\frac{x^+}{x^-}=-e^{2t}.
\end{equation}
Thus
\begin{equation}
r=\frac{1+x^-x^+}{1-x^-x^+}, \hspace{20pt} r'(0) = 2.
\end{equation}
The pure de Sitter metric is
\begin{equation}
\d s^2=-\frac{4}{(1-x^-x^+)^2}\d x^-\d x^+
+\left(\frac{1+x^-x^+}{1-x^-x^+}\right)^2\d\Omega_{d-2}^2.
\end{equation}
It is convenient to use the embedding space coordinates $-(Y^0)^2 + Y^i Y^i = 1$ and the linear combinations $Y^\pm = (Y^0\pm Y^1)/2$. The relevant embedding coordinates are
\begin{equation}
Y^\pm=\frac{x^\pm}{1-x^-x^+},
\hspace{20pt}
Y^a=\frac{1+x^-x^+}{1-x^-x^+}n^a,
\hspace{20pt}
n^a n^a=1,
\end{equation}
where $a=2,\dots,d$ labels the sphere directions. If the recoil is in the $Y^3$ direction, then $Y^3=\frac{1+x^-x^+}{1-x^-x^+}\cos\vartheta$, so on the $x^-=0$ horizon we have $Y^3=\cos\vartheta$ and $\delta(Y^-)(\d Y^-)^2=\delta(x^-)(\d x^-)^2$.

Starting from de Sitter space with a static observer, a moving observer can be obtained from a boost $e^{-v J_{03}}$ in e.g.~the $Y^0,Y^3$ plane, where $J_{03} = Y^3\partial_0 + Y^0\partial_3.$
To derive the shockwave form, we want the recoil event to take place far in the past, so we should conjugate by a large boost (time translation of the static patch) in the $Y^\pm$ plane
\begin{equation}
Y^- \to \lambda Y^-,
\qquad
Y^+ \to \lambda^{-1}Y^+.
\end{equation}
The net effect is a boost by $e^{-\frac{v\lambda}{2}(J_{03} + J_{13})}$, which is a diffeomorphism by the vector field
\begin{equation}
    \xi = -\frac{v\lambda}{2}(J_{03} + J_{13}).
\end{equation}
The metric for the recoil event corresponds applying this diffeomorphism only to the future:
\begin{equation}\label{recoilapply}
    \d s^2_{\rm recoil} = \Theta(-Y^-)\d s^2 + \Theta(Y^-)\left(\d s^2+\mathcal{L}_\xi \d s^2\right)+O(\xi^2).
\end{equation}
This can be simplified by ``undoing'' the diffeomorphism to the future, applying the diffemorphism $-\Theta(Y^-)\xi$ to the entire spacetime (\ref{recoilapply}). The resulting metric looks like the original one on both sides of $Y^- = 0$, but has a shock wave piece coming from the singular part of the Lie derivative of $-\Theta(Y^-)\xi$. To linear order,
\begin{align}
\d s^2_{\rm recoil}
%&= \d s^2-2\delta(Y^-)\,\xi_A\,\d Y^- \d Y^A \\ 
%&= \d s^2 -2v\lambda Y^3\delta(Y^-)\d (Y^-)^2 \\
&= \d s^2 -2v\lambda\cos\vartheta\,\delta(x^-)(\d x^-)^2.
\label{eq: C20}
\end{align}
This matches (\ref{recoillight}) provided $v\lambda = \frac{p_-}{m}$, which is indeed the equation for conservation of momentum in the recoil event.

\section{Single-particle wave functions for de Sitter}

Let us work in empty dS$_d$ with de Sitter radius one, and consider a scalar field $\varphi$ with
\begin{equation}
m_\varphi^2=\frac{(d-1)^2}{4}+\nu^2.
\end{equation}
As a function of the geodesic distance between points $\vartheta$, the propagator on $S^d$ is
\begin{equation}
G_d(\vartheta)
=
\frac{\Gamma(\frac{d-1}{2})}{2^{\frac{d+1}{2}}\pi^{\frac{d+1}{2}}}
\int_0^\infty\d s\,\frac{\cos(\nu s)}{(\cosh s-\cos\vartheta)^{\frac{d-1}{2}}},
\end{equation}
The propagator between two points $\theta_2,\theta_4$ along the observer's worldline can be decomposed along the $x^- = 0$ horizon as
\begin{align}
\langle \varphi(\theta_4)\varphi(\theta_2)\rangle
&= 
2\i\int_{-\infty}^{\infty}\d x^+\int_{S^{d-2}}\d\Omega\,
\langle \varphi(\theta_4)\varphi(x^+,\Omega)\rangle\partial_+\langle \varphi(x^+,\Omega)\varphi(\theta_2)\rangle.
\end{align}
The relevant propagators are
\begin{equation}
\begin{aligned}
\langle \varphi(x^+,\Omega)\varphi(\theta_2)\rangle=G_d(\vartheta_2),\hspace{35pt}\cos\vartheta_2&=x^+e^{\i\theta_2},
\\
\langle \varphi(\theta_4)\varphi(x^+,\Omega)\rangle=G_d(\vartheta_4),\hspace{35pt}\cos\vartheta_4&=x^+e^{\i\theta_4}.
\end{aligned}
\end{equation}
This can be written in momentum space for $p_+$ as
\begin{equation}
\langle \varphi(\theta_4)\varphi(\theta_2)\rangle
=
2\int_{-\infty}^{0}\frac{(-p_+)\d p_+}{2\pi}\int_{S^{d-2}}\d\Omega\,
\Psi_4(\theta_4,-p_+)\Psi_2(\theta_2,p_+).
\end{equation}
The $\Psi_2$ wave function is
\begin{align}
\Psi_2(\theta_2,p_+)
&=
\int\d x^+e^{-\i x^+p_+}\langle \varphi(x^+,\Omega)\varphi(\theta_2)\rangle
\notag\\
&=
\frac{\Gamma(\frac{d-1}{2})}{2^{\frac{d+1}{2}}\pi^{\frac{d+1}{2}}}\int_0^\infty\d s\,\cos(\nu s)\int_{-\infty}^{\infty}\d x^+\frac{e^{-\i x^+p_+}}{(\cosh s-x^+e^{\i\theta_2})^{\frac{d-1}{2}}}
\notag\\
&=
\theta(-p_+)\frac{2\pi(-\i e^{-\i\theta_2})^{\frac{d-1}{2}}}{2^{\frac{d+1}{2}}\pi^{\frac{d+1}{2}}}\left(-p_+\right)^{\frac{d-3}{2}}\int_0^\infty\d s\,\cos(\nu s)e^{-\i e^{-\i\theta_2}p_+\cosh s}
\notag\\
&=
\theta(-p_+)\frac{(-\i e^{-\i\theta_2})^{\frac{d-1}{2}}}{2\pi}\left(\frac{-p_+}{2\pi}\right)^{\frac{d-3}{2}}K_{\i\nu}(\i e^{-\i\theta_2}p_+).
\end{align}
Similarly,
\begin{equation}
\Psi_4(\theta_4,-p_+)
=
\theta(-p_+)\frac{(\i e^{-\i\theta_4})^{\frac{d-1}{2}}}{2\pi}\left(\frac{-p_+}{2\pi}\right)^{\frac{d-3}{2}}K_{\i\nu}(-\i e^{-\i\theta_4}p_+).
\end{equation}

\section{Pure recoil eikonal OTOCs in higher dimensions}
\label{app:recoil in higher d}
In this section we compute the pure-recoil limit of the eikonal OTOCs in $d$ dimensions, following the approach of Kolchmeyer and Liu~\cite{Kolchmeyer:2024fly}.

As a preliminary step, we decompose the Wightman function in the flat slicing into comoving momentum mode functions. In the future Poincar\'e patch,
\begin{equation}
\d s^2=-\d t^2+e^{2t}\d\vec y^{\,2},\ \ \ \ \vec y\in\mathbb R^{d-1}
\end{equation}
\begin{equation}
G_\nu^{s}(1,2)=\int\frac
{\d^{d-1}\vec k}{(2\pi)^{d-1}}e^{\i\vec k\cdot(\vec y_1-\vec y_2)}\cdot \frac{1}{\pi}e^{-(\frac{d-1}{2})(t_1+t_2)}K_{\i\nu}(-\i s |\vec k|e^{-t_1}+0^+)K_{\i\nu}(\i s|\vec k|e^{-t_2}+0^+)
\label{eq: mode function, future}
\end{equation}
where $s\in\{\pm\}$ labels the operator ordering,
$G_\nu^+(1,2)=\lr{0|\varphi(1)\varphi(2)|0}$ and $G_\nu^-(1,2)=\lr{0|\varphi(2)\varphi(1)|0}$, with $|0\rangle$ the Bunch--Davies vacuum. The $0^+$ is included since Bessel function $K_{\i\nu}(z)$ has a branch cut on negative real axis.  Similarly, in the past Poincar\'e patch,
\begin{equation}
\d s^2=-\d t^2+e^{-2t}\d\vec y^{\,2},\ \ \ \ \vec y\in\mathbb R^{d-1}
\end{equation}
\begin{equation}
G_\nu^{s}(1,2)=\int\frac
{\d^{d-1}\vec k}{(2\pi)^{d-1}}e^{\i\vec k\cdot (\vec y_1-\vec y_2)}\cdot \frac{1}{\pi}e^{(\frac{d-1}{2})(t_1+t_2)}K_{\i\nu}(\i s|\vec k|e^{t_1}+0^+)K_{\i\nu}(-\i s|\vec k|e^{t_2}+0^+)
\label{eq: mode function, past}
\end{equation}

We now evaluate the QFT OTOC dressed by a heavy observer worldline with action $S_\text{worldline}$, in the configuration
\begin{align}
\begin{tikzpicture}[scale=3.5, baseline=(current bounding box.center)]
\coordinate (T1) at (0,0);
\coordinate (T2) at (0,1);
\coordinate (T3) at (1,1);
\coordinate (T4) at (1,0);
\coordinate (T5) at (0.25,1);
\coordinate (T6) at (0.15,0);
\draw[line width=1pt, draw=gray!50] (T1) -- (T2) -- (T3) -- (T4) -- cycle;
\draw[line width=1pt, draw=gray!50, dashed] (0,0) -- (1,1);
\draw[line width=1pt, draw=gray!50, dashed] (1,0) -- (0,1);
\draw[line width=1pt, draw=black, line cap=round] (T1) -- (T2);
\draw[line width=1pt, draw=black, line cap=round] (T1) .. controls +(90:0.4) and +(270:0.4) .. (T5);
\draw[line width=1pt, draw=black, line cap=round] (T5) .. controls +(270:0.4) and +(90:0.4) .. (T6);
\draw[line width=1pt, draw=black, line cap=round] (T6) .. controls +(90:0.4) and +(270:0.4) .. (T2);
\draw[line width=1pt, draw=blue] (T1)--(T6);
\draw[line width=1pt, draw=black, ->] ([xshift=0.5, yshift=-2]T1) -- node[midway, below] {$\vec y_-$} ([xshift=-0.3, yshift=-2]T6);
\draw[line width=1pt, draw=red] (T2)--(T5);
\draw[line width=1pt, draw=black, ->] ([xshift=0.5, yshift=2]T2) -- node[midway, above] {$\vec y_+$} ([xshift=-0.3, yshift=2]T5);
\fill[blue] (T1) circle (0.02) node[left] {$W_2$};
\fill[blue] (T6) circle (0.02) node[right] {$W_4$};
\fill[red] (T2) circle (0.02) node[left] {$V_1$};
\fill[red] (T5) circle (0.02) node[right] {$V_3$};
\end{tikzpicture}
\simeq\int \d^{d-1}\vec y_+\d^{d-1}\vec y_- e^{\i S_\text{worldline}}G^{s}_{\nu}(2,4)G^{s'}_{\nu'}(1,3)
\label{eq: recoil F in higher dim}
\end{align}
We begin with an integral over four insertion points on dS$_d$. After gauge-fixing we fix the location of $W_2$. In the heavy-worldline limit we may also fix $V_1$ and integrate over the transverse fluctuations $\vec y_{\pm}$ of the remaining insertions $V_3$ and $W_4$; these fluctuations encode recoil. In the eikonal regime $\vec y_{\pm}$ is small, and the worldline action (minus the mass times the sum of future-directed proper times of the four black segments in the figure) reduces to
\begin{equation}
S_\text{worldline}=-m\int \d\tau\approx2 m (\vec y_+\cdot \vec y_+)
\end{equation}
Substituting \eqref{eq: mode function, future} and \eqref{eq: mode function, past} into \eqref{eq: recoil F in higher dim}, the $\vec y_{\pm}$ integrals become Gaussian and can be performed explicitly.

The operator-ordering choices $s,s'$ in \eqref{eq: recoil F in higher dim} are fixed by the placement of the projector $\Pi_m\otimes|0\rangle\langle0|$ (depicted by the triangles in below) in the trace over $\mathcal{H}_\text{obs}\otimes\mathcal{H}_\text{QFT}$, which enforces the presence of a semiclassical observer. By invariance under a common shift of all $\theta_i$, the final answer depends only on the product $ss'$:
\begin{align}
ss'=1:\ \ \ \ \eqref{eq: recoil F in higher dim}=\mathcal{F}_{\text{recoil},12}(\theta_1,\theta_2,\theta_3,\theta_4|m)=\begin{tikzpicture}[scale=1, baseline=(current bounding box.center)]
% Main circle
\draw[thick] (0,0) circle (1);
% Blue line
\draw[blue] (160:1) -- (-20:1);
% Red lines
\draw[red] (200:1) -- (200:.15);
\draw[red] (20:1) -- (20:.15);
% Points and labels
\fill[blue] (160:1) circle (0.08) node[left] {$\theta_4$};
\fill[blue] (-20:1) circle (0.08) node[right] {$\theta_2$};
\fill[red] (20:1) circle (0.08) node[right] {$\theta_3$};
\fill[red] (200:1) circle (0.08) node[left] {$\theta_1$};
\node at (-35:.7) {${\color{blue}\nu}$};
\node at (33:.7) {${\color{red}\nu'}$};
% =====================================================
% Triangle drawn on top
% Modify the following three coordinates to move vertices
% =====================================================
\coordinate (T1) at (1+0.08,-0.15);  % first vertex of triangle
\coordinate (T2) at (1-0.08,-0.15);   % second vertex of triangle
\coordinate (T3) at (1,0);      % third vertex of triangle
\draw[thick, draw=black, fill=white] (T1) -- (T2) -- (T3) -- cycle;
% =====================================================
\coordinate (T4) at (1+0.08,0.15);  % first vertex of triangle
\coordinate (T5) at (1-0.08,0.15);   % second vertex of triangle
\coordinate (T6) at (1,0);      % third vertex of triangle
\draw[thick, draw=black, fill=white] (T4) -- (T5) -- (T6) -- cycle;
\end{tikzpicture}=
\begin{tikzpicture}[scale=1, baseline=(current bounding box.center)]
% Main circle
\draw[thick] (0,0) circle (1);
% Blue line
\draw[blue] (160:1) -- (-20:1);
% Red lines
\draw[red] (200:1) -- (200:.15);
\draw[red] (20:1) -- (20:.15);
% Points and labels
\fill[blue] (160:1) circle (0.08) node[left] {$\theta_4$};
\fill[blue] (-20:1) circle (0.08) node[right] {$\theta_2$};
\fill[red] (20:1) circle (0.08) node[right] {$\theta_3$};
\fill[red] (200:1) circle (0.08) node[left] {$\theta_1$};
\node at (-35:.7) {${\color{blue}\nu}$};
\node at (33:.7) {${\color{red}\nu'}$};
% =====================================================
% Triangle drawn on top
% Modify the following three coordinates to move vertices
% =====================================================
\coordinate (T1) at (-1-0.08,0.15);  % first vertex of triangle
\coordinate (T2) at (-1+0.08,0.15);   % second vertex of triangle
\coordinate (T3) at (-1,0);      % third vertex of triangle
\draw[thick, draw=black, fill=white] (T1) -- (T2) -- (T3) -- cycle;
% =====================================================
\coordinate (T4) at (-1-0.08,-0.15);  % first vertex of triangle
\coordinate (T5) at (-1+0.08,-0.15);   % second vertex of triangle
\coordinate (T6) at (-1,0);      % third vertex of triangle
\draw[thick, draw=black, fill=white] (T4) -- (T5) -- (T6) -- cycle;
\end{tikzpicture}
\end{align}
\begin{align}
ss'=-1:\ \ \ \eqref{eq: recoil F in higher dim}=\mathcal{F}_{\text{recoil},14}(\theta_1,\theta_2,\theta_3,\theta_4|m)=
\begin{tikzpicture}[scale=1, baseline=(current bounding box.center)]
% Main circle
\draw[thick] (0,0) circle (1);
% Blue line
\draw[blue] (160:1) -- (-20:1);
% Red lines
\draw[red] (200:1) -- (200:.15);
\draw[red] (20:1) -- (20:.15);
% Points and labels
\fill[blue] (160:1) circle (0.08) node[left] {$\theta_4$};
\fill[blue] (-20:1) circle (0.08) node[right] {$\theta_2$};
\fill[red] (20:1) circle (0.08) node[right] {$\theta_3$};
\fill[red] (200:1) circle (0.08) node[left] {$\theta_1$};
\node at (-35:.7) {${\color{blue}\nu}$};
\node at (33:.7) {${\color{red}\nu'}$};
% =====================================================
% Triangle drawn on top
% Modify the following three coordinates to move vertices
% =====================================================
\coordinate (T1) at (-0.15,1+0.08);  % first vertex of triangle
\coordinate (T2) at (-0.15,1-0.08);   % second vertex of triangle
\coordinate (T3) at (0,1);      % third vertex of triangle
\draw[thick, draw=black, fill=white] (T1) -- (T2) -- (T3) -- cycle;
% =====================================================
\coordinate (T4) at (0.15,1+0.08);  % first vertex of triangle
\coordinate (T5) at (0.15,1-0.08);   % second vertex of triangle
\coordinate (T6) at (0,1);      % third vertex of triangle
\draw[thick, draw=black, fill=white] (T4) -- (T5) -- (T6) -- cycle;
\end{tikzpicture}=
\begin{tikzpicture}[scale=1, baseline=(current bounding box.center)]
% Main circle
\draw[thick] (0,0) circle (1);
% Blue line
\draw[blue] (160:1) -- (-20:1);
% Red lines
\draw[red] (200:1) -- (200:.15);
\draw[red] (20:1) -- (20:.15);
% Points and labels
\fill[blue] (160:1) circle (0.08) node[left] {$\theta_4$};
\fill[blue] (-20:1) circle (0.08) node[right] {$\theta_2$};
\fill[red] (20:1) circle (0.08) node[right] {$\theta_3$};
\fill[red] (200:1) circle (0.08) node[left] {$\theta_1$};
\node at (-35:.7) {${\color{blue}\nu}$};
\node at (33:.7) {${\color{red}\nu'}$};
% =====================================================
% Triangle drawn on top
% Modify the following three coordinates to move vertices
% =====================================================
\coordinate (T1) at (0.15,-1-0.08);  % first vertex of triangle
\coordinate (T2) at (0.15,-1+0.08);   % second vertex of triangle
\coordinate (T3) at (0,-1);      % third vertex of triangle
\draw[thick, draw=black, fill=white] (T1) -- (T2) -- (T3) -- cycle;
% =====================================================
\coordinate (T4) at (-0.15,-1-0.08);  % first vertex of triangle
\coordinate (T5) at (-0.15,-1+0.08);   % second vertex of triangle
\coordinate (T6) at (0,-1);      % third vertex of triangle
\draw[thick, draw=black, fill=white] (T4) -- (T5) -- (T6) -- cycle;
\end{tikzpicture}
\end{align}
As in \eqref{summing}, each contribution splits into $t<0$ and $t>0$ pieces. The resulting expressions are
\begin{equation}
\begin{aligned}
\mathcal{F}^{t<0}_{\text{recoil}}(\theta_1,\theta_2,\theta_3,\theta_4|m)=C_0& e^{-\i\left(\frac{d-1}{2}\right)(\theta_2+\theta_4)}\int_{-\infty}^0p_+^{d-2}\d p_+K_{\i\nu}(\i sp_+e^{-\i\theta_4})K_{\i\nu}(-\i sp_+e^{-\i\theta_2})\\
\times&e^{\i\left(\frac{d-1}{2}\right)(\theta_1+\theta_3)}\int_{-\infty}^0p_-^{d-2}\d p_-K_{\i\nu'}(-\i s'p_-e^{\i\theta_3})K_{\i\nu'}(\i s'p_-e^{\i\theta_1})\\
\times&\operatorname{vol}(S^{d-3})\int_0^{\pi}(\sin\phi)^{d-3}\d\phi \cdot \exp\left(\i p_+p_-\frac{\cos\phi}{2m}\right)
\end{aligned}
\end{equation}
\begin{equation}
\begin{aligned}
\mathcal{F}^{t>0}_{\text{recoil}}(\theta_1,\theta_2,\theta_3,\theta_4|m)=C_0& e^{\i\left(\frac{d-1}{2}\right)(\theta_2+\theta_4)}\int_{-\infty}^0p_+^{d-2}\d p_+K_{\i\nu}(-\i sp_+e^{\i\theta_4})K_{\i\nu}(\i sp_+e^{\i\theta_2})\\
\times&e^{-\i\left(\frac{d-1}{2}\right)(\theta_1+\theta_3)}\int_{-\infty}^0p_-^{d-2}\d p_-K_{\i\nu'}(\i s'p_-e^{-\i\theta_3})K_{\i\nu'}(-\i s'p_-e^{-\i\theta_1})\\
\times&\operatorname{vol}(S^{d-3})\int_0^{\pi}(\sin\phi)^{d-3}\d\phi \cdot \exp\left(\i p_+p_-\frac{\cos\phi}{2m}\right)
\end{aligned}
\label{eq: E10}
\end{equation}
where $C_0$ is a  constant independent of $m,\theta_i$. Here, $\phi$ is relative angle between comoving momentum $\vec k_{\pm}$ and $-p_\pm$ is the modulus. For $d=3$, in the $G\to0$ limit and with the corresponding choice of $s,s'$, this reproduces $\mathcal{F}^{t\lessgtr 0}_{12}, \mathcal{F}^{t\lessgtr 0}_{14}$.

From the last line of \eqref{eq: E10}, we can read out $h(\phi)=(\cos\phi)/(2m)$, defined as the coefficient in front of $\i p_+p_-$ in the Eikonal phase. It matches the geometry computation in \eqref{eq: C20}, where $\phi$ should be identified as $\vartheta$ there. By Taylor expand $\exp(\i p_+p_-h(\phi))$, the perturbation series will only contain even powers of $e^t/m$, consistent with $d=3$ case. 

\section{Tracial OTOC}\label{app: tracial}
\begin{figure}[H]
\centering
\begin{minipage}[t]{0.49\textwidth}
\centering
{\small dS\(_3\)}\par\smallskip
\includegraphics[width=\linewidth]{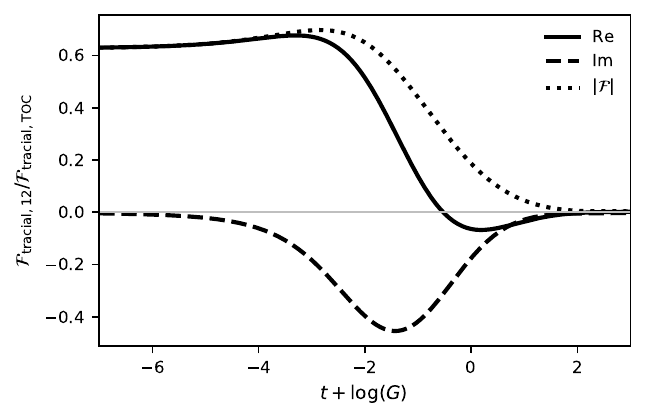}
\end{minipage}
\hfill
\begin{minipage}[t]{0.49\textwidth}
\centering
{\small dS JT}\par\smallskip
\includegraphics[width=\linewidth]{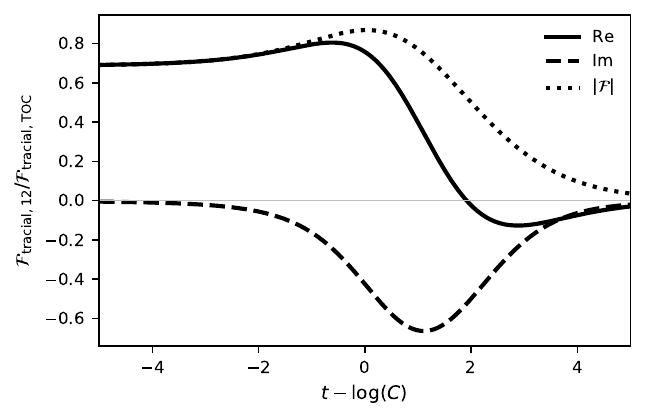}
\end{minipage}
\caption{Here we plot $\bm{\mathcal{F}}_{\text{tracial}}(t,0,t,0)$ normalized by the analogous ``tracial" time ordered four point function formula by inserting the TOC in the $t'$ integral \eqref{eqn:tracialOTOC}. An interesting feature of the plot is that $\bm{\mathcal{F}}_{\text{tracial}}$ has initial growth in time, but it is always bounded by the TOC. At early time, the difference between TOC and OTOC comes from the constraint of the mass $m\geq m_0$, which causes operators fail to commute. If the initial state of the observer is highly excited, this difference becomes exponentially small. At left we have the dS$_3$ case with $\nu = \nu' = 1$ and $\alpha = 3\pi/2$. At right we have the dS JT case with $\Delta = \Delta' = 1$.}
\label{fig:tracial}
\end{figure}

In this paper, we mainly considered an OTOC with Boltzmann factors included, and a constraint on the average mass. But from the perspective of \cite{Chandrasekaran:2022cip,Witten:2023xze}, the more natural quantity is the infinite temperature or tracial version of the OTOC. To compute this, we can generalize the discussion of the two point function in \eqref{Git1stline}:
\begin{align}\label{eqn:tracialOTOC}
    \bm{\mathcal{F}}_{\text{tracial}}(t_1,t_2,t_3,t_4)&\equiv\int_{m_0}^\infty \d^4 m_i\,
e^{\i\sum_i m_i(t_i-t_{i-1})}\bm{\mathcal{F}}(m_1,m_2,m_3,m_4)\\
&=Z_{\rm sphere}(m_0)\int \d\omega_i \delta(\sum_i \omega_i)e^{\i \sum_i \omega_i t_i}\int_{m_0-\min_i s_i}\d m e^{-2\pi (m-m_0)}\mathcal{F}(\omega_i|m)\notag\\
&\approx Z_{\rm sphere}(m_0)\int \d\omega_i \delta(\sum_i \omega_i) e^{\i \sum_i \omega_i t_i}
\frac{e^{2\pi\min_i s_i}}{2\pi}\,
\mathcal{F}(\omega_i|m_0)\\
&=\frac{Z_{\rm sphere}(m_0)}{(2\pi)^3}
\int\frac{\d^4t'_i}{\text{vol}(\mathbb{R})}\,
\prod_{i=1}^4
\frac{1}{\frac{\pi}{2}+\i[(t'_i-t_i)-(t'_{i-1}-t_{i-1})]}\\
&\hspace{20pt}\times \mathcal{F}({\color{red}-\tfrac{\pi}{2}+\i t_1'},{\color{blue}\i t_2'}, {\color{red}\tfrac{\pi}{2} + \i t_3'},{\color{blue}\pi+\i t_4'}|m_0).
\end{align}
Here the variables are defined in \eqref{eqn:FtobF} with:
\be
m_i = m + s_i,\hspace{20pt}
s_i = \tfrac{1}{4}(2\omega_i+\omega_{i+1}-\omega_{i-1}).
\ee
The quotient by $\text{vol}(\mathbb{R})$ removes the zero mode associated with the time translation that shifts all the $t'_i$s. As in the case of the two point function, the tracial four point function is smeared out in time. Plugging in $\mathcal{F}_{12}$ or $\mathcal{F}_{14}$ leads to bounded expressions that are complex conjugate to each other, and one could obtain a real tracial OTOC by taking the average. See figure \ref{fig:tracial} for plots with $\mathcal{F}_{12}$. 

\section{The conformally coupled case for dS$_3$}
\begin{figure}[H]
\centering
\begin{minipage}[t]{0.48\textwidth}
\centering
\begin{tikzpicture}[scale=0.85,baseline=(current bounding box.center)]
\def\epsang{5.73}
\path[use as bounding box] (-1.55,-1.15) rectangle (1.55,1.15);
\draw[dotted, thick] (180+\epsang/2:1.5)--(\epsang/2:1.5);
\draw[thick] (0,0) circle (1);
\fill[blue] (1,0) circle (0.08);
\fill[blue] (-1,0) circle (0.08);
\fill[red] (\epsang:1) circle (0.08);
\fill[red] (180+\epsang:1) circle (0.08);
\end{tikzpicture}
\par\vspace{-0.15em}
\includegraphics[width=\linewidth,height=0.55\linewidth]{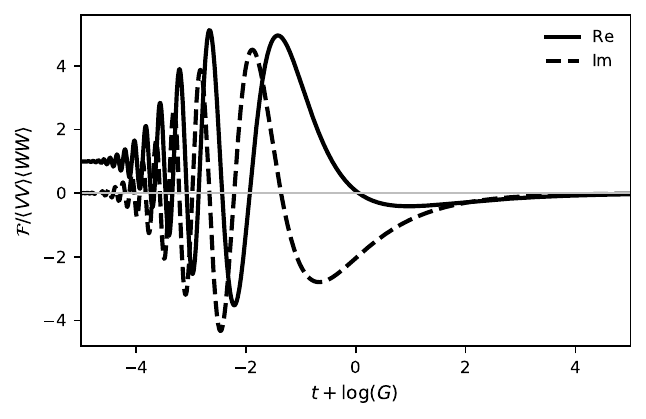}
\end{minipage}
\hfill
\begin{minipage}[t]{0.48\textwidth}
\centering
\begin{tikzpicture}[scale=0.85,baseline=(current bounding box.center)]
\path[use as bounding box] (-1.55,-1.15) rectangle (1.55,1.15);
\draw[thick] (0,0) circle (1);
\draw[dotted, thick] (190:1.5)--(10:1.5);
\fill[blue] (1,0) circle (0.08);
\fill[blue] (-1,0) circle (0.08);
\fill[red] (0.9239,0.3827) circle (0.08);
\fill[red] (-0.9239,-0.3827) circle (0.08);
\end{tikzpicture}
\par\vspace{-0.15em}
\includegraphics[width=\linewidth,height=0.55\linewidth]{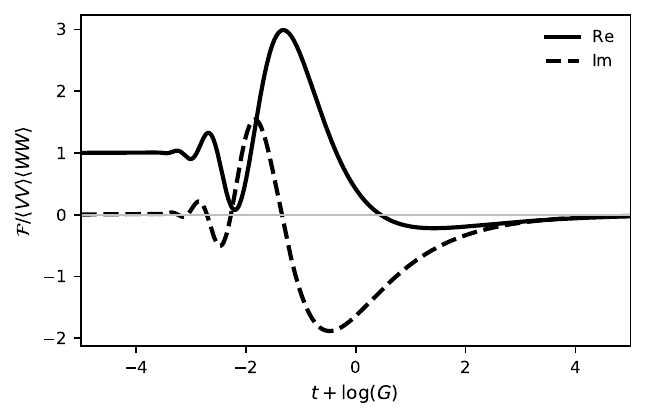}
\end{minipage}

\par\vspace{0.35em}

\begin{minipage}[t]{0.48\textwidth}
\centering
\begin{tikzpicture}[scale=0.85,baseline=(current bounding box.center)]
\path[use as bounding box] (-1.55,-1.15) rectangle (1.55,1.15);
\draw[dotted, thick] (190:1.5)--(10:1.5);
\draw[thick] (0,0) circle (1);
\fill[blue] (1,0) circle (0.08);
\fill[blue] (-1,0) circle (0.08);
\fill[red] (0,1) circle (0.08);
\fill[red] (0,-1) circle (0.08);
\end{tikzpicture}
\par\vspace{-0.15em}
\includegraphics[width=\linewidth,height=0.55\linewidth]{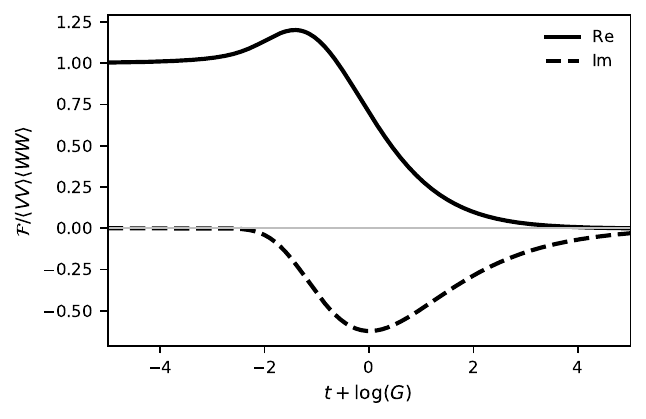}
\end{minipage}
\hfill
\begin{minipage}[t]{0.48\textwidth}
\centering
\begin{tikzpicture}[scale=0.85,baseline=(current bounding box.center)]
\def\epsang{0}
\path[use as bounding box] (-1.55,-1.15) rectangle (1.55,1.15);
\draw[thick] (0,0) circle (1);
\draw[dotted, thick] (190:1.5)--(10:1.5);
\fill[purple] (-\epsang:1) circle (0.08);
\fill[purple] (180-\epsang:1) circle (0.08);
\end{tikzpicture}
\par\vspace{-0.15em}
\includegraphics[width=\linewidth,height=0.55\linewidth]{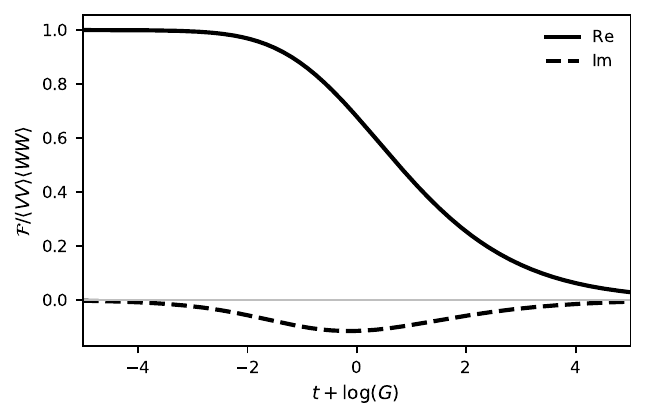}
\end{minipage}

\vspace{-0.25em}
\caption{$\mathcal{F}_{12}(\theta_i|m)$ is plotted for dS$_3$, with $\alpha = 3\pi/2$ and $\nu = \nu' = \i/2$. The peak in the lower left panel is a smoothed out version of the signalling singularity associated to causal contact between the left and right antipodes. For fields of larger mass, this feature becomes more dramatic.}
\label{fig:regularized OTOC}
\end{figure}
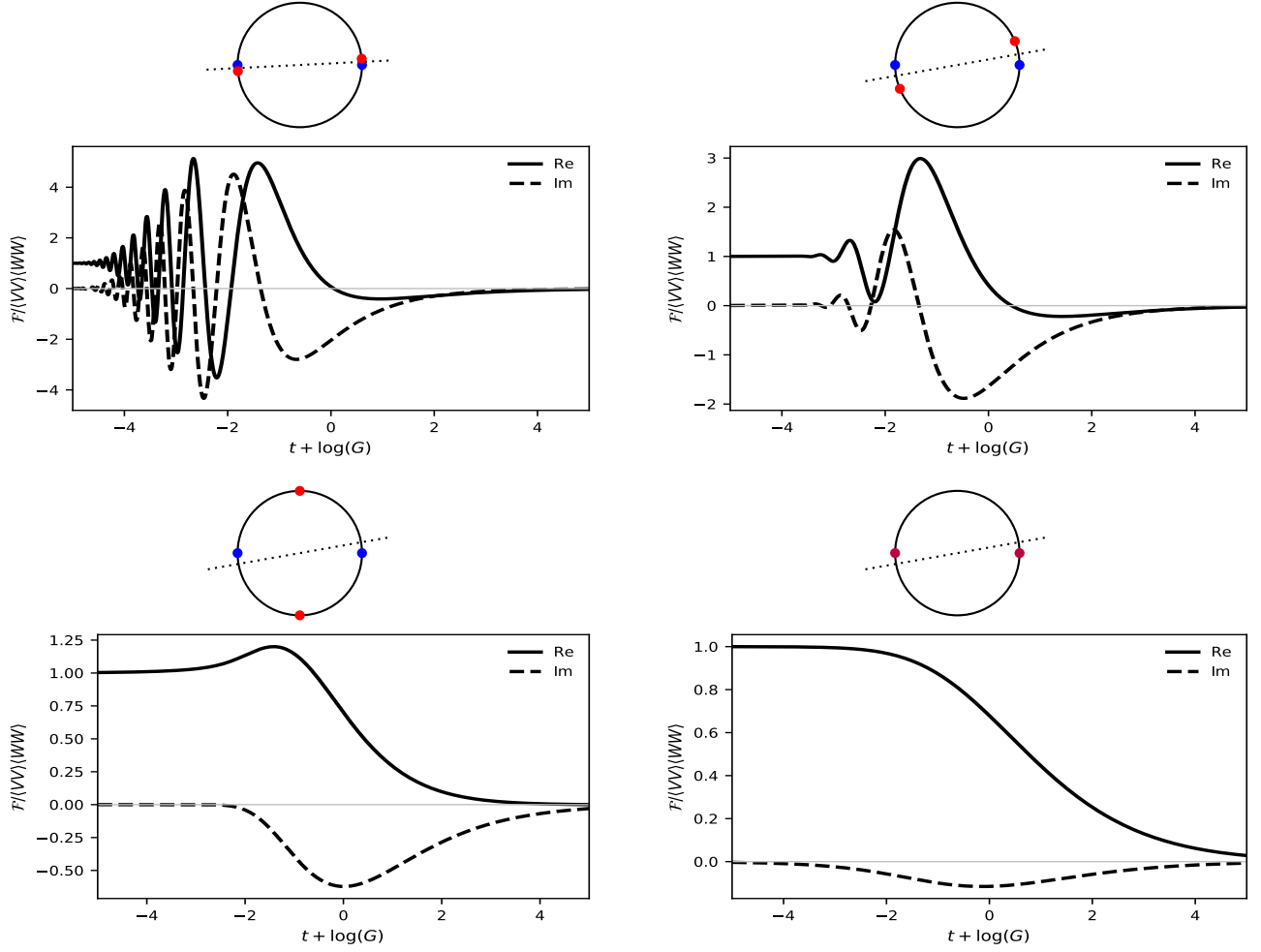

For the case of conformally coupled scalars, $\nu = \i/2$, the wave functions simplify because
\begin{equation}
K_{\i\nu}(z)\Big|_{\nu = \i/2}=\sqrt{\frac{\pi}{2z}}e^{-z}.
\end{equation}
Substituting this into (\ref{eq:F12-time}) gives e.g.
\begin{align}
\mathcal{F}_{12}^{t<0}
&=
\frac{e^{\i\frac{\theta_1+\theta_3-\theta_2-\theta_4}{2}}}
{(2\alpha)^2}
\int_{-\pi}^{\pi}\frac{\d\phi}{2\pi}
\int_{-\infty}^0 \d p_+\,\d p_-\,
\exp\Big\{
\i p_+(e^{-\i\theta_4} - e^{-\i\theta_2})
+ \i p_-(e^{\i\theta_1} - e^{\i\theta_3})
+ \i p_+ p_- h(\phi)
\Big\}\notag\\
&=
\frac{e^{\i\frac{\theta_1+\theta_3-\theta_2-\theta_4}{2}}}
{(2\alpha)^2}
\int_{-\pi}^{\pi}\frac{\d\phi}{2\pi}
\int_{-\infty}^0 \d p_-\,
\frac{
\exp\Big\{
\i p_-(e^{\i\theta_1} - e^{\i\theta_3})
\Big\}}
{\i(e^{-\i\theta_4} - e^{-\i\theta_2} + p_-h(\phi))}.
\label{conformalGeneral:label}
\end{align}
Interestingly, this is an average over $\phi$ of the JT answer with $C\propto h(\phi)$ and $\Delta = 1/2$. In figure \eqref{fig:regularized OTOC}, we plot this function for configurations of the form $\mathcal{F}(-\tfrac{\pi}{2}+\i t + \tau,0,\tfrac{\pi}{2}+\i t + \tau,\pi|m)$ with different $\tau$.

\bibliography{references}

\bibliographystyle{utphys}

\end{document}